\documentclass[journal=jctcce,manuscript=article,layout=traditional]{achemso}
\setkeys{acs}{articletitle = true}

\usepackage{natbib}
\usepackage{array}
\usepackage{stfloats}
\usepackage{graphicx}
\usepackage{color}
\usepackage{dcolumn} % Align table columns on decimal point
\usepackage{bm}      % bold math
\usepackage{amssymb}
\usepackage{amsmath}
\usepackage{amsfonts}
\usepackage{xr-hyper}
\usepackage[colorlinks,citecolor=red]{hyperref}
\usepackage{url}
\usepackage[T1]{fontenc}
\usepackage{multirow}
\usepackage{booktabs}

\makeatletter
\newcommand*{\addFileDependency}[1]{% argument=file name and extension
  \typeout{(#1)}
  \@addtofilelist{#1}
  \IfFileExists{#1}{}{\typeout{No file #1.}}
}
\makeatother

\newcommand*{\myexternaldocument}[1]{%
    \externaldocument{#1}%
    \addFileDependency{#1.tex}%
    \addFileDependency{#1.aux}%
}
\myexternaldocument{supplement}

\usepackage{tikz} %circle numbers
\newcolumntype{L}[1]{>{\raggedright\let\newline\\\arraybackslash\hspace{0pt}}m{#1}}
\newcolumntype{C}[1]{>{\centering\let\newline\\\arraybackslash\hspace{0pt}}m{#1}}
\newcolumntype{R}[1]{>{\raggedleft\let\newline\\\arraybackslash\hspace{0pt}}m{#1}}
\newcolumntype{P}[1]{>{\centering\arraybackslash}p{#1}}
\newcolumntype{M}[1]{>{\centering\arraybackslash}m{#1}}

\include{MyCommand}

\newcommand{\YL}[1]{{\color{black}{#1}}}

\author{Yuan Liu}
\affiliation{Department of Chemistry, Brown University, Providence, RI 02912}
\author{Tong Shen}
\affiliation{Department of Chemistry, Brown University, Providence, RI 02912}
\author{Hang Zhang}
\affiliation{Department of Chemistry, Princeton University, Princeton, NJ 08544}
\author{Brenda Rubenstein}
\email{brenda\_rubenstein@brown.edu}
\affiliation{Department of Chemistry, Brown University, Providence, RI 02912}

\title{Unveiling the Finite Temperature Physics of Hydrogen Chains via Auxiliary Field Quantum Monte Carlo}

\begin{document}

\begin{abstract}
The ability to accurately predict the finite temperature properties and phase diagrams of realistic quantum solids is central to uncovering new phases and engineering materials with novel properties ripe for device applications. Nonetheless, there remain comparatively few many-body techniques capable of elucidating the finite temperature physics of solids from first principles. In this work, we take a significant step towards developing such a technique by generalizing our previous, \YL{\textit{exact}} fully \textit{ab initio} finite temperature Auxiliary Field Quantum Monte Carlo (FT-AFQMC) method to model periodic solids and employing it to uncover the finite temperature physics of periodic hydrogen chains. \YL{Our chains' unit cells consist of 10 hydrogen atoms modeled in a minimal basis and we sample 5 k-points from the first Brillouin zone to arrive at a supercell consisting of 50 orbitals and 50 electrons.} Based upon our calculations of these chains' many-body energies, free energies, entropies, heat capacities, double and natural occupancies, and charge and spin correlation functions, we outline their metal-insulator and magnetic ordering as a function of both H-H bond distance and temperature. At low temperatures approaching the ground state, we observe both metal-insulator and ferromagnetic-antiferromagnetic crossovers at bond lengths between 0.5 and 0.75 \r{A}. We then demonstrate how this low-temperature ordering evolves into a metallic phase with decreasing magnetic order at higher temperatures. In order to contextualize our results, we compare the features we observe to those previously seen in one-dimensional, half-filled Hubbard models at finite temperature and in ground state hydrogen chains. Interestingly, we identify signatures of the Pomeranchuk effect in hydrogen chains for the first time and show that spin and charge excitations that typically arise at distinct temperatures in the Hubbard model are indistinguishably coupled in these systems. Beyond qualitatively revealing the many-body phase behavior of hydrogen chains \YL{in a numerically exact manner without invoking the phaseless approximation}, our efforts shed light on the further theoretical developments that will be required to construct the phase diagrams of the more complex transition metal, lanthanide, and actinide solids of longstanding interest to physicists. 
\end{abstract}

% --------------------------------------------------------------------------------------
\section{Introduction}

One of the central pursuits of modern condensed matter physics has been to predict and explain 
%the 
%often complex array of microscopic interactions that underpin 
%contours of 
material phase diagrams. While phase transitions may be driven by changing a variety of external conditions, some of the most experimentally accessible phase transitions are those that occur upon varying temperature and include thermally-induced structural,\cite{Ramirez_PRB_1971} metal-insulator,\cite{Mott_VO_PRB_1975, Imada_RMP_1998} magnetic,\cite{Lichtenstein_PRL_2001} and superconducting transitions,\cite{Orenstein_Science_2000} as well as involute combinations thereof. Despite decades of intensive research, however, much remains to be understood about such transitions in many materials, 
%very prominently including $d$- and $f$-electron-containing materials such as 
including the transition metal oxides,\cite{Imada_RMP_1998,Georges_AnnRev_2013} lanthanides,\cite{Ramirez_PRB_1971} and actinides.\cite{Zhu_NatComm_2013} Indeed, one of the longest standing questions in all of physics has been the mechanism that underlies the transition to high-temperature superconductivity in the cuprates\cite{Orenstein_Science_2000} and other more recently discovered superconductors, including the pnictides,\cite{Wen_AnnRev_2011} ruthenates,\cite{Maeno_Nature_1994} and sulfur hydrides.\cite{Drozdov_Nature_2015} %The mechanisms that underlie the volume collapse transitions seen in the early lanthanides and actinides have similarly evaded a clear understanding.\cite{McMahan1998} 

While modern electronic structure methods should in principle be able to shed a clarifying light on the mechanisms that underlie these transitions, reliable \textit{first-principles} calculations of materials 
%-- based upon their \textit{ab initio} Hamiltonians -- 
at finite temperatures remain a formidable challenge. Even leaving the proper description of phonons aside,\cite{Louie_PRL_2007} this is in large part because being able to model finite temperature phase transitions not only requires being able to capture thermal effects and handle all of the numerous terms of an \textit{ab initio} Hamiltonian, but also being able to correctly treat all forms of electron correlation and the wide range of energy scales that underlie most material properties. 
%the large eigenvalue spread that accompanies realistic Hamiltonians. 
It is because of these complexities 
%associated with simulating realistic Hamiltonians 
that most finite temperature formalisms, including many diagrammatic\cite{Kozik_2010} and determinant-based\cite{zhang1999finite} quantum Monte Carlo (QMC) techniques, finite temperature Density Matrix Renormalization Group (DMRG) theory,\cite{Wang_PRB_1997} Dynamical Mean Field Theory (DMFT),\cite{Georges_RMP_1996} and the Dynamical Cluster Approximation\cite{Maier_RMP_2005} were initially developed with low-energy effective models, such as the Hubbard or $t$-$J$ models, in mind. Among the most successful of the theories explicitly constructed to accommodate realistic Hamiltonians are embedding theories such as combinations of DMFT and electronic structure techniques like Density Functional Theory (DFT+DMFT) \cite{Savrasov_PRB_2004,Kotliar_RMP_2006} and the GW method (GW+DMFT),\cite{Sun_PRB_2002} finite temperature, second order Green's function theory (FT-GF2),\cite{Walden_JCP_2016,Kananenka_JCTC_2016,Kananenka_JCTC_2016_2} and, most recently, finite temperature Density Matrix Embedding Theory (FT-DMET).\cite{Sun_arXiv_2019} Nevertheless, these theories typically struggle to describe long-range electron correlation that extends beyond the confines of their embedded fragments. 

While the quest for a finite temperature theory for solids may therefore seem treacherous, two key developments make the landscape more hospitable than it initially seems. First and foremost, over the past few decades, monumental strides have been made in the physics and chemistry communities developing \textit{ab initio} theories capable of accurately describing correlation in the ground state. Although many of these theories were originally aimed at describing molecules, in recent years, techniques such as Moller-Plesset Perturbation (MPX) Theory,\cite{Marsman_JCP_2009} Coupled Cluster (CC) Theory,\cite{McClain_JCTC_2017} and Full Configuration Interaction Quantum Monte Carlo\cite{Booth_Nature_2013} have been generalized and successfully applied to \textit{ab initio} solids, tantalizingly opening the door to ``chemically accurate'' descriptions of materials at relatively low electronic temperatures. At the same time, a number of novel finite temperature generalizations of ground state electronic structure theories applicable to \textit{ab initio} Hamiltonians have emerged in recent years, including finite temperature MPX,\cite{Jha_arXiv_2019} \YL{CC,\cite{White_JCTC_2018,hummel2018finite,harsha2019thermofield,white2020finite}} and stochastic algorithms, such as Density Matrix Quantum Monte Carlo,\cite{petras2020using} Path Integral Monte Carlo,\cite{Militzer_PRL_2015} and our own Auxiliary Field Quantum Monte Carlo.\cite{Liu_JCTC_2018,rubenstein2012finite} To date, the majority of these finite temperature theories have been benchmarked against small molecules, but taken together, these advances beg the questions: \textit{How accurately can these finite temperature theories be generalized to solids? And, can they reveal any fundamental insights into the physics of heretofore obscure phase transitions?}   

One enlightening proving ground for beginning to resolve these questions are periodic hydrogen chains. As the first element on the periodic table that can be reasonably represented using a compact basis, one may at first dismiss hydrogen as being too simplistic. Nevertheless, 
%as is known from introductory quantum chemistry courses, 
developing theories capable of accurately modeling the ground state dissociation of even the hydrogen dimer has been a historical challenge because of its multireference character.\cite{szabo1996modern}
%electron-phonon coupling aside.\cite{szabo1996modern} 
Indeed, without corrections, perturbation, and even coupled cluster, theories struggle to capture the correct ground state energy of the dimer at large bond lengths. These struggles are magnified within periodic hydrogen chains and networks, the most simplistic models of solids. In recent years, some of the most powerful electronic structure methods currently available, including Diffusion Monte Carlo,\cite{Stella_PRB_2011} DMRG,\cite{Hachman_2006} and Multireference Configuration Interaction, have been brought to bear on the ground states of varying length one-dimensional hydrogen chains, demonstrating that, despite their seeming simplicity, they exhibit fascinating metal-insulator, ferromagnetic-antiferromagnetic, and dimerization transitions.\cite{motta2019ground,motta2017towards,Sinitskiy_JCP_2010,Stella_PRB_2011,Hachman_2006} Given the complexity of their ground state physics and the limited set of finite temperature methods available, even less is known about these hydrogen chains at finite temperatures. In two seminal studies, Zgid \textit{et al.} employed the temperature-dependent GF2 (FT-GF2) method to show that, at low temperatures, short bond length hydrogen chains exhibit metallic behavior, while long bond length chains exhibit insulating behavior. However, due to the method's intrinsic nonlinearities, FT-GF2 was found to yield two solutions with different, sometimes conflicting properties at intermediate and long bond lengths, thus complicating the ultimate interpretation of these results.\cite{rusakov2016self,welden2016exploring} Beyond this work, most of what is known is based upon analytical derivations \cite{lieb1994absence,schulz1993wigner,gebhard1994charge} and simulations of one-dimensional variants of the Hubbard model, \cite{hirsch1984charge,kuroki1994phase,glocke2007half} which lack the array of long-range Coulomb and hopping interactions that make hydrogen physics unique. The possibility of uncovering thus far unexplored finite temperature hydrogen physics thus makes these chains an alluring target, even beyond their utility for benchmarking. 

In this paper, we extend our previous research developing a finite temperature Auxiliary Field Quantum Monte Carlo (FT-AFQMC) method capable of treating arbitrary \emph{ab initio} Hamiltonians\cite{Liu_JCTC_2018} to the treatment of periodic solids. %QMC methods are naturally capable of sampling the abundance of electronic states that may be populated at finite temperatures and have time-and-again been proven to achieve chemical accuracy both in ground state and finite temperature simulations.\cite{Motta_Review_2018} In a recent work, we adapted the FT-AFQMC formalism\cite{zhang1999finite} to model \emph{ab initio} Hamiltonians and established a set of energy benchmarks for molecules at finite temperature.\cite{Liu_JCTC_2018} Here, we further extend this formalism to its more natural setting of solids by
Using periodic Gaussian-type orbitals (p-GTOs),\cite{Sun_JCP_2017} we employ our algorithm to study metal-insulator and magnetic crossovers in periodic H$_{10}$ chains. 
%We do so by exploiting periodic-Gaussian-type orbitals (p-GTOs)\cite{Sun_JCP_2017} and then employ our algorithm to study metal-insulator and magnetic crossovers in periodic H$_{10}$ chains.
To characterize the charge and magnetic ordering within our chains as a function of temperature and H-H bond distances, we evaluate the chains' free energies, entropies, heat capacities, double and natural occupancies, and spin and charge correlation functions within the FT-AFQMC formalism. Based upon these measures, at temperatures that approach the ground state, we find clear evidence of a crossover from a metallic, ferromagnetic to an insulating, antiferromagnetic phase as the inter-H distance is increased from 0.5 \r{A} to 0.75 \r{A}, in qualitative agreement with recent ground state studies.\cite{motta2019ground} Our calculations moreover illustrate how the chains become increasingly metallic, while losing their magnetic ordering upon increasing the temperature. Interestingly, unlike in studies of 1D Hubbard models which manifest two peaks, we observe only one peak in these chains' heat capacities as a function of temperature, suggesting that realistic, off-diagonal interactions couple charge and spin fluctuations together in a manner not observed in models with only short-range interactions. We moreover find the first evidence of the Pomeranchuk effect in a realistic one-dimensional solid. Our work therefore takes a significant step toward the fully \emph{ab initio} modeling of the thermodynamic phase transitions of solids while also exemplifying the distinctive physics such \emph{ab initio} modeling reveals. 

We begin describing our findings with a discussion of our methodology, including how we integrate periodic Gaussians into our FT-AFQMC formalism and compute a wide variety of many-body observables in Section \ref{sec:methods}. We then present our results for 
%the energies, free energies, entropies, heat capacities, and charge and spin correlation functions of hydrogen 
chains of varying lengths as a function of temperature in Section \ref{sec:results}. In order to contextualize our findings,
%the phase diagram that emerges from our results while also highlighting the unique physics that arises from treating \emph{ab initio} Hamiltonians, 
we contrast our results with long-standing results for one-dimensional Hubbard models as well as more recent simulations of ground state hydrogen chains. Lastly, we conclude with a discussion of future innovations that will improve the accuracy of our current results and extend the applicability of our formalism to larger, more complex multidimensional solids in Section \ref{sec:conclusions}.

\section{\label{sec:methods}Method: FT-AFQMC in a Periodic Gaussian Basis}

\subsection{\emph{Ab Initio} Finite Temperature AFQMC}
In this section, we summarize the salient points of the \emph{ab initio} FT-AFQMC algorithm; for a more comprehensive description, please refer to our previous publication on the topic.\cite{Liu_JCTC_2018}

In equilibrium quantum statistical mechanics, the key quantity of interest from which all other quantities derive is the partition function. As is traditional in many finite temperature formalisms, in FT-AFQMC,\cite{white1988algorithm,zhang1999finite} focus is placed on sampling the grand canonical partition function 
\begin{equation}
\Xi \equiv Tr(e^{-\beta(\hat{H}-\mu\hat{N})}), \label{partition_function_def}
\end{equation}
where $\hat{H}$ is the system's Hamiltonian, $\mu$ is the chemical potential, $\hat{N}$ is the electron number operator, and $\beta=1/k_{B}T$ is the inverse temperature. The most straightforward way of evaluating this partition function is to evaluate the trace by explicitly summing over all states in Fock space, as is done in exact diagonalization. Since the size of Fock space grows exponentially with the number of basis functions employed, however, this approach very rapidly becomes intractable for all but the smallest of toy systems.  
 
In order to circumvent this exponential blockade
%, rather than explicitly evaluating the grand canonical partition function, 
FT-AFQMC samples the grand canonical partition function.\cite{zhang1999finite,white1988algorithm} To do so, the partition function 
%at inverse temperature $\beta$ 
is first discretized into $L$ smaller imaginary time pieces, or propagators 
\begin{equation}
Tr(e^{-\beta(\hat{H}-\mu\hat{N})}) = Tr\Big(\lim_{\triangle\tau\to0}\prod_{l=1}^{L}e^{-\triangle\tau(\hat{H}-\mu\hat{N})}\Big), 
\label{PartitionFunction}
\end{equation}
with $\Delta \tau = \beta/L$ and where $l$ labels the imaginary time slice. In order to further simplify these short-time propagators, the Hamiltonian is broken into one-, $\hat{H}_{1}$, and two-body, $\hat{H}_{2}$, pieces (note that, while the chemical potential term may be combined into the one-body piece, we separate it in the following exposition). It can be proven that any \emph{ab initio} Hamiltonian can be recast as\cite{Motta_Review_2018, Liu_JCTC_2018}
\begin{align}
    \hat{H} = \hat{H}_1 + \hat{H}_{2}  = \hat{H}_{1} + \frac{1}{2}\sum_{\gamma}\lambda_{\gamma}\hat{v}_{\gamma}^2,
\label{Decomposed_Ham} 
\end{align}
where $\hat{v}_{\gamma}$ represents a one-body operator, $\lambda_{\gamma}$ is a complex constant, and $\gamma$ is an index that runs over all such operators. This final expression for the Hamiltonian may then be substituted into Equation \eqref{PartitionFunction}. While the exponentials of the $\hat{H}_{1}$ terms may be directly evaluated because they are one-body in nature,\cite{Hirsch_PRB_1985} the exponentials of the two-body $v_{\gamma}^{2}$ terms must be simplified into exponentials of one-body terms via the continuous Hubbard-Stratonovich transformation\cite{Hirsch_PRB_1983}
\begin{equation}
e^{-\Delta \tau \lambda_{\gamma} \hat{v}_{\gamma}^{2}/2} = \frac{1}{\sqrt{2\pi}}\int_{-\infty}^{\infty} d\phi_{\gamma} ~ e^{-\phi_{\gamma}^{2}/2} e^{\phi_{\gamma} \sqrt{-\lambda_{\gamma} \Delta \tau} \hat{v}_{\gamma}}, 
\label{HS}
\end{equation}
in which $\phi_{\gamma}$ denotes the auxiliary field associated with the $\gamma$-th one-body operator. Combining all of these exponentials together, the propagator at each imaginary time slice may be expressed as 
\begin{align}
e^{-\Delta \tau (\hat{H} - \mu \hat{N})} = \int_{-\infty}^{\infty} d \bm{\phi} ~p(\bm{\phi}) \hat{B}(\bm{\phi}),
\label{short_time_propagator}
\end{align}
where 
\begin{equation}
\hat{B}(\bm{\phi}) = e^{-\frac{\Delta \tau}{2} (\hat{H}_1 - \mu\hat{N})} \left[ \prod_{\gamma} e^{ \phi_{\gamma} \sqrt{-\lambda_{\gamma} \triangle\tau } \hat{v}_{\gamma}} \right] e^{-\frac{\Delta \tau}{2} (\hat{H}_1 - \mu\hat{N})}
\label{Operator_Equation}
\end{equation}
is a one-body propagator that is a function of the multidimensional auxiliary field vector $\bm{\phi}$ and $p(\bm{\phi})$ is a multidimensional normal distribution over $\bm{\phi}$. The one-body propagator can be further expressed as the product of spin up and down contributions as $\hat{B}(\bm{\phi})=\hat{B}_{\uparrow}(\bm{\phi}) \hat{B}_{\downarrow}(\bm{\phi})$ assuming the Hamiltonian does not contain any spin-orbit or relativistic terms that flip spins. The partition function $\Xi$ may thus be written as a trace over one-body \YL{propagators} that are functions of auxiliary fields in a high-dimensional auxiliary field space
\begin{equation}
\Xi \approx Tr\Big[\prod_{l=1}^{L}\int_{-\infty}^{\infty}d\bm{\phi}_{l} ~p(\bm{\phi}_{l})\hat{B}_{\uparrow}(\bm{\phi}_{l}) \hat{B}_{\downarrow}(\bm{\phi}_{l}) \Big].    
\label{partition_function_trace}
\end{equation}
The trace over fermions may be evaluated analytically to yield the final expression for the partition function as an integral over determinants\cite{hirsch1982monte}
\begin{align}
\Xi &\approx \int_{-\infty}^{\infty} d\bm{\phi} ~ p(\bm{\phi}) \text{Det}[I + B_{\uparrow}(\bm{\phi}_{L})...B_{\uparrow}(\bm{\phi}_{2}) B_{\uparrow}(\bm{\phi}_{1})] \nonumber \\
&\times \text{Det}[I + B_{\downarrow}(\bm{\phi}_{L})...B_{\downarrow}(\bm{\phi}_{2}) B_{\downarrow}(\bm{\phi}_{1})],
\label{partition_function}
\end{align}
where $B(\cdot)$ is the corresponding matrix representation of the one-body \YL{propagator} $\hat{B}(\cdot)$. It is only in the limit that $L \to \infty$ (or $\Delta \tau \to 0$) that the exact partition function is recovered. 

In order to evaluate finite temperature observables, such as those described in Section \ref{section:properties}, FT-AFQMC samples the partition function by sampling the fields in Equation \eqref{partition_function}. As sampling any arbitrary set of fields may produce complex-valued determinants that correspond to complex probabilities, a phase problem in which complex probabilities that cancel one another during averaging is likely to emerge.\cite{Zhang_PRL_2003} To mitigate this problem, we first background subtract a reasonable approximant to the full Hamiltonian and recover the remaining portion by sampling auxiliary fields in a step by step and orbital by orbital fashion.\cite{zhang1999finite,Motta_Review_2018} In principle, the subtracted background can be arbitrary and will only affect the variance of the observables while leaving their expectation values unaltered. However, the closer the background subtracted Hamiltonian is to the exact solution, the milder the phase problem will be. Here, a mean field background is subtracted because of the ease with which observables may be evaluated in mean field theory (see the Supplemental Information for further details).   

More specifically, we begin by initializing the weight of each walker (i.e., random sample) to 1 and each determinant to $\text{Det}[I + B_{T}...B_{T} B_{T}]$. At each time slice and orbital, we then sample a new auxiliary field and replace the corresponding trial one-body propagator $\hat{B}_T$ with an updated one-body propagator $\hat{B}(\bm{\phi})$. If we denote the last field sampled at time slice $k$ for orbital $i$ as $\phi_{ik}$, the resulting determinant $M_{ik}^{\alpha}$ will be
\YL{
\begin{equation}
M^{\alpha}_{ik}=Det\Big[I+\Big(B_{\alpha}^{T}\Big)^{L-k} B_{\alpha}(\phi_{ik}...\phi_{1k})...B_{\alpha}(\bm{\phi}_{1})\Big],
\end{equation}
}
where $\alpha$ denotes the spin. As each field is sampled, the walker weight is multiplied by a factor $W(\phi_{ik})$, defined as the ratio of the newly updated determinants to the previous determinants
\begin{equation}
W(\phi_{ik})=\frac{M^{\uparrow}_{ik}M^{\downarrow}_{ik}}{M^{\uparrow}_{(i-1)k}M^{\downarrow}_{(i-1)k}}.
\label{weight}
\end{equation}
Once all fields are sampled, each walker's observables may be computed based upon its final determinant. A weighted average may then be obtained over all walkers. This process of replacing all of the trial density matrices in the initial expression for the determinants may subsequently be repeated until observables of interest are computed with sufficiently small error bars.

The phaseless approximation may additionally be applied on top of this formalism, as is conventionally done in the ground state AFQMC formalism.\cite{Motta_Review_2018} In this work, however, we forego using the phaseless approximation and instead average over the phases that arise so as to eliminate the uncontrolled bias the phaseless approximation introduces. This implies that the results presented here are exact and not biased by the use of trial density matrices with specific charge or magnetic order. 
%density matrices with specific order at the cost of not being able to simulate at the very lowest of temperatures.  

\subsection{Periodic Gaussians}
%\BR{This introduction to periodic gaussians is currently too long, since we didn't personally invent them. Will abbreviate.} 

%\BR{check tense}
In order to capture the periodicity of the hydrogen chains we seek to model, we construct our Hamiltonians using a periodic Gaussian basis.\cite{mcclain2017gaussian} While Gaussian-type orbitals (GTOs) have served as a remarkably successful basis for the construction of localized molecular wave functions in quantum chemistry,\cite{szabo1996modern} the simulation of periodic solids typically calls for the use of a delocalized plane wave basis.\cite{martin2004electronic} \YL{Plane waves, however, often converge painfully slowly with respect to the cutoff energy when localized states are involved.} This convergence may be dramatically accelerated by representing localized core electron states with pseudopotentials,\cite{martin2004electronic} but at the inevitable price of introducing pseudopotential errors. \YL{At the high temperatures we aim for our methodology to model, electrons will be excited from the core states pseudopotentials are designed to replace, meaning that pseudopotential methods will inherently be doomed to fail and one must instead turn to all-electron calculations.}

One particularly promising basis set for reconciling this intrinsic tension between localization and delocalization is the periodic Gaussian basis set. This basis set uses conventional Gaussian basis functions to represent atomic orbitals locally and then translates them periodically to capture the periodicity of the system. Translational-symmetry-adapted periodic Gaussian basis functions (p-GTOs), $\phi_{n\bm{k}}(\bm{r})$, may therefore be generated by translating a single GTO,  $\widetilde{\phi}_{n}(\bm{r})$, according to the formula:\cite{mcclain2017gaussian}     
\begin{equation}
    \phi_{n\bm{k}}(\bm{r}) = \sum_{\bm{R}} e^{i\bm{k}\cdot\bm{R}} \widetilde{\phi}_{n}(\bm{r}-\bm{R}) \equiv e^{i\bm{k}\cdot \bm{r}} u_{n\bm{k}}(\bm{r}). 
\label{GTO_Expression}
\end{equation}
In the above, $n$ denotes the original GTO index, ${\bm{k}}$ represents the translational momentum vector, which can be sampled from the first Brillouin zone, and $\bm{R}$ is a lattice translation vector. According to Bloch's Theorem, p-GTOs may also be expressed as products of plane waves,  $e^{i\bm{k}\cdot\bm{r}}$, and Bloch functions, $u_{n\bm{k}}(\bm{r})$, leading to the second equality in Equation \eqref{GTO_Expression}. 
%In the p-GTO basis, localized states are thus efficiently captured by Gaussians, while delocalized states are efficiently captured by plane waves. 
Based upon this construction, $N_{o}$ GTOs and $N_{k}$ reduced k-points in the first Brillouin zone yield a total of $N_b = N_{o} \times N_{k}$ basis functions, which also defines the dimensionality of the problem (i.e., all propagators will be of dimension $N_{b}^{2}$). 

Because the Bloch function $u_{n\bm{k}}(\bm{r})$ is periodic, each p-GTO can be Fourier transformed by a set of auxiliary plane waves \cite{mcclain2017gaussian}
\begin{align}
    \phi_{n\bm{k}}(\bm{r}) &= \sum_{\bm{G}} \phi_{n\bm{k}}(\bm{G})e^{i(\bm{k}+\bm{G})\cdot\bm{r}} \label{pgto-ft1} \\
    \phi_{n\bm{k}}(\bm{G}) &= \frac{1}{\Omega}\int_{\Omega} d\bm{r}~ \phi_{n\bm{k}}(\bm{r})e^{-i(\bm{k}+\bm{G})\cdot\bm{r}},
    \label{pgto-ft2}
\end{align}
where $\bm{G}$ is a reciprocal lattice vector. These Fourier components are used to calculate the one- and two-electron integrals in the p-GTO basis. Density-fitting is used to improve the efficiency with which the two-electron integrals are represented.\cite{sun2017gaussian} Based upon these one- and two-electron integrals, the second-quantized Hamiltonian for any  solid in the p-GTO basis may be expressed as
\begin{align}
    \hat{H} =& \sum_{m\bm{k_m},n,\bm{k_n},\alpha} T_{m\bm{k_m}\alpha,n\bm{k_n}\alpha} \hat{c}^{\dagger}_{m\bm{k_m}\alpha} \hat{c}_{n\bm{k_n}\alpha} \nonumber \\ 
    +& \frac{1}{2} \sum'_{\substack{p\bm{k_p}q\bm{k_q} \\ m\bm{k_m}n\bm{k_n}\\ \alpha \beta}} V_{p\bm{k_p},q\bm{k_q},m\bm{k_m},n\bm{k_n}}^{\alpha\beta\alpha\beta} \hat{c}^{\dagger}_{p\bm{k_p}\alpha} \hat{c}^{\dagger}_{q\bm{k_q}\beta} \hat{c}_{n\bm{k_n}\beta} \hat{c}_{m\bm{k_m}\alpha}, \label{Hamiltonian} 
\end{align}
where $\hat{c}_{p\bm{k}_p\alpha}^{\dagger}$/$\hat{c}_{p\bm{k}_p\alpha}$ respectively creates/annihilates an electron with GTO index $p$, reduced momentum vector ${\bm{k_{p}}}$, and spin $\alpha$, $T_{m {\bm k_m} \alpha, n {\bm k_n} \alpha}$ denotes a one-body matrix element, and $V_{p {\bm k_{p}}, q {\bm k_{q}}, m {\bm k_{m}}, n {\bm k_{n}}}^{\alpha\beta\alpha\beta}$ denotes a two-body matrix element. Note that the prime on the second summation over two-body matrix elements indicates that the expression must conserve the crystal momentum such that 
\begin{align}
    \bm{k_p}+\bm{k_q}-\bm{k_m}-\bm{k_n} &= \bm{0} ~~(\text{mod }\bm{G}). \label{momentum_conserv}
\end{align}
\YL{Similarly, crystal momentum conservation requires that $\bm{k_m} - \bm{k_n} = \bm{0} ~(\text{mod} ~\bm{G})$ for the one-body term $T_{m {\bm k_m} \alpha, n {\bm k_n} \alpha}$, a condition that reduces to $\bm{k_m} = \bm{k_n}$ because all reduced momenta are within the first Brillouin zone.} Since the p-GTOs are complex orbitals, the Hamiltonian matrix elements are also complex in general. It is this Hamiltonian that is decomposed according to Equation \ref{Decomposed_Ham} in our implementation of FT-AFQMC for solids. 

\subsection{Calculation of Thermodynamic Properties and Correlation Functions}
\label{section:properties} 

Thermodynamic quantities and correlation functions are essential for characterizing finite temperature phases. Computationally accessing these quantities at a many-body level can be challenging. In this section, we describe how these quantities are obtained from our FT-AFQMC simulations. In the case of the grand potential and entropy, we will describe how we obtain their mean field approximants and then discuss the difficulties associated with directly calculating them at a many-body level instead of relying upon expressions based upon the internal energy. 

\subsubsection{Internal Energy}
\label{section:internal_energy}
The central quantity based upon which virtually all other equilibrium thermodynamic quantities can be expressed is the equal-time, one-electron  Green's function, which may be written as
\begin{align}
G_{m\bm{k_m}, n\bm{k_n}}^{\alpha} &=\frac{Tr[\hat{c}_{m\bm{k_m}\alpha}\hat{c}_{n\bm{k_n}\alpha}^{\dagger}\hat{B}_{\alpha}(\bm{\phi}_{L})\hat{B}_{\alpha}(\bm{\phi}_{L-1})...\hat{B}_{\alpha}(\bm{\phi}_{1})]}{Tr[\hat{B}_{\alpha}(\bm{\phi}_{L})\hat{B}_{\alpha}(\bm{\phi}_{L-1})...\hat{B}_{\alpha}(\bm{\phi}_{1})]} \nonumber \\
&=\Big[\frac{I}{I+B_{\alpha}(\bm{\phi}_{L})B_{\alpha}(\bm{\phi}_{L-1})...B_{\alpha}(\bm{\phi}_{1})}\Big]_{m\bm{k_m}, n\bm{k_n}},
\label{GF}
\end{align}
where $\alpha$ and $\bm{k}$ label the spin and reduced momentum of the electron, respectively.\cite{Hirsch_PRB_1985} The one-body density matrix may be obtained from the Green's function according to the expression $D_{m\bm{k_m},n\bm{k_n}}^{\alpha} \equiv \langle \hat{c}_{m\bm{k_m}\alpha}^{\dagger} \hat{c}_{n\bm{k_n}\alpha} \rangle =  \delta_{m\bm{k_m},n\bm{k_n}} - G_{n\bm{k_n},m\bm{k_m}}^{\alpha}$. From Wick's Theorem, the total internal energy, $U$, may then be expressed as
%\begin{widetext}
\begin{align}
U = \big\langle\hat{H}\big\rangle &=\sum_{m\bm{k_m},n\bm{k_n},\alpha}T_{m\bm{k_m}\alpha,n\bm{k_n}\alpha} D_{m\bm{k_m},n\bm{k_n}}^{\alpha}+
\frac{1}{2}\sum_{\alpha\ne\beta}\sum'_{\substack{pqmn \\ \bm{k_p k_q k_m k_n}}} V_{p\bm{k_p},q\bm{k_q},m\bm{k_m},n\bm{k_n}}^{\alpha\beta\alpha\beta} D_{p\bm{k_p}m\bm{k_m}}^{\alpha} D_{q\bm{k_q}n\bm{k_n}}^{\beta} \nonumber \\
&+\frac{1}{2}\sum_{\alpha}\sum'_{\substack{pqmn \\ \bm{k_p k_q k_m k_n}}} V_{p\bm{k_p},q\bm{k_q},m\bm{k_m},n\bm{k_n}}^{\alpha\alpha\alpha\alpha} (D_{p\bm{k_p}m\bm{k_m}}^{\alpha}D_{q\bm{k_q}n\bm{k_n}}^{\alpha} - D_{p\bm{k_p}n\bm{k_n}}^{\alpha} D_{q\bm{k_q}m\bm{k_m}}^{\alpha}). 
\label{internal_energy}
\end{align}
%\end{widetext}

\subsubsection{Heat Capacity}
\label{method_cv}

The heat capacity quantifies how much internal energy is needed to raise the temperature of a system by one unit. Distinct features in the temperature dependence of the heat capacity thus directly correspond to distinct physical processes or phases transitions. One straightforward way of calculating the constant-volume heat capacity, $C_{V}(T)$, is by differentiating the internal energy $U$ with respect to the temperature
\begin{align}
    C_{V}(T) = \frac{dU}{dT}.
    \label{cv_diff}
\end{align}
If the internal energy can be computed at discrete temperatures, as is usual in numerical simulations, the differentiation in Equation \eqref{cv_diff} may be approximated by finite differences.\cite{predescu2003heat} Alternatively, one may first fit a continuous interpolation curve between the discrete internal energy points and then perform analytic differentiation on the interpolation function. In our simulations, we employed the second approach by first obtaining discrete internal energy points across a wide temperature range (from 0.05 to 10 Hartree/$k_B$; unless otherwise noted, we expressed all quantities in terms of atomic units) with a spacing of about 0.1 for temperatures below 1 and 0.5 for temperatures above 1. We subsequently interpolated among the discrete data points using cubic splines and performed analytic differentiation to obtain our $C_{V}(T)$ curves. It should be noted that, because Monte Carlo data possess intrinsic statistical fluctuations, the specific heat curves obtained in this fashion will typically manifest small nonphysical fluctuations whose magnitudes depend on the error bars on the original internal energy points. \YL{We propagated the error bars from the internal energy to the heat capacity as described in Sec. \ref{sec:cv-error} of the Supplemental Information.}

%could remove
%In the path integral formalism, by mapping a quantum system with $d$ spatial dimensions and one imaginary time dimension to a $(d+1)$-dimensional classical system, \cite{hirsch1981efficient} one can also link the heat capacity $C_V$ to fluctuations in the internal energy according to
%\begin{align}
%    C_V = N k_B \beta^2 (\langle U^2 \rangle - \langle U \rangle^2) + \text{corrections}.
%\end{align}
%The path integral expression for $C_{V}$ may therefore be viewed as a sum of the classical fluctuation expression plus quantum corrections.\cite{marx1992calculation,neirotti2000heat} However, for an \textit{ab initio} Hamiltonian with two-body interactions, evaluation of the $\langle U^2 \rangle$ term would require three- and four-body density matrices, which are costly to obtain. We therefore note this approach to inform future explorations, but do not employ it here. 

\subsubsection{The Grand Potential and Entropy}

The grand potential is the characteristic state function for the grand canonical ensemble in which we perform our simulations and may be defined as\cite{fetter2012quantum}
\begin{align}
     \Omega &= -k_B T \text{ln}\Xi. 
     \label{omega_def}
\end{align}
The standard thermodynamic relation\cite{callen1985thermodynamics} for the grand potential may be manipulated to indirectly yield an expression for the entropy 
\begin{align}
    S = \frac{U - \Omega - \mu N}{T},
    \label{entropy_indirect}
\end{align}
where $\mu$ is the chemical potential and $N$ is the total number of electrons. In our FT-AFQMC formalism, by combining Equations \eqref{omega_def} and \eqref{entropy_indirect} with Equation \eqref{partition_function}, both the entropy and the grand potential can, in principle, be written as an average over auxiliary field configurations. However, this would require evaluating the absolute value of the partition function (instead of only weighted observables like the Green's function), \YL{which cannot be accomplished using traditional Monte Carlo sampling techniques. \cite{Malone_PRL_2016}}
% As a result, this approach would work only at high temperature regimes where only very limited physics happens.
Another way to directly calculate the entropy is to start from its von Neumann definition, $S = -\text{Tr}\left[\hat{\rho} \text{ln} \hat{\rho}\right]$, where $\hat{\rho}$ is the density matrix operator. Note that in the above definition, the nonlinear logarithmic operation as well as the multiplicative form of $\hat{\rho} \text{ln} \hat{\rho}$ inevitably leads to contributions from high-order reduced density matrices, \YL{which are difficult to access at the many-body level.\cite{toldin2018entanglement,Blunt_PRB_2014} }
%Progress has recently been made in this direction by calculating the entanglment entropy of an interacting system of fermions within AFQMC by restricting the density matrix operator to a subsystem and then explicitly calculating higher-order contributions from the nonlinear logarithmic operation on the density matrices.\cite{toldin2018entanglement}

Due to the challenges associated with the aforementioned approaches, in this work, we indirectly calculate the many-body entropy by taking the integral over the specific heat
\begin{align}
    S(T) = S(T_0) + \int_{T_0}^{T} \frac{C_{V}(T)}{T} dT,
\end{align}
where $S(T_0)$ is the entropy at $T=T_0$. We choose $T_0$ to be our lowest simulation temperature, 0.05 Hartree/$k_B$, and integrate to obtain $S(T)$ at higher temperatures. In our figures, we align all of our entropy-temperature curves at different bond lengths so that they overlap at high temperatures, at which the entropy is expected to approach $10 \ln(4)$ (in the high temperature limit, there are $4^{10}$ equiprobable states in the grand canonical ensemble, stemming from having 10 spatial orbitals in the supercell, each of which can be occupied by 0, 1 up, 1 down, and 2 electrons = 4 states). 

%\BR{I am confused at the following presentation. Can't you do this for ANY Hamiltonian as long as you diagonalize it? Not just mean field theory? The difficulty lies in the number of states needed for diagonalization...also the assumption of only 1 electron or 0 for each single particle state} \YL{Yes, the difficulty lies in the diagonalization of many states. For the statement of occupancy of only 0 or 1, I meant to illustrate how to take the trace of the mean field one-body Hamiltonian by realizing the eigenvalues of the density operator per site is 0 or 1. I deleted that statement.}

\subsubsection{Double Occupancy}
\label{subsubsec:docc}
In order to characterize the insulating behavior of the system, it is useful to calculate the double occupancy, a measure of how likely two electrons with opposite spins are to occupy the same hydrogen atom on average.  We define the average double occupancy as $\mathcal{D}$, given by
\begin{align}
     \mathcal{D}
     &= \frac{1}{N_{o}} \frac{1}{N_k^2} \sum_{m}\sum'_{\bm{kk'k''k'''}} \langle \hat{c}_{m\bm{k}\uparrow}^{\dagger} \hat{c}_{m\bm{k'}\uparrow} \hat{c}_{m\bm{k''}\downarrow}^{\dagger} \hat{c}_{m\bm{k'''}\downarrow} \rangle,
     \label{double_occupancy}
\end{align}
where the $'$ in the second summation denotes the crystal momentum conservation condition of Equation \eqref{momentum_conserv}. Note that, in order to obtain the per atom occupancy, we have summed over both spatial orbitals and k-point indices in the above. 
%By definition, $\mathcal{D}_{\bm{k}}$ characterizes how likely it is for one spatial orbital (or hydrogen atom, in our system; site in lattice models) to be occupied by two electrons with opposite spins but the same reduced momentum, $\bm{k}$, while $\mathcal{D}$ is an average over $\mathcal{D}_{\bm{k}}$ for all reduced momenta.
According to Wick's Theorem, the double occupancy in Equation \eqref{double_occupancy} may effectively be decoupled into a product of one-body density matrices $\langle \hat{c}_{m\bm{k}\uparrow}^{\dagger} \hat{c}_{m\bm{k'}\uparrow} \hat{c}_{m\bm{k''}\downarrow}^{\dagger} \hat{c}_{m\bm{k'''}\downarrow} \rangle = D_{m\bm{k},m\bm{k'}}^{\uparrow} D_{m\bm{k''},m\bm{k'''}}^{\downarrow} $, where the one-body density matrix $D$ is defined in Section \ref{section:internal_energy}. To acquire physically meaningful results for chains, the double occupancies were calculated in the orthogonal atomic orbital (oAO) basis, instead of the molecular orbital (MO) basis. To accomplish this, we first simulated the system in the MO basis, and performed a basis transformation to obtain the oAO basis density matrices from the MO ones. See the Supplemental Information for further details regarding this transformation. For a half-filled system, it is easy to show that the double occupancy can only range from 0.0 to 0.5. The former value is realized when all sites are occupied by a single electron, as in the case of antiferromagnetic (AFM) order, while the latter is realized when all electrons are paired and occupy half of the total sites, such as in charge density waves. An intermediate value of 0.25 may be realized when each site is occupied by an average of half an up electron and half a down electron, as would occur in the non-interacting or high temperature limit.

\subsubsection{Spin-Spin and Charge-Charge Correlation Functions}
\label{method:corrfun}
To better characterize the magnetic and charge order of the hydrogen chains, we also calculate their equal-time correlation functions
\begin{align}
C_{ss}(i) &= \sum'_{\substack{\bm{kk'} \\ \bm{k''k'''}}} \langle (\hat{n}_{0\bm{kk'}}^{\uparrow} - \hat{n}_{0\bm{kk'}}^{\downarrow})(\hat{n}_{i\bm{k''k'''}}^{\uparrow} - \hat{n}_{i\bm{k''k'''}}^{\downarrow}) \rangle / N_k^2  \label{Css} \\
C_{cc}(i) &= \sum'_{\substack{\bm{kk'}\bm{k''k'''}}} \big[\langle (\hat{n}_{0\bm{kk'}}^{\uparrow} + \hat{n}_{0\bm{kk'}}^{\downarrow})(\hat{n}_{i\bm{k''k'''}}^{\uparrow} + \hat{n}_{i\bm{k''k'''}}^{\downarrow}) \rangle \nonumber \\ 
&- \langle \hat{n}_{0\bm{kk'}}^{\uparrow} +\hat{n}_{0\bm{kk'}}^{\downarrow} \rangle \langle\hat{n}_{i\bm{k''k'''}}^{\uparrow} +\hat{n}_{i\bm{k''k'''}}^{\downarrow} \rangle \big] / N_k^2,  \label{Ccc}
\end{align}
where $C_{ss}(i)$ and $C_{cc}(i)$ are the spin-spin and charge-charge correlation functions between atoms 0 and $i$, respectively. $\hat{n}_{i\bm{kk'}}^{\sigma} = \hat{c}_{i\bm{k}\sigma}^{\dagger} \hat{c}_{i\bm{k'}\sigma}$ is the density operator describing an electron with spin $\sigma$ at atom $i$ scattered from reduced wave vector $\bm{k'}$ to $\bm{k}$. 
%From Wick's Theorem, these correlation functions may be rewritten in terms of one-body density matrices. 
Note that we have subtracted the average charge per atom in Equation \eqref{Ccc} to reduce background charge fluctuations; the analogous spin term averages to 0 and therefore does not need to be explicitly subtracted. As was done for the double occupancies, we also compute these correlation functions in the oAO basis. 

\subsection{Mean Field Thermodynamic Quantities}

In the following, we also present results for the internal energy, entropy, and grand potential at the mean field level for reference. In mean field theory, the Hamiltonian is first approximated as $\hat{H}_{MF}$, which contains only one-body terms (see the Supplemental Information for further details). Let the mean field Hamiltonian's eigenvalues and eigenvectors be denoted by  $\varepsilon_{\bm{k}i\sigma}$ and $\phi_{\bm{k}i\sigma}$, respectively, where $\bm{k}$ represents the reduced momentum in the first Brillouin zone, $i$ is the orbital index which runs from 1 to the total number of orbitals $N_o$, and $\sigma$ labels the spin. Substituting the mean field Hamiltonian $\hat{H}_{MF}$ into Equation \eqref{partition_function_def}, the mean field partition function reduces to 
\begin{align}
    \Xi_{MF} = \bigg(\prod_{\bm{k},\sigma}\prod_{i=1}^{N_o} [1+e^{-\beta (\varepsilon_{\bm{k}i\sigma}-\mu)}]\bigg)^{1/N_k}.
\end{align}
Then, substituting this into Equation \eqref{omega_def} yields the mean field grand potential
\begin{align}
    \Omega_{MF} = - \frac{1}{N_k} k_B T \sum_{\bm{k},
    \sigma} \sum_{i=1}^{N_o} \ln{[1+e^{-\beta (\varepsilon_{\bm{k}i\sigma}-\mu)}]}.
\end{align}
Furthermore, the mean field internal energy may be calculated as a weighted average over all single-particle energies 
\begin{align}
        U_{MF} = \frac{\sum_{\bm{k},\sigma}\sum_{i=1}^{N_o} \varepsilon_{\bm{k}i\sigma} e^{-\beta (\varepsilon_{\bm{k}i\sigma}-\mu)}}{\sum_{\bm{k},\sigma}\sum_{i=1}^{N_o} e^{-\beta (\varepsilon_{\bm{k}i\sigma}-\mu)}}. 
\end{align}
These expressions may finally be combined according to Equation \eqref{entropy_indirect} to obtain the mean field entropy.

\subsection{Simulation Details}

In order to analyze our algorithm's ability to accurately capture the thermodynamics of a realistic solid, we model a 1D periodic hydrogen chain. The unit cell in our simulations contain ten hydrogen atoms, each possessing a single \textit{s}-type orbital (i.e., we work in the STO-6G basis). In order to observe how our chains' thermodynamic properties vary as a function of H-H bond length, we evenly adjust the distances among the hydrogen atoms in our chains from short (0.5 \AA) to long (2.5 \AA) bond lengths and study the temperature dependence of their properties from $T=0.05$ to $T=10$ Hartree/$k_B$ \YL{($k_B$ is taken as 1 throughout this work)}. Altogether, we study nine different chains with H-H bond lengths, $R$, of 0.5, 0.75, 1.0, 1.25, 1.5, 1.75, 2.0, 2.25, and 2.50 \AA. \YL{At each bond length and temperature, we separately tune the chemical potential to guarantee an average occupancy of 10 is reached in the unit cell.} \YL{Because our unit cell exists in 3D space, we also preserve 10 \AA~of vacuum in the lateral directions around our hydrogen chains to avoid any fictitious lateral interactions between adjacent unit cells (please see the Supplementary Information for information regarding the convergence of our energies with respect to the vacuum employed).}
%The convergence should be slower since finite temperature states are more extended in space. However, due to the finite spatial extension of the minimal basis employed in the present work, we expect the finite temperature states also converge similarly as the ground state.)} 
We generate our one- and two-electron integrals from PySCF\cite{sun2018pyscf} using from 2-7 k-points in the first Brillouin zone of our chains \YL{with unrestricted periodic Hartree-Fock orbitals.} PySCF employs a Gaussian and plane-wave mixed density fitting technique to accurately and efficiently evaluate integral matrix elements.\cite{sun2017gaussian} \YL{All of the related integral files and data used to generate this paper's figures can be found in this work's Brown Digital Repository.\cite{BDR-integrals, BDR-figures-data}} 

\section{\label{sec:results} Results and Discussion}

\subsection{Energy Convergence with Respect to k-Points}

\begin{figure}[ht]
\includegraphics[width=0.55\textwidth]{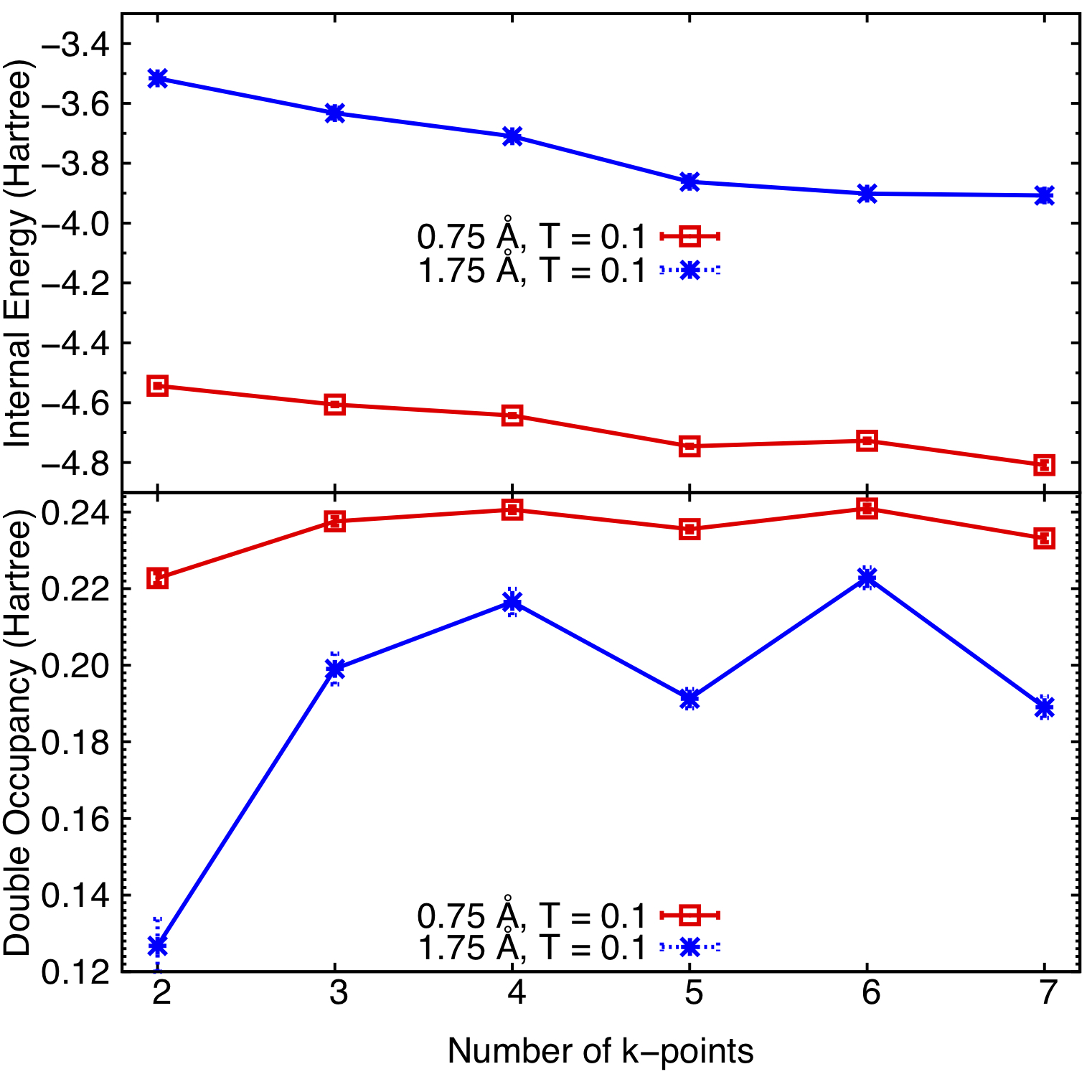}
\caption{Internal energies (upper panel) and double occupancies (lower panel) of hydrogen chains at H-H bond lengths of 0.75 (red empty squares) and 1.75 (blue dot crosses) \r{A} as a function of the number of k-points sampled in the first Brillouin zone. All of the results are calculated at a temperature of $T=0.1$ Hartree/$k_B$. 
%Convergence with respect to the number of k-points was not fully established even at 7 k-points.
}
\label{fig:convergence}
\end{figure}

Before studying the phase behavior of our hydrogen chains, we begin by examining how some of their representative properties converge with respect to the number of k-points sampled. \YL{In Figure \ref{fig:convergence}, we plot the internal energies and double occupancies of hydrogen chains with H-H bond lengths of 0.75 and 1.75~{\AA} at $T=0.1$ Hartree/$k_{B}$ as a function of the number of k-points (see the Supplemental Information for additional plots of the k-point convergence at $T=1.0$ and for odd and even numbers of k-points).} We see that the internal energies decrease as we increase the number of k-points sampled because more k-points stabilize the system by more accurately capturing the interactions among periodic simulation cells. In contrast, the double occupancies oscillate roughly around the same value upon increasing the number of k-points sampled beyond three. Thus, at seven k-points, the largest number of k-points we explored, convergence of the internal energy and the double occupancy is still not rigorously established. Due to computational cost, all of the results in the following sections are obtained with five k-points in the first Brillouin zone, which we believe to be a reasonable number of k-points for drawing physically meaningful conclusions. No special treatment is applied to correct for finite size errors; we therefore do not make any claims about the thermodynamic limit in this \YL{initial}, qualitative study. A discussion of the finite size errors we expect to accompany our current calculations is provided in the Supplemental Information. 

\subsection{\label{sec:effint} Effective Interaction Strengths and Temperatures of the Hydrogen Chains}

In order to analyze our simulation results, it is insightful to first contextualize the effective interaction strengths of our hydrogen chains in terms of 1D Hubbard \cite{hubbard1963electron} and extended Hubbard \cite{ghosh1970electron} model effective interaction strengths. While hydrogen chains obviously possess long-range interactions that span their length and distinguish them from Hubbard models, to develop intuition, we investigate the magnitudes of their on-site and nearest-neighbor inter-site Coulomb repulsions, as well as their hopping matrix elements. 

\begin{figure}[ht]
\includegraphics[width=0.55\textwidth]{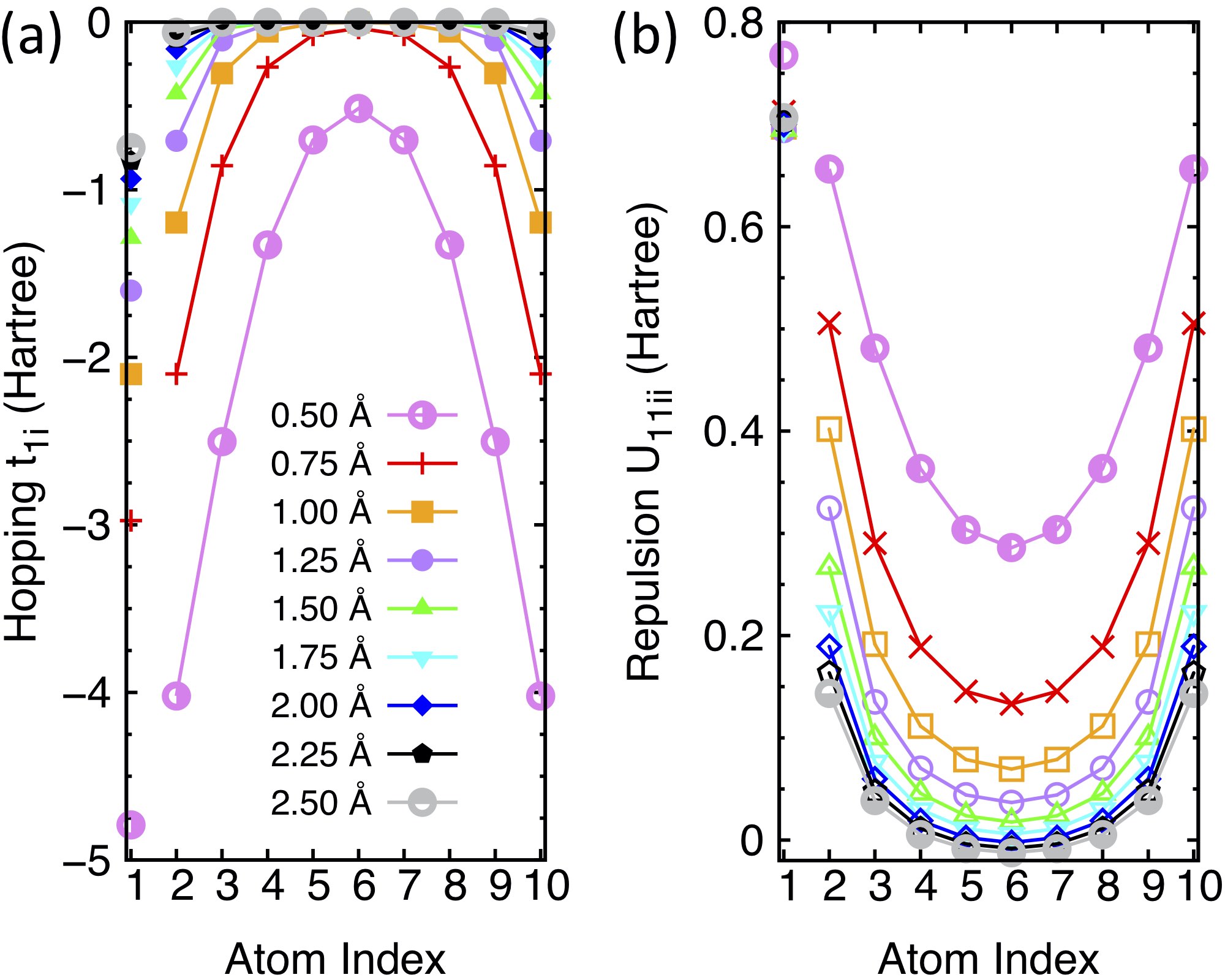}
\caption{Interaction strengths of selected (a) one-body hopping terms, $t_{1i}$, and (b) two-body Coulomb repulsions, $U_{11ii}$, of hydrogen chains for different bond lengths. $t_{1i}$ denotes the hopping matrix element between sites 1 and $i$ at the $\bm{\Gamma}$ point, while $U_{11ii}$ represents the density-density Coulomb interaction between sites 1 and $i$ at the $\bm{\Gamma}$ point. Note that the diagonal terms, $t_{11}$ and $U_{1111}$, are purposefully plotted separately on the left.}
\label{fig:Ut}
\end{figure}

\YL{To do so, in Figure \ref{fig:Ut}, we plot the inter-site (a) one-body hopping matrix elements, $T_{1\bm{\Gamma}\uparrow,i\bm{\Gamma}\uparrow}$ (denoted as $t_{1i}$ in the following) and (b) density-density Coulomb repulsions, $V_{1\bm{\Gamma},i\bm{\Gamma},1\bm{\Gamma},i\bm{\Gamma}}^{\uparrow\uparrow\uparrow\uparrow}$ (denoted as $U_{11ii}$), between atoms 1 and $i$ as a function of atom label $i$ 
%(interatomic distances between atoms can then be obtained based upon the bond lengths) 
at the $\bm{\Gamma}$ point for two up electrons.} From the plot, it is evident that chains with shorter bond lengths have larger magnitude hopping matrix elements and Coulomb repulsions than chains with longer bond lengths, which is consistent with the larger overlap between $1s$ wave functions on more closely spaced hydrogen atoms. Nevertheless, while the magnitudes of the $t_{11}$ matrix elements appreciably change with bond distance, the $U_{1111}$ matrix elements remain roughly constant with bond length because they are largely determined by the shape of the on-site $1s$ wave function, which remains the same irrespective of bond length.
Indeed, at a particular bond length, the inter-site hopping matrix elements decrease roughly exponentially as the distance between two atoms increases, while the inter-site Coulomb repulsions only decay as roughly $1/r$ where $r$ is the distance between two atoms. Note that the smallest magnitude interaction strengths appear between atoms 1 and 6 \YL{at the $\bm{\Gamma}$ point} because all of the atoms beyond atom 6 are actually closer to atom 1 due to the periodic boundary conditions we imposed on our models. 
%\YL{When other k-points beyond the $\bm{\Gamma}$ point are also included, such periodicity should be extended to even longer length scale because calculations done at many k-points for a small unit cell are formally equivalent to single k-points calculations on a much larger unit cell.}

Because the effective interaction strengths of our chains are more physically meaningful than their absolute interaction strengths, we calculate our chains' effective interaction strengths, $U/t$, where $U$ is the Hubbard on-site repulsion and $t$ is the Hubbard hopping parameter, in the following way. First, the effective hopping, $t_{\text{eff}}$, may be defined as the magnitude of the nearest-neighbor hopping amplitude in the hydrogen chain, $t_{\text{eff}} = |t_{\langle i,j\rangle}|$. Second, since the on- and inter-site Coulomb repulsions scale differently with the bond length, we define two different effective Coulomb interaction strengths, $U_{\text{on}}$ and $U_{\text{inter}}$, where the former is taken as the on-site repulsion matrix element, $U_{1111}$, and the latter is obtained by fitting the decay of $U_{11ii}$ in Figure \ref{fig:Ut} to the form $U_{\text{inter}}/|1-i|$. The ratios of $U_{\text{on}}/t_{\text{eff}}$ and $U_{\text{inter}}/t_{\text{eff}}$ are finally evaluated and denoted as $\tilde{U}_{\text{on}}$ and $\tilde{U}_{\text{inter}}$, respectively. The resulting effective interactions are summarized in Table \ref{tab:Ut} for chains of different bond lengths. 

\begin{table}[h]
\centering
\begin{tabular}{C{1.65cm}C{1cm}|C{1cm}C{1cm}|C{1cm}C{1cm}|C{1cm}}
\hline \hline
$R_{H-H}$ (\AA) & $t_{\text{eff}}$ & $U_{\text{inter}}$ & $\tilde{U}_{\text{inter}}$ & $U_{\text{on}}$ & $\tilde{U}_{\text{on}}$ & $\tilde{T}_{\text{min}}$ \\ \hline
0.50  & 4.79    & 0.47 &	0.10    & 0.77 &  0.16   & 0.01  \\
0.75  & 2.10	& 0.47 &    0.23	& 0.71 &  0.34   & 0.02  \\
1.00  & 1.20	& 0.43 &	0.36	& 0.70 &  0.58   & 0.04  \\
1.25  & 0.71	& 0.37 &	0.52	& 0.69 &  0.98   & 0.07  \\
1.50  & 0.43    & 0.32 &	0.74	& 0.69 &  1.62   & 0.12  \\
1.75  & 0.26    & 0.28 &	1.06	& 0.70 &  2.67   & 0.19  \\
2.00  & 0.16    & 0.25 &	1.54	& 0.70 &  4.38   & 0.31  \\
2.25  & 0.10    & 0.22 &	2.24	& 0.70 &  7.18   & 0.51  \\
2.50  & 0.06    & 0.20 &	3.32	& 0.71 &  11.78  & 0.83  \\
\hline \hline
\end{tabular}
\caption{Interaction strengths of hydrogen chains with different bond lengths as quantified by the effective inter-site repulsion, $\tilde{U}_{\text{inter}} = U_{\text{inter}}/t_{\text{eff}}$, and on-site repulsion, $\tilde{U}_{\text{on}} = U_{\text{on}}/t_{\text{eff}}$. $t_{\text{eff}}$ is the nearest-neighbor hopping interaction from the original \textit{ab initio} Hamiltonian. The effective temperature, $\tilde{T}_{\text{min}} = T_{\text{min}} / t_{\text{eff}}$, where $T_{\text{min}}=0.05$ Hartree/$k_B$, for each bond length is also listed in the last column.}
\label{tab:Ut}
\end{table}

Recall that the standard Hubbard model only accounts for on-site repulsion, so the $\tilde{U}_{\text{on}}$ given in Table \ref{tab:Ut} may serve as natural points of comparison between the Hubbard model and the chains studied here. Similarly, $\tilde{U}_{\text{inter}}$ plays the role of the nearest-neighbor interaction, $V$, in the extended Hubbard model. As can be observed from Table \ref{tab:Ut}, the $t_{\text{eff}}$ decreases in magnitude with bond length, while $\tilde{U}_{\text{inter}}$, $\tilde{U}_{\text{on}}$, and $\tilde{T}_{min}$ all increase with bond length. That said, the shortest-length chains ($R \leq 1.25$ \AA) essentially correspond to the 1D Hubbard model in the small $U/t$ regime ($U/t < 1$). The 2.00 \r{A} chain borders upon being strongly correlated (in one dimension) with a $U/t \approx 4$, while the $R \geq 2.25$ \r{A} chains are very strongly correlated with $U/t \geq 7$. For all of the bond lengths studied, however, the inter-site interactions provided in the third and fourth columns of Table \ref{tab:Ut} maintain significant values not accounted for in the standard Hubbard model. Moreover, the strength of the inter-site repulsion relative to the on-site repulsion decreases as the bond length is increased. In particular, the $R = 1.25$ \AA$~$chain marks a bond length beyond which $V/U$ dips below 0.5. This signifies that the longest-length chains more closely approximate the standard Hubbard model, while the shortest-length chains more closely approximate the extended Hubbard model. Since $\tilde{T}_{\text{min}}$ increases with bond length, this furthermore implies that the longest chains are effectively modeled at higher temperatures than the shortest chains. %\YL{See the Supplementary Information for a plot of the up spin eigenvalue spectrum from the mean field theory across all chains.} 
We will frequently return to these effective interactions and temperatures in the subsequent analysis of our chains' physics.

\subsection{Internal Energy Crossover}

\begin{figure}[ht]
\includegraphics[width=0.55\textwidth]{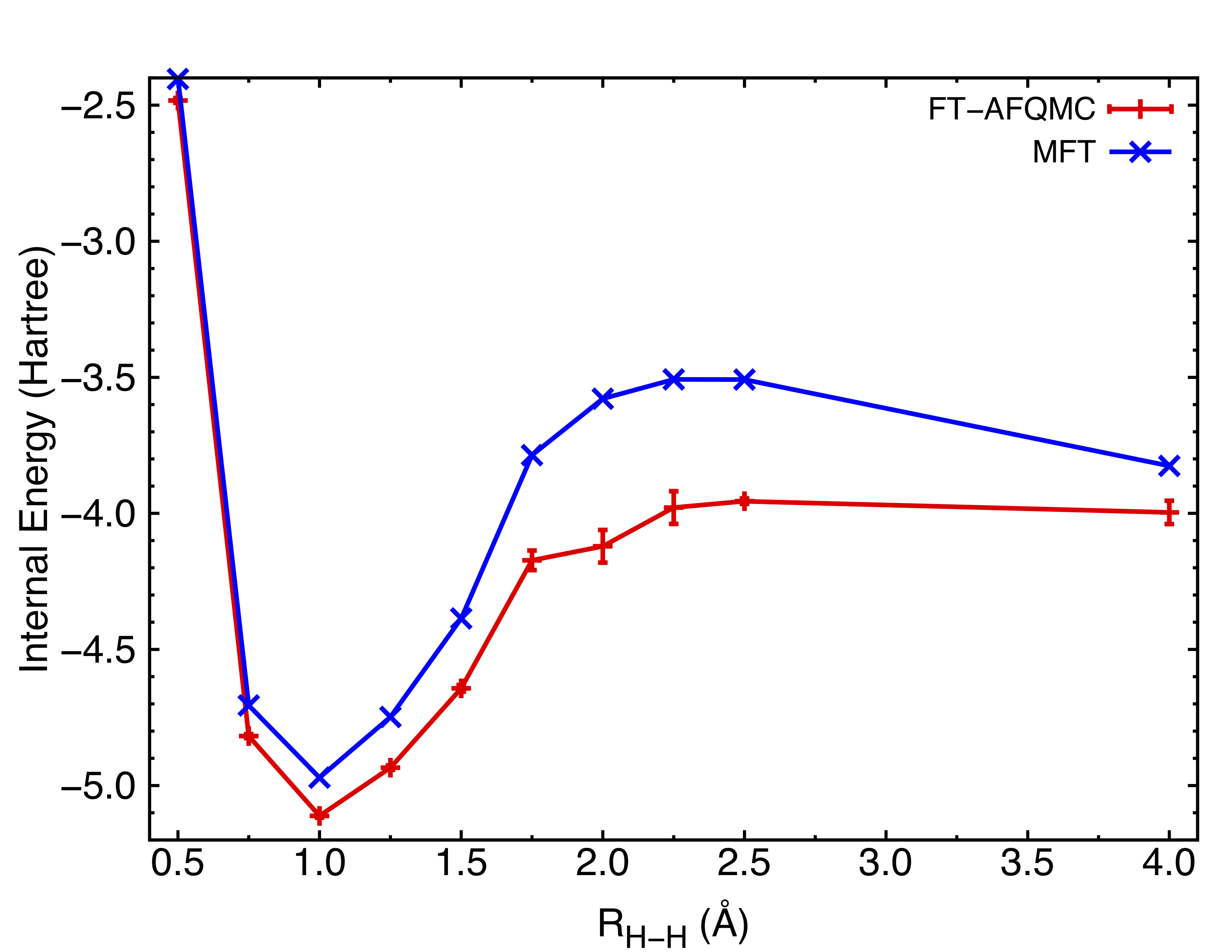}
\caption{Internal energy of our hydrogen chains as a function of H-H bond length at $T = 0.05$ Hartree/$k_B$, the lowest temperature employed in this study. The FT-AFQMC results are denoted by red $+$'s, while the MFT results are denoted by blue $\times$'s. The internal energy is minimized around 1.0 \AA.} 
\label{fig:eng-r}
\end{figure}

As an initial \YL{entry} into the physics of our hydrogen chains, we first study how their internal energies vary as a function of temperature. We take $T = 10$ Hartrees/$k_B$ as the upper temperature limit in our studies because 10 Hartrees/$k_B$ is larger than the effective $U/t$ for all but our 2.5 \r{A} hydrogen chain.
%, where, in analogy with the half-filled Hubbard model, $U$ is the energy required to generate excitations to doubly occupied states and thus defines our energy scale and $t$ is the hopping energy (see Section \ref{sec:effint} for more specific comparisons to the Hubbard model). 
On the other hand, we take $T = 0.05$ Hartrees/$k_B$ as the lower temperature limit in our studies because our hydrogen chains approach their ground state behavior at this temperature. Indeed, as depicted in Figure \ref{fig:eng-r}, at $T = 0.05$ Hartrees/$k_B$, the internal energy according to both MFT and FT-AFQMC is lowest at a bond length of around 1.0 \r{A}, which is comparable to the equilibrium ground state bond length of 0.945 \r{A} found in previous work.\cite{motta2017towards,rusakov2018self} 
We moreover find that the FT-AFQMC energy at this near-equilibrium bond length approaches the 0.542 Hartrees/atom expected in the ground state.\cite{motta2017towards} Additionally notice that mean field theory predicts a false dissociation barrier around 2.0 \r{A} which also appears in the Hartree-Fock dissociation curve of H$_2$ molecules.\cite{szabo1996modern} In contrast, our many-body AFQMC results give the correct dissociative behavior. Thus, while modeling at $T = 0.05$ Hartrees/$k_B$ is certainly not tantamount to modeling the ground state, our results suggest that this temperature is low enough to approximate many ground state features. 

%\BR{Further discussion of bands and also expectations for crossover based upon previous works should go here...} 
%\Edits{The equidistant hydrogen chain can be viewed as translating a unit cell containing one hydrogen atom by a fixed distance R (the bond length) in one dimension infinitely many times. Assuming each hydrogen atom only is represented by a single $1s$ orbital, according to conventional band theory, this will result in only one band. In our simulation with ten hydrogen atoms in one super cell, there should be ten bands. However, these are equivalent to the one band picture in the sense that the ten bands can be generated by folding the original one band into a smaller Brillouin zone that is 1/10 the size of the original one. For simplicity, we will use the one band picture in the following discussion.} 

\begin{figure}[ht]
\includegraphics[width=0.55\textwidth]{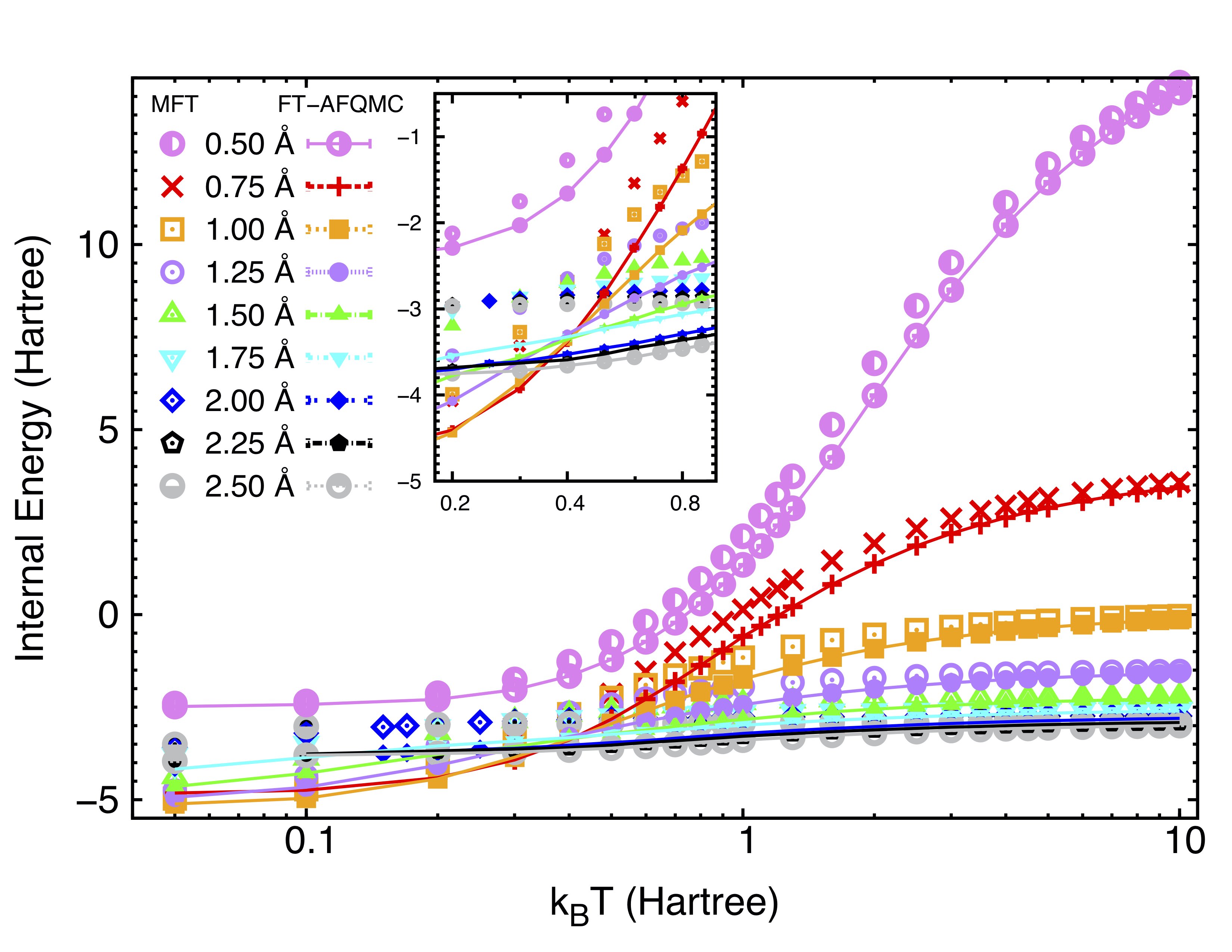}
\caption{Internal energy of the hydrogen chains for different H-H bond lengths over a range of temperatures. The FT-AFQMC results are denoted by the solid symbols, which are interconnected by solid lines as a guide to the eye, while the MFT results are denoted by the open symbols. Different colors correspond to different bond lengths. Note that the horizontal axis is plotted on a log scale in order to emphasize the low-temperature behavior. The inset highlights the crossover regime at intermediate temperatures and is plotted on a linear scale.}
\label{fig:eng}
\end{figure}

In Figure \ref{fig:eng}, we therefore turn to depicting how the internal energy varies for chains with differing bond lengths between these two limiting temperatures. 
As is clear from the figure, the 0.5 {\AA} chain undergoes the largest change in internal energy of several Hartrees from low to high temperatures; this change monotonically decreases with increasing bond length, with the 2.5 {\AA} chain manifesting a change of barely a Hartree. \YL{These differences between the energies of the low and high temperature states may be directly linked to the energy spread of the many-body eigenstates of the respective chains.} As is well-known from the archetypal example of H$_{2}$ dissociation,\cite{szabo1996modern} varying the bond length between atoms in a hydrogen chain varies the overlap among the hydrogens' atomic wave functions, modulating their \YL{energy spread} in turn. \YL{More specifically, when the H-H bond length is decreased, the highest energy state is pushed toward higher energies, while the lowest energy state is pushed toward lower energies, resulting in an overall increase in the energy spread -- and vice-versa.} By extension, the large change in internal energy with varying temperatures for the 0.5 \r{A} chain is indicative of \YL{its states having a large energy spread}, while the much smaller change for the 2.5 \r{A} chain is indicative of \YL{its states being more concentrated in energy space (\textit{i.e.}, it has a larger density of states).} \YL{Please see the Supplemental Information for a plot depicting the spread of the mean field states for all of the chains studied in this work for reference.} Because of these varying \YL{distributions of states}, a crossover is observed around a temperature of 0.4 Hartree/$k_{B}$, at which point the $R \geq 0.75$ \r{A} chains reverse their energetic ordering. 
%the $0.75 \leq R \leq 1.25$ \r{A} chains go from exhibiting a smaller average internal energy than their larger bond length counterparts at low temperatures to the reverse
%, \YL{except for the 0.5 \AA~chain due to its significantly larger ground state energy}. 
This is because the average energy for all of the chains is pushed upwards by the inclusion of the higher energy states that are accessed at high temperatures and these high energy states are even higher in energy for the small bond length chains. What is also clear from Figure \ref{fig:eng} is that the mean field energies are universally greater than the many-body, FT-AFQMC energies. The correlation energy missed by mean field theory is generally expected to increase the attractive interactions among the atoms, leading to lower many-body energies, particularly at the intermediate temperatures (\YL{see the inset, Table \ref{tab:mft-error}, and Fig. \ref{fig:eng-corr} in the Supplemental Information for more details}) at which previous work also found mean field discrepancies to be largest.\cite{Liu_JCTC_2018}
%Explanation for 0.5 A

%In the one band picture, since the hydrogen chain is half-filled, the Fermi level will lie halfway in the 1st Brillouin zone, $k_F = \pm \frac{\pi}{2R}$, where the band is continuous. Under this theory, the 1D hydrogen chain should be metallic. However, as Peierls noted, this configuration is not stable \cite{}. A band gap tends to open right at the Fermi level $k_F = \pm \frac{\pi}{2R}$ via periodic lattice distortion by dimmerization, which doubles the size of the unit cell while pushing the filled band to lower energy and stabilizes the whole system. Due to the existence of the Peierls band gap, this distorted version of 1D hydrogen chain will be an insulator at zero temperature, and partially conducting current at finite temperature by electrons/holes created through thermal excitation. However, this is beyond the scoop of the current work, and we will only discuss equidistant hydrogen chains.

\subsection{Heat Capacity Features}
As discussed in Section \ref{method_cv}, heat capacity curves provide more detailed signatures of the temperature dependence of different physical properties. To obtain the heat capacities, $C_{V}(T)$, we thus differentiate the internal energies provided in Figure \ref{fig:eng} with respect to the temperature at a variety of bond lengths, as depicted in Figure \ref{fig:cv}. Since quantum Monte Carlo data possess intrinsic fluctuations, we attribute the small fluctuations in the heat capacity curves to numerical fluctuations introduced by the internal energy data and ignore them in our interpretations. \YL{This interpretation is supported by the fact that the largest fluctuations appear around exactly the temperatures which possess the largest statistical error bars (see Secs. \ref{sec:cv-error} and \ref{ssec:cv-interp-compare} of the Supplemental Information for a more thorough discussion of how these error bars were obtained and their consequences.)} 
\YL{
\begin{figure}[ht]
\includegraphics[width=0.55\textwidth]{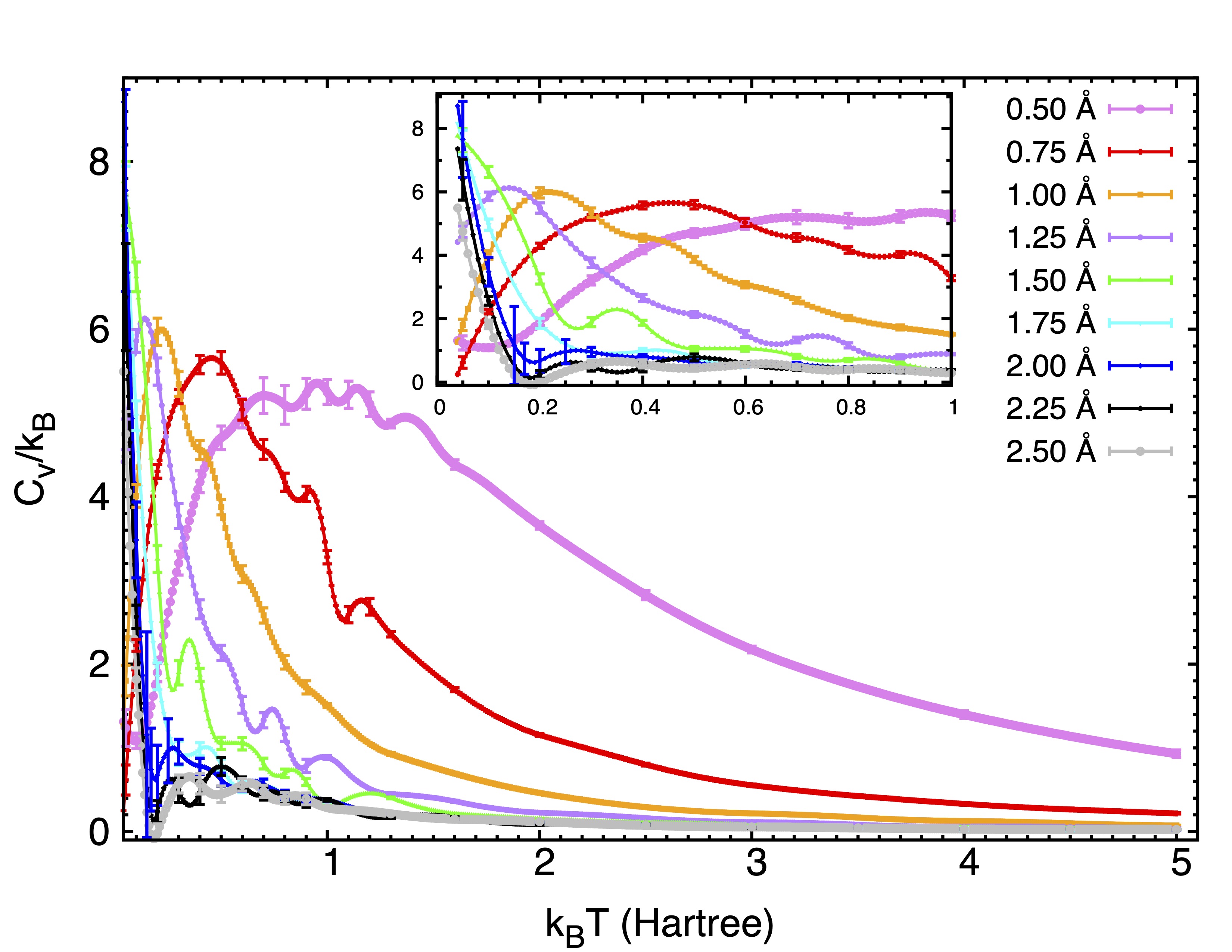}
\caption{\color{black} The heat capacity of the hydrogen chains at various bond lengths as a function of temperature. See Section \ref{method_cv} for more details on how the heat capacities were computed. The inset expands the low temperature regime from $T = 0.05$ to 1 to ease identification of the most important heat capacity features. Note that the horizontal temperature axis is plotted on a linear scale. Error bars are depicted on the temperatures used in our interpolation.}
\label{fig:cv}
\end{figure}
}

For the $R = 0.75$ \r{A} chain, a single broad feature rapidly rises from zero at low temperatures, peaks around $T = 0.5$, and then gradually decays to zero again in the high temperature limit. 
%The zero offset at low temperatures means that the chain is almost entirely in its ground state at $T = 0.05$. 
%\BR{The linear scaling of the heat capacity at sufficiently low temperatures is characteristic of fermions in any dimension. \cite{voit1995one}} 
Similar peaks are also observed in the $R = 0.5, 1.0$ and $1.25$ \r{A} chains, but as the bond length is increased (decreased), the corresponding heat capacity peak widths shrink (expand) while their heights increase (decrease), indicating a more rapid (gradual) response to temperature changes. For chains with bond lengths greater than 1.5 \r{A}, the heat capacity peak is pushed to temperatures lower than those surveyed in this work. The fact that this peak shifts to lower temperatures with increasing bond lengths indicates that the excitation responsible for this peak requires less thermal energy to initiate at longer bond lengths. 

We note that the results presented here generally agree with earlier numerical results obtained by Shiba \cite{shiba1972thermodynamic} for the 1D Hubbard chain at finite temperature with a small number of sites in the grand canonical ensemble. However, there is one significant difference in the way our heat capacities vary with temperature. In the 1D Hubbard model, the heat capacity possesses a single broad feature in the small $U/t$ regime, but splits into two peaks for $U/t \ge 4$, where one of the two peaks lies between $0 < T/t < 0.5$ and is sharp, while the other lies in the high temperature regime between $0.5 < T/t < 5$ and is broad. According to Shiba, the sharp feature at low temperatures is due to a collective spin excitation from the antiferromagnetic ground state of the Hubbard model, while the broad peak in the high-temperature regime originates from a charge excitation across the Mott gap. 

If our chains' behavior is analogous to that seen in the Hubbard model, based upon their effective on-site interactions, we anticipate two peaks to be observed in the heat capacities of the 2.0, 2.25, and 2.50 \r{A} chains. According to previous 1D Hubbard model results,\cite{shiba1972thermodynamic} we thus estimate the 2.00, 2.25, and 2.50 \r{A} chains to have spin excitation peaks around 0.054, 0.022, and 0.0084 Hartree/$k_{B}$ and charge excitation peaks around 0.17, 0.14, and 0.15 Hartree/$k_{B}$ (see the Supplemental Information for further details regarding the Hubbard excitation temperatures). The expected spin excitation temperatures for the 2.25 and 2.5 \r{A} chains thus fall below the temperatures simulated in the present work, but the spin excitation temperature for the 2.0 \r{A} chain should be observable within our data. Nevertheless, we only observe one feature in the low temperature regime for the 2.0 \r{A} chain. The expected high-temperature, broad feature is absent from our results for this chain, even though its effective U is greater than 4. Similarly, the high temperature peaks expected to occur around 0.14 and 0.15 Hartree/$k_B$ for the 2.25 and 2.5 \r{A} chains also do not appear in our data. However, the decreasing width and increasing sharpness of the low-temperature peaks in our results agree with what Shiba observed of the low temperature peaks in the Hubbard model. It makes no sense to assume that the absence of the high-temperature feature in our simulations is due to the absence of charge excitations, because we know that, in the grand canonical ensemble, charge and spin excitations should both appear. As this comparatively simple model may only exhibit flavors of spin and charge ordering, this peak must be due to the emergence of either or both of these orders. So as to precisely characterize this order, we therefore proceed to analyze observables that are bellwethers of charge- and spin-related ordering in the following sections. 

%At the longest bond length of 4.0 \r{A}, the hydrogen atoms in the chain are virtually isolated from one another. Given one Gaussian-type orbital per atom, there is no way to produce any orbital excitations on a single atom, so the heat capacity at 4.0 \r{A} nearly vanishes.  

%\BR{Presumably, at short bond lengths, it is harder to overcome the Coulomb repulsion to reach excitations that distribute electrons over sites. Thus, it requires a higher temperature to excite the accessible states. However, at longer bond lengths, there is less Coulomb repulsion to overcome, so excitations to these accessible states can occur at lower temperatures.}

\subsection{Trends in the Double Occupancy and a Metal-Insulator Crossover}
 
In order to determine whether the heat capacity peaks stem from a form of charge ordering, we first examine the double occupancies of our hydrogen chains as a function of temperature. As discussed in Mott's seminal paper, if the Coulomb repulsion among electrons in a material is stronger than the electrons' propensity to hop among atoms,
%, as is the case for chains with sufficiently long H-H bonds, 
periodic lattices will manifest Mott insulating behavior in which charges remain localized even though they should delocalize according to conventional band theory.\cite{mott2004metal} Since the Coulomb repulsion among electrons in hydrogen chains can greatly exceed their propensity to hop at sufficiently long H-H bond lengths (see Section \ref{sec:effint} for a more explicit discussion), Mott was thus able to show that 3D hydrogen lattices undergo a discontinous metal-Mott-insulator transition at bond lengths around 1.3 \r{A}. Similar results were obtained using extended DMFT for two-dimensional systems.\cite{chitra2000effect} We therefore anticipate a Mott insulator to emerge with increasing H-H bond distances in the 1D hydrogen chains studied here.\cite{poilblanc1997insulator} 
%\BR{Explain why a band insulator cannot emerge.} 
%\YL{The term "band insulator" seems to be an artifact of band structure theory, which is one-particle theory. In a many-particle theory, there is no band, so I guess this word simply doesn't apply anymore. We can discuss whether it is an metal or insulator, but it seems difficult to classify it as a band insulator. As a side note, in 2D and 3D hydrogen lattice, there is not band insulator; simply a 1st order transition from metal to Mott insulator.} 
A simple gauge of the emergence of insulating behavior is the double occupancy, which measures the fraction of, in this case, atoms, that are occupied by two electrons.\cite{Jordens_Nature,Scarola_PRL_2009} Smaller double occupancies are indicative of correlated Mott insulating behavior in which strong electron-electron interactions suppress two electrons from occupying the same orbital (in this case, H, since each atom possesses only a single GTO), while larger double occupancies are indicative of less correlated, more metallic behavior in which electrons are free to occupy virtually any atom.
%In \BR{metals} with minimal electron-electron repulsion, electrons can freely occupy virtually any atom, regardless of whether a second electron already resides on that atom, leading to finite double occupancies. In contrast, in insulators with stronger electron-electron repulsions, electrons will attempt to minimize their energy by occupying different atoms, leading to double occupancies that tend toward 0. 

\begin{figure}[ht]
\includegraphics[width=0.55\textwidth]{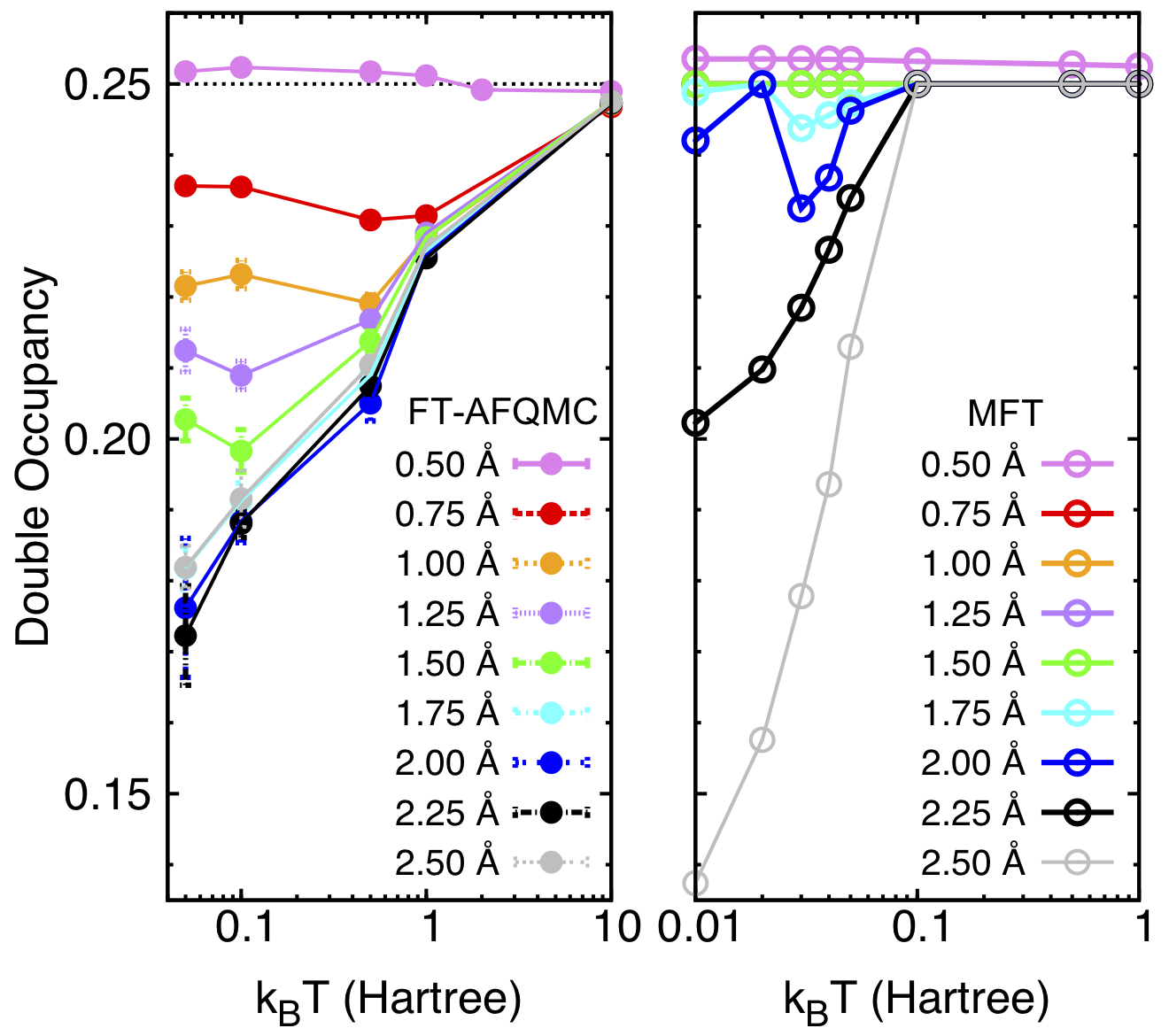}
\caption{Double occupancy per atom of the hydrogen chains at different bond lengths over a range of temperatures according to (left) FT-AFQMC and (right) MFT. Different colors represent different bond lengths. Note that the ranges of the horizontal axes (in log scale) differ between the left and right panels.}
\label{fig:docc}
\end{figure}

As depicted in Figure \ref{fig:docc}, we therefore calculated the average double occupancy per atom for our chains as a function of temperature using both FT-AFQMC (Figure \ref{fig:docc}a) and MFT (Figure \ref{fig:docc}b). We find that almost irrespective of the temperature, the double occupancy at shorter bond lengths is greater than that at longer bond lengths, which implies that our chains at shorter bond lengths are more metallic than those with longer bond lengths. In the high temperature limit ($T = 10$), all electronic states are equally populated, implying that these hot electrons essentially exhibit classical behavior. As a result, electrons are as likely to inhabit an already occupied atom as an unoccupied atom. At half-filling, this signifies that half of the atoms will be randomly populated by up electrons and half will be randomly populated by down electrons, meaning that, as calculated, a quarter (0.25) of  all atoms will be doubly occupied. 

In contrast, in the low-temperature limit, different chains manifest different metal-insulating behavior as a function of $R$. 
%degree of correlation
At the shortest bond length, $R = 0.5$ \AA, the double occupancy rises above 0.25 when the temperature is lowered, suggesting that the ground state of the $R=0.5$ \AA ~chain is metallic. In contrast, when $R \ge 0.75$ \AA, the double occupancy is suppressed as the temperature is lowered due to the increased Coulomb repulsion between electrons, which favors the formation of local moments. The fact that the double occupancy increases for the 0.5 \r{A} chain, yet decreases for the 0.75 \r{A} chain is consistent with recent work on ground state hydrogen chains which predicts a first order metal-insulator transition to occur around a bond length of roughly 1.5 Bohr = 0.794 \r{A}.\cite{motta2019ground} 

Interestingly, for $0.75$ \r{A} $\le R \le 1.5$ \AA, a minimum in the double occupancy may be observed to occur between 0.1 and 1 Hartree/$k_{B}$. The somewhat counterintuitive negative-sloping double occupancy curve that emerges at low temperatures stems from an effect analogous to the Pomeranchuk effect in the Hubbard model:\cite{georges1992numerical,georges1993physical,sciolla2013competition} as the temperature is increased through the low temperature regime, the formation of local spin moments (indicated by decreasing double occupancies) will increase the entropy of the spin degrees of freedom, thus decreasing the free energy and stabilizing the system. To the best of our knowledge, this is the first observation of such a phenomenon in the presence of long-range Coulomb interactions. As $R$ is further increased beyond 1.75 \r{A}, the effects of strong correlation dominate over the entire temperature range explored, resulting in only positive-sloping double occupancies, which further corroborate Mott-insulating behavior. 

We furthermore note that MFT fairly accurately predicts the double occupancy at high temperatures, but significantly overestimates it (thus also overestimating the metallic character of the chains) at low temperatures, particularly for strongly correlated chains with $1.50 \leq R \leq 2.25$ \AA. This is consistent with the well-known fact that mean field theories, such as Density Functional Theory, often wrongly predict metallic behavior for strongly correlated insulators such as NiO.\cite{shen1991electronic,kunevs2007nio}

\begin{figure}[ht]
\includegraphics[width=0.55\textwidth]{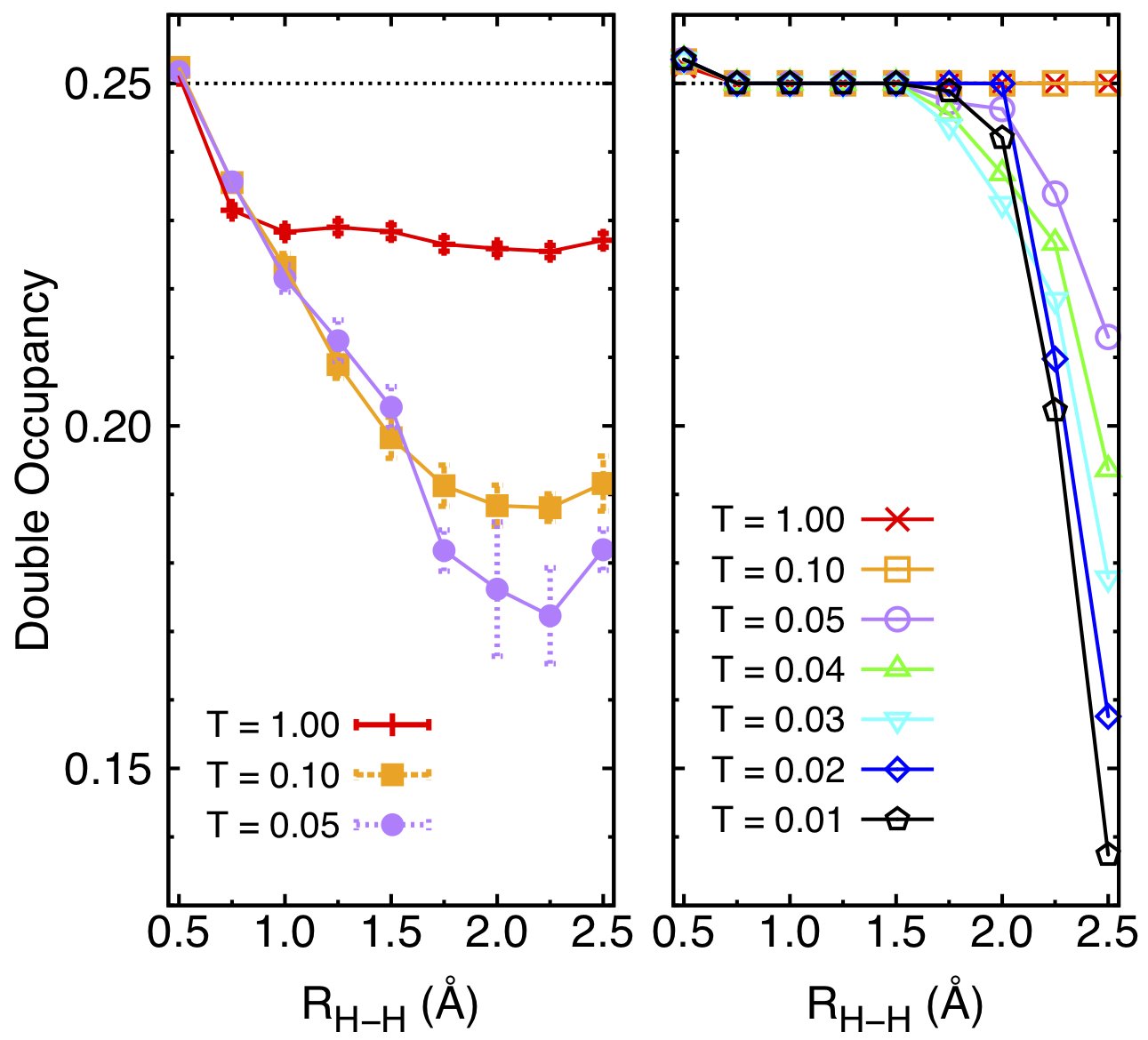}
\caption{Double occupancy per atom of the hydrogen chains as a function of bond length for several temperatures according to (left) FT-AFQMC and (right) MFT calculations. Different colors represent different temperatures.}
\label{fig:docc-r}
\end{figure}

Lastly, it is illuminating to plot the double occupancy as a function of $R$ at low temperatures, as shown in Figure \ref{fig:docc-r}. As discussed above, we can clearly see a crossover from metallic to insulating behavior between $R=0.5$ and $R=0.75$ \AA. Comparing the left and right panels of Figure \ref{fig:docc-r}, it is also evident that FT-AFQMC captures the effects of correlation irrespective of the bond length, while MFT seems to be ignorant to the increasing degree of correlation up until $R = 1.75$ \AA, after which it predicts a sharp drop in the double occupancy at temperatures lower than 0.05 Hartree/$k_B$. For temperatures greater than that, MFT predicts a double occupancy of 0.25, which is the same as that expected in the high temperature limit. The sudden drop in the double occupancy at low temperatures in the right panel suggests that MFT predicts a finite temperature crossover into the Mott insulator regime at around $R=1.75$ \r{A}. 
%BR: if a ref asks, we can discuss this, but we might be getting into the weeds to discuss it here...
%Note that the leveling off of the FT-AFQMC results (left panel) at large bond lengths for each temperature is due to the rapid decreasing of the magnitude of the hopping coupling term. A given fixed physical temperature $T$ would correspond to a higher effective temperature $\tilde{T} = T/t$ for smaller hopping term in longer bond length chains.

\subsection{Magnetic and Insulating Crossovers as Revealed by Spin-Spin and Charge-Charge Correlation Functions}

\begin{figure}[ht]
\includegraphics[width=0.55\textwidth]{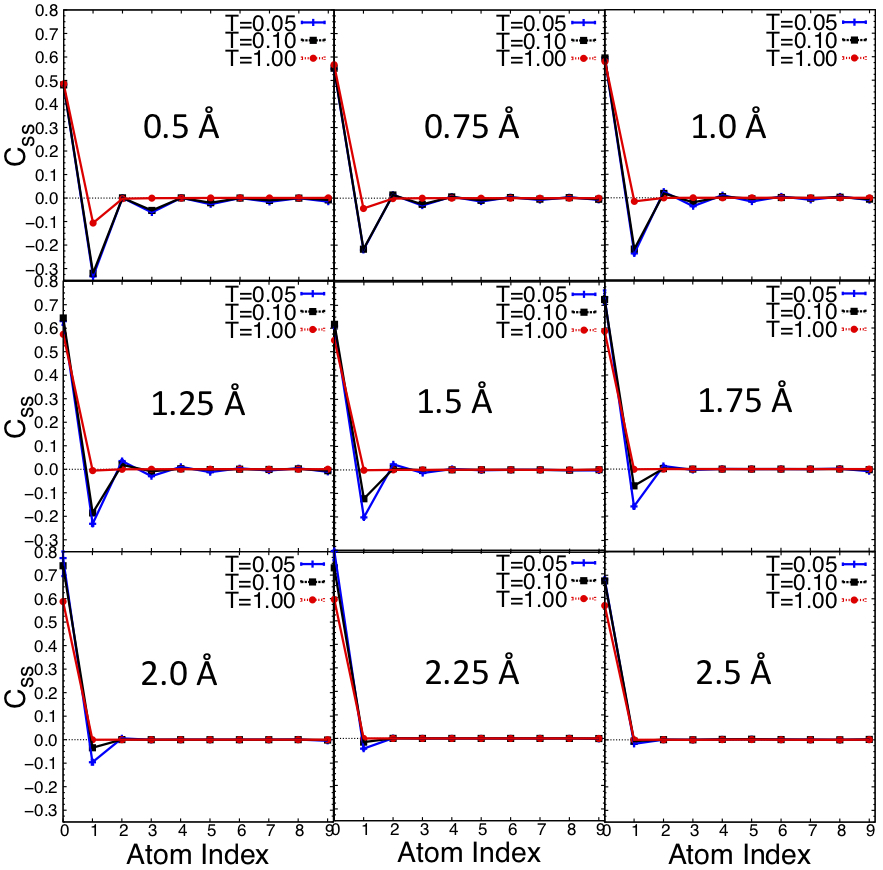}
\caption{The spin-spin correlation functions, $C_{ss}$, of the hydrogen chains at different bond lengths and temperatures ($T =$ 0.05, 0.1, and 1 Hartree/$k_{B}$) calculated using FT-AFQMC. The horizontal dashed line in each panel at $C_{ss} = 0$ is a guide to the eye.}
\label{fig:corrfun_ss}
\end{figure}

\begin{figure}[ht]
\includegraphics[width=0.55\textwidth]{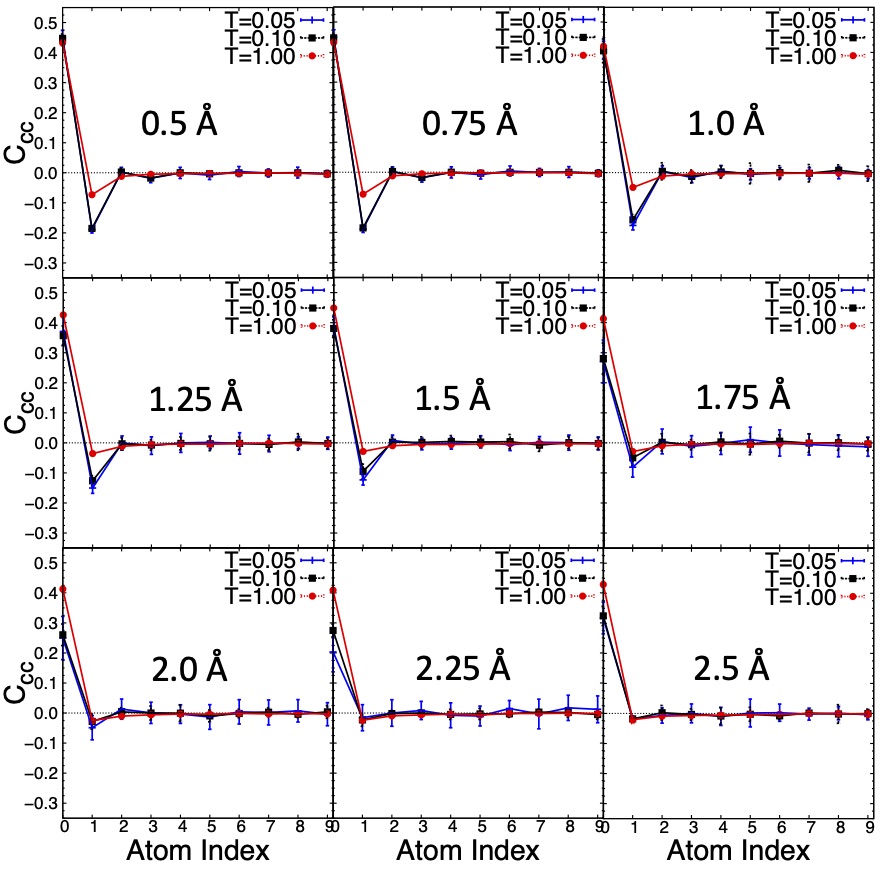}
\caption{\color{black} The charge-charge correlation function, $C_{cc}$, of the hydrogen chains at different bond lengths and temperatures ($T =$ 0.05, 0.1, and 1 Hartree/$k_{B}$) calculated using FT-AFQMC. The horizontal dashed line in each panel at $C_{cc} = 0$ is a guide to the eye.}
\label{fig:corrfun_cc}
\end{figure}

In order to characterize the spin and charge excitation contributions to the heat capacity of our hydrogen chains, we compute spin-spin, $C_{ss}$, and charge-charge, $C_{cc}$, correlation functions for all chains at $T=1.0$, 0.1, and 0.05 Hartree/$k_{B}$, as shown in Figures \ref{fig:corrfun_ss} and \ref{fig:corrfun_cc}, respectively. 
As described in Section \ref{method:corrfun}, all of the correlation functions presented were computed in the orthogonal atomic orbital basis. The positivity of the correlation functions at site 0 is guaranteed since charges and spins should be most correlated with themselves, while the large dips in all of the charge-charge and spin-spin correlation functions at site 1 are a direct consequence of correlation and exchange holes, respectively. In Figure \ref{fig:corrfun_ss}, at $T = 0.05$, AFM order clearly emerges for chains at bond lengths of 0.75, 1.0, 1.25, and 1.5 \AA, as can be seen from the alternating positive and negative values of $C_{ss}$.  %\BR{and $C_{cc}$.} 
For $R=0.5$ \AA, however, the correlation functions remain negative for all atoms greater than 1. This is indicative of ferromagnetic (FM) ordering. The sign change of the correlation functions from $R=0.5$ to 0.75 \r{A} suggests a 
%quantum fluctuation driven 
crossover from FM order to AFM order at low, but finite temperatures.  For chains with longer bond lengths ($\ge 1.75$ \AA), AFM order dissipates at the temperatures explored because of their smaller effective exchange coupling strengths. At still longer bond lengths ($\ge 2.25$ \AA), even the exchange hole is barely visible, indicating the vanishing effective exchange at these temperatures. \YL{Note that the correlation functions do not exhibit the same periodicity as seen in Fig.~\ref{fig:Ut} because our hydrogen chain results have been averaged over 5 k-points (not just the $\Gamma$ point), which would be equivalent to performing simulations on a supercell of 50 hydrogen atoms at the $\Gamma$ point.}

The charge-charge correlation functions also exhibit oscillations between positive and negative values at low temperatures for $R \leq 1.25$ \r{A}. This is indicative of charge ordering, but over a slightly smaller range of bond distances than AFM or FM order is observed in the $C_{ss}$. The emergence of charge order for the shorter bond length chains is likely due to their larger ratio of nearest-neighbor to on-site Coulomb repulsion, as discussed in Section \ref{sec:effint}.  Indeed, previous work \cite{hirsch1984charge,voit1992phase,glocke2007half} on the extended Hubbard model predicts the formation of a charge density wave (CDW) phase when $U$ and $V$ satisfy $V/U > c$, where $c$ is roughly 0.5, and a spin density wave (SDW) phase for $V/U$ smaller than that. Table \ref{tab:Ut} clearly suggests that for $R \le 1.25$ \AA, the effective inter- and on-site repulsion strengths, $\tilde{U}_{\text{inter}}$ and $\tilde{U}_{\text{on}}$, satisfy the criteria for favoring CDW formation. Chains with $R \ge 1.5$ \r{A} exhibit no significant charge order except for their correlation holes. \YL{Note that the error bars on these charge-charge correlation functions are larger than those on our spin-spin correlation functions because the charge-charge correlation functions involve subtracting off background terms that possess their own error bars. The lowest temperature charge-charge correlation functions possess the largest error bars because they are most strongly correlated.} 
%Due to the higher effective temperatures in the 2.25 and 2.5 \AA~chains, their error bars are smaller as compared to the 2.0 \AA~chain at the same temperature of $T = 0.05 \text{ Hartree}/k_B$.}

Combining this information with our previous observations about the double occupancies, our results suggest that the 0.5 \r{A} chain will be a ferromagnetic metal, while the 0.75 \r{A} chain will be an antiferromagnetic insulator in the ground state, as was observed in recent zero temperature studies.\cite{motta2019ground} In this work, we did not model chains with bond lengths shorter than 0.5 \r{A}, but it would be interesting to examine the remnants of this ground state quantum phase transition at temperatures slightly above zero, at which both quantum and thermal fluctuations can play important roles. We leave this to future explorations. Our simulations thus reveal two metal to insulator crossovers: a ground state quantum phase transition that occurs upon varying bond length, and as described in Section \ref{subsubsec:docc}, a finite temperature crossover that may be observed in all chains with $R \geq 0.75$ \r{A}. 

Our correlation functions also reveal an interesting interplay between the spin and charge degrees of freedom in hydrogen chains. Firstly, the existence of both spin and charge orders at low temperatures and the absence of both orders at higher temperatures indicates that the spin and charge degrees of freedom are coupled together in hydrogen chains. Unlike in the 1D Hubbard model, spin and charge excitations do not separately appear at different temperatures. Secondly, note that the value of $C_{ss}$ at site 0 (which measures the autocorrelation between an atom and itself) in Figure \ref{fig:corrfun_ss} increases from around $C_{ss} = 0.5$ at $R=0.5$ \r{A} to 0.8 at $R=2.5$ \r{A} at low temperatures, while $C_{cc}$ in Figure $\ref{fig:corrfun_cc}$ displays the opposite trend. The increasing local spin moment, yet decreasing charge-charge repulsion on a single atom corroborates the gradual development of stronger and stronger AFM order as $R$ is increased. Indeed, in a perfect AFM phase, $C_{ss}$ and $C_{cc}$ on atom 0 should assume a value of 1 and 0 respectively, since there should only be one electron occupying each atom and spins on adjacent atoms should anti-align. The coupling of spin and charge degrees of freedom is most likely mediated by the long-range Coulomb interactions inherent to such realistic systems.

\subsection{Natural Occupancies in the Orthogonal AO Basis}
\label{sec:natocc}

\begin{figure}[ht]
\includegraphics[width=0.55\textwidth]{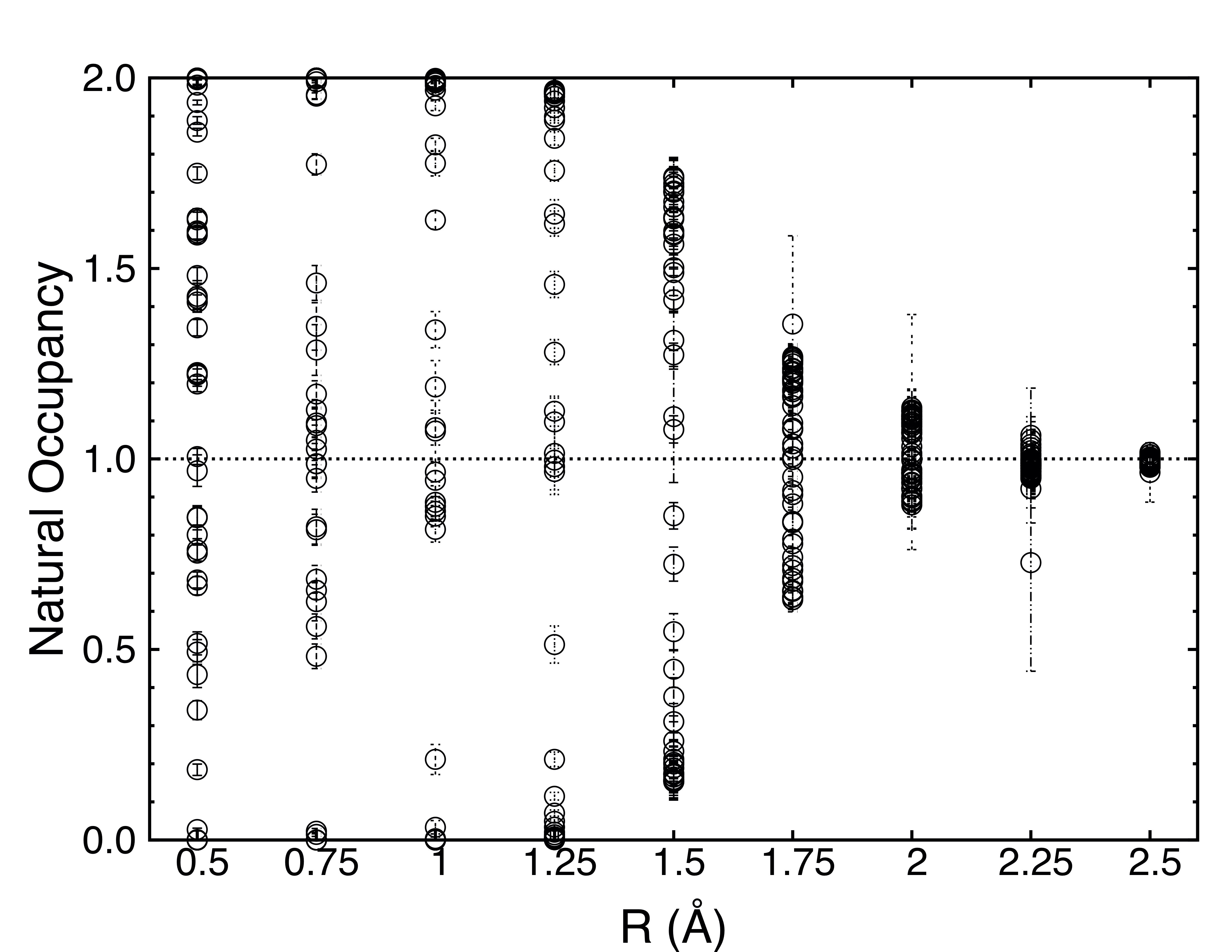}
\caption{Natural occupancies of the natural orbitals of the hydrogen chains for all bond lengths studied at $T$ = 0.05 Hartree/$k_B$. 50 natural occupancies, which stem from the 50 p-GTOs employed in these simulations, are plotted for each bond length. The natural occupancies are obtained by diagonalizing the density matrix expressed in terms of the orthogonal, periodic AO basis.} 
\label{fig:nao-r}
\end{figure}

To obtain further support for the ordering assignments we make above, we additionally computed the chains' natural occupancies and orbitals by diagonalizing their density matrices in the orthogonal, periodic AO basis (see the Supplemental Information for more information about this basis and process).\cite{lowdin1955quantum} In Figure \ref{fig:nao-r}, we plot the natural occupancies of all of the orbitals of all of our chains at the lowest temperature in our simulation, $T$ = 0.05 Hartree/$k_B$. 
For $R = 0.5$ \r{A}, we see that the natural occupancies are scattered across a wide range of values, many of which are fractional, corroborating that the chain is metallic. As the bond length is increased to 0.75, 1.0, and 1.25 \r{A}, the natural occupancy tends to concentrate in empty, singly-occupied, and doubly-occupied natural orbitals, with little in between. As the chains are further stretched to 1.5 \r{A}, the double occupancy (and, accordingly, zero occupancy) is further suppressed and the natural occupancies further concentrate around one, marking the onset of Mottness. Indeed, at 2.5 \r{A}, all natural orbitals are singly occupied, which is indicative of a perfect Mott insulator.

The qualitative change of the distribution of natural occupancies from the 0.5 \r{A} chain to the 0.75 \r{A} chain, choruses with the metal to insulator crossover manifested in Figure \ref{fig:docc-r}, and the disappearance of the natural occupancy cluster around one from the 1.25 \r{A} chain to the 1.5 \r{A} chain coincides with the disappearance of charge order in Figure \ref{fig:corrfun_cc}. 
%due to the decreasing ratio of inter versus on-site Coulomb repulsion. 
We further note that, although the natural occupancies presented here are calculated at a low temperature of $T=0.05$ Hartree$/k_B$, contributions from thermal excitations may still exist, especially for chains with longer bond lengths and correspondingly larger effective temperatures. 
%I rather avoid direct comparison here in case Dominika ends up reffing the paper, which there is a high probability of
%Moreover, our results here also suggest that the two solutions obtained in GF2 method for hydrogen chains \cite{rusakov2016self} have to be combined together to acquire a complete physical picture, and each one of the solution only tells half of the story.

\begin{figure}[ht]
\includegraphics[width=0.55\textwidth]{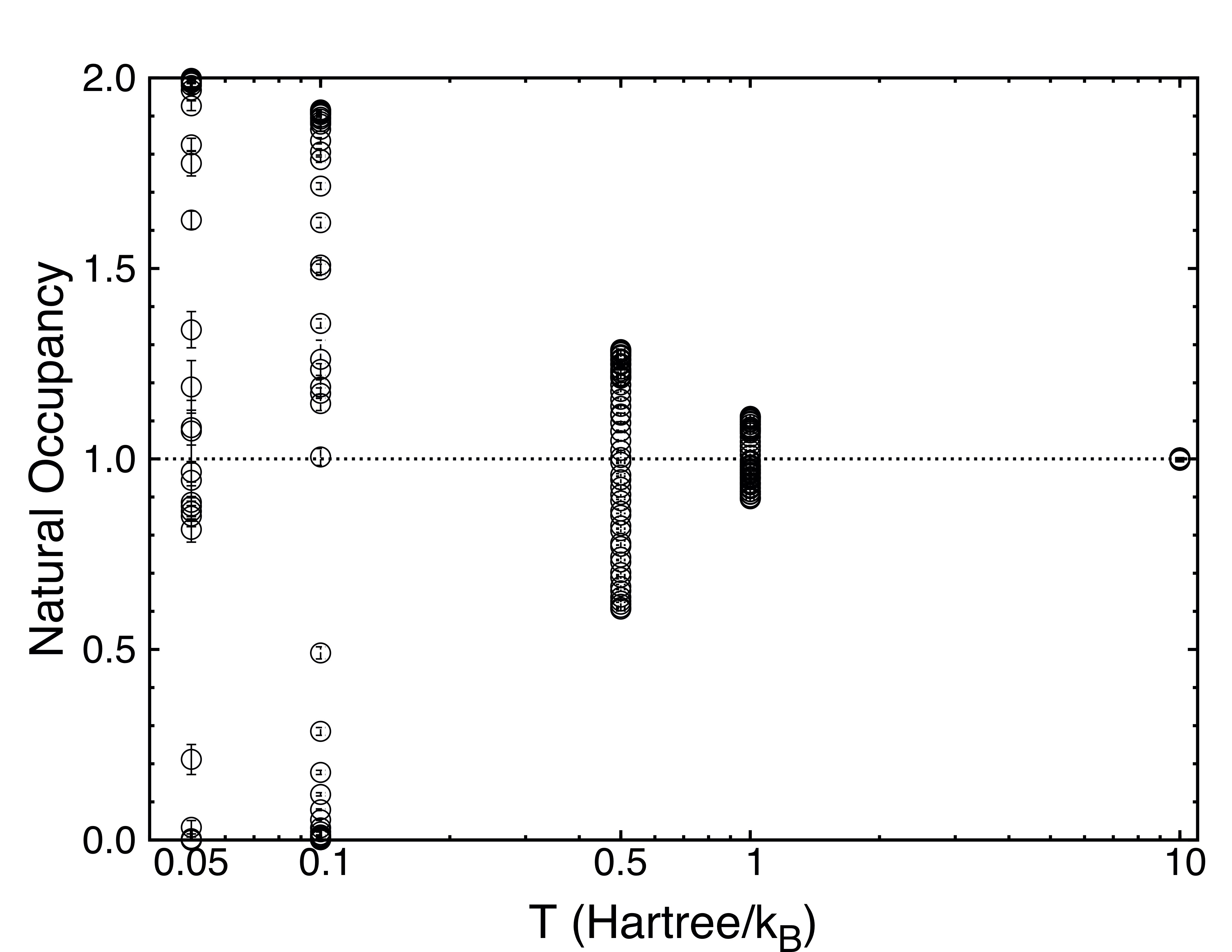}
\caption{Natural occupancies of the 1.0 \r{A} chain at $T =$ 0.05, 0.1, 0.5, 1, and 10 Hartree/$k_{B}$ in the orthogonal, periodic AO basis.}
\label{fig:nao-T}
\end{figure}

At a given bond length, we can imagine that by varying the temperature, electrons can be thermally excited to higher lying bands which can change the occupancy of the chain. To further investigate such temperature effects, we plot the natural occupancies for the 1.0 \r{A} chain as a function of temperature in Figure \ref{fig:nao-T}. 
From this plot, we can observe that, as the temperature is increased, the spread of the natural occupancies tends to decrease and the natural occupancies all concentrate around one. This is because, as the temperature is increased, the kinetic energy of the electrons increases, making it easier for electrons that previously inhabited empty and doubly-occupied natural orbitals to instead inhabit singly-occupied natural orbitals. Note that the reason for the suppression of empty and double occupancy at high temperatures in Figure \ref{fig:nao-T} differs from that for Figure \ref{fig:nao-r} in long bond length chains, as the latter is due to Mottness induced by strong on-site Coulomb repulsions. The two scenarios can be easily distinguished by their completely different double occupancies as shown in Figure \ref{fig:docc-r}.

\subsection{The Grand Potential and Entropy}

\begin{figure}[ht]
\includegraphics[width=0.55\textwidth]{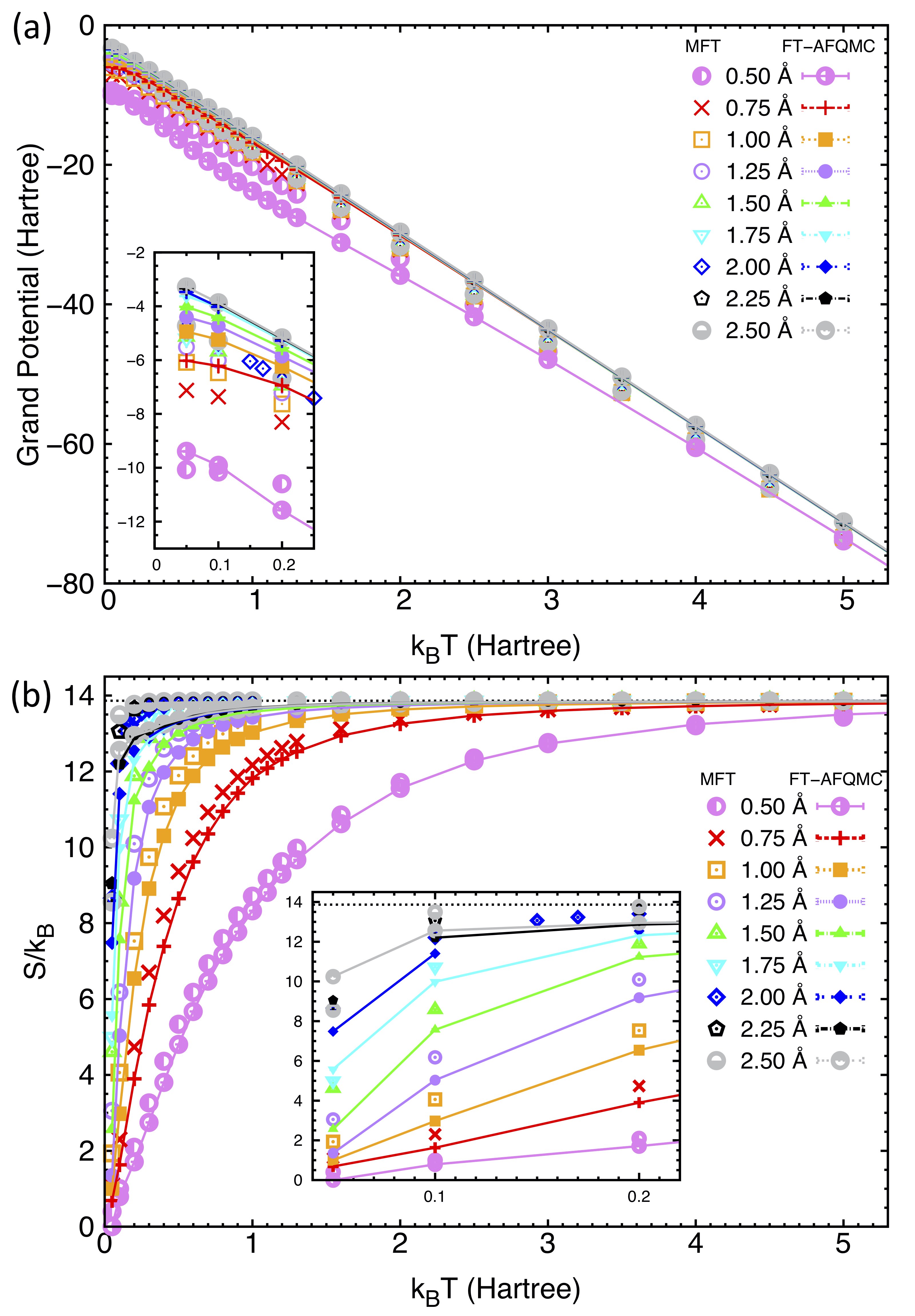}
\caption{The grand potential (a) and entropy (b) of hydrogen chains at different bond lengths calculated using both FT-AFQMC and MFT. The horizontal black dashed line in (b) is the high temperature limit of the exact entropy for a ten-orbital system, which is $10ln4$. }
\label{fig:omg-ent}
\end{figure}

Ultimately, finite temperature phase transitions are dictated by changes to a system's free energy, in this case, the grand potential. Unfortunately, the Mermin-Wagner Theorem dictates that thermal fluctuations at even infinitesimally small, yet non-zero temperatures will break any long-range order within one-dimensional systems,\cite{mermin1966absence,ghosh1971nonexistence} so finite temperature phase transitions cannot rigorously occur in the chains modeled here. Even at zero temperature, only  \emph{quasi}-long-range (instead of \emph{truly} long-range) AFM order was observed in hydrogen chains.\cite{motta2019ground} Therefore, in this Section, we present the grand potentials and entropies of our chains in Figure \ref{fig:omg-ent} followed by only a very brief discussion of their physical significance. 

As presented in panel (a) of Figure \ref{fig:omg-ent}, at high temperatures, the grand potential depends linearly on the temperature, consistent with Equation \eqref{omega_def}, since in the high-temperature limit, all states are equally populated and the grand canonical partition function is constant. The grand potential remains linear at intermediate temperatures, illustrating the importance of the entropic, $-TS$ contribution to the partition function. At the low temperatures magnified in the inset, we observe that the grand potential plateaus for the 0.5 and 0.75 \r{A} chains. This plateau is delayed to even lower temperatures not presented here for chains with longer bond lengths and hence larger effective temperatures. From panel (b), it is clear that the entropy for all chains in the high temperature limit assumes a value of $10ln4$, which is the theoretical high temperature limit of the entropy for a system with 10 spatial orbitals (and 4 potential electronic states per orbital) in the grand canonical ensemble. As the temperature is lowered, the entropy of all chains starts to decrease at different characteristic temperatures roughly determined by the chains' band widths. In the low temperature limit, the entropy of all chains tends to converge to zero, as can be seen from the inset of panel (b). For the chains with bond lengths of 0.5, 0.75, 1.0, and 1.25 \r{A}, for $T < 0.2$, we observe that the entropy scales linearly with temperature and extrapolates to zero at zero temperature. The same linear behavior is likely to be observed in the longer bond length chains at temperatures below $T=0.05$ because they possess larger effective temperatures. Compared with the AFQMC results, MFT systematically underestimates the grand potential and overestimates the entropy. 

We should also note that, from Figure \ref{fig:omg-ent}, the grand potentials and entropies of chains with shorter bond lengths are consistently smaller than those with longer bond lengths irrespective of the temperature. The entropy is smallest for shorter chains because these chains possess larger band widths, meaning that fewer of their states will be accessible at any single temperature. As is also presented in Table \ref{tab:Ut}, shorter bond length chains possess lower effective temperatures than longer bond length chains at the same physical temperatures. The ordering of the grand potentials may be explained by the magnitudes of their chemical potential terms, $-\mu N$. Shorter bond length hydrogen chains possess larger average electron densities. Larger electron densities lead to larger Coulomb repulsions among electrons, which means that larger magnitude chemical potentials are required to maintain a fixed average number of electrons. As a result, the larger negative contribution from the chemical potential term drives down the grand potential for shorter bond length chains.

\section{\label{sec:conclusions} Conclusions}

%\BR{What about size scaling with respect to k points...need to include a phrase in second paragraph on this} 

In conclusion, we have generalized our previous fully \emph{ab initio} finite temperature Auxiliary Field Quantum Monte Carlo method to solids and employed it to study the many-body thermodynamics of periodic hydrogen chains. Despite being the simplest possible models of \emph{ab initio} solids \emph{in principle}, hydrogen chains manifest an elaborate array of interrelated metal-insulator and magnetic orders, making them a meaningful testbed for studying the phase behavior of more complex solids.
%of ultimate interest to the community. 
By calculating these chains' many-body internal energies, free energies, entropies, double and natural occupancies, spin and charge correlation functions, and heat capacities as a function of H-H bond length and temperature, we demonstrated that, at low temperatures, they undergo metal to  insulator and ferromagnetic to antiferromagnetic crossovers at bond lengths between 0.5 and 0.75 \r{A}. Charge ordering is shown to emerge in chains with bond lengths less than or equal to 1.25 \r{A} and a Mott insulating phase is predicted for bond lengths greater than 1.5 \r{A}. Upon increasing the temperature beyond 1 Hartree/k$_{B}$, we observe the emergence of metallic character and the decay of any accompanying magnetic order in all of the chains studied. At intermediate bond lengths, we moreover see signatures of a Pomeranchuk effect in both our AFQMC and MFT results, the first time this has been identified in a realistic one-dimensional solid. In contrast with the heat capacity curves of one-dimensional Hubbard chains which exhibit two peaks for sufficiently large $U$ values, we find that the heat capacity curves of the chains analyzed here possess just a single peak, which we attribute to their distinctive coupling of spin and charge degrees of freedom. Such results highlight the important role long-range Coulomb interactions -- and techniques capable of treating them -- assume in determining the phase diagrams of realistic solids.  

As this study has served as an initial foray into the finite temperature physics of hydrogen chains, several key aspects of our modeling can be improved in follow-up work. First and foremost, for computational expediency and to facilitate comparisons with the single-band Hubbard model, we employed a minimal STO-6G GTO basis set with limited k-point sampling, meaning that we have not fully converged our calculations to the thermodynamic limit. Increasing the size of our GTO basis will predominantly help converge localized states, while increasing our k-point sampling will help converge delocalized states and alleviate finite size effects. 
%\BR{In the thermodynamic limit, one would expect...physical consequences...} \YL{In the thermodynamic limit, we would expect that the required number of basis tends to increase as temperature is increased because more high-lying states are accessed. As a result, thermodynamic limit should be pursued with respect to every particular temperatures, which provides a great challenge for further exploration. On the other hand, at the high temperature limit, the system becomes classically homogeneous where the required k-point number to reach thermodynamic limit may decrease.}
As discussed in the Supplemental Information, the path to converging to the thermodynamic limit at finite temperatures, however, is less well-paved than for the ground state. This is because finite temperature calculations inherently involve contributions from high-lying electronic states that converge to the thermodynamic limit differently than the ground state and often require significantly large bases. \YL{Indeed, the form of the finite size scaling for the uniform electron gas has been found to be temperature-dependent. \cite{Malone_Foulkes_JCP_2015,Malone_PRL_2016,White_JCTC_2018}} Achieving the thermodynamic limit for solids in the ground state is most often accomplished by adding both one- and two-body finite size corrections to finite size energy calculations.\cite{Kwee_PRL_2008,Ma_PRB_2011} \YL{Recently, accurate finite size corrections for the uniform electron gas at finite temperatures were derived by combining quantum Monte Carlo simulation results with corrections obtained from random-phase or Singwi-Tosi-Land-Sjölander approximations at small momentum $k$. \cite{dornheim2016ab,brown2013path}} Analogous finite temperature corrections \YL{for \textit{ab initio} solids} have yet to be methodically developed, making rigorous convergence to the thermodynamic limit beyond the scope of this work. \YL{Upon including finite size corrections, it may turn out that additional features may emerge in our heat capacity and other curves. This would be a useful point for future works to resolve.} We therefore encourage future community efforts aimed at deriving such corrections.   

Convergence challenges aside, our work brings into focus a number of other algorithmic improvements that will need to be developed to map the finite temperature phase diagrams of more complex, \emph{ab initio} solids. Lacking better, less costly ways of estimating many-body thermodynamic quantities, in this work, we chose to compute entropies and free energies based upon internal energy data. Nevertheless, the many-body entropy can be directly obtained by taking the logarithm of the many-body density matrix, which requires expanding the logarithm into an infinite number of terms involving increasingly higher powers of the density matrix.\cite{toldin2018entanglement} Future thought should therefore be devoted to deriving more computationally accessible estimators of thermodynamic quantities for \emph{ab initio} treatments that would enable the more accurate characterization of phase diagrams and provide a means of independently validating estimates based upon the energy alone. Moreover, in this work, we simulated down to effective temperatures of 0.01 for the 0.5 \r{A} chain and 0.83 for the 2.50 \r{A} chain (see the last column of Table \ref{tab:Ut}), which precluded us from observing the spin excitations one would expect to observe at even lower temperatures for chains with $R > 2.25$ \AA~ based upon Hubbard model simulations.\cite{shiba1972thermodynamic} To study temperatures immediately lower than those surveyed here, our methodology may be extended by iteratively improving the quality of our trial density matrices without necessarily introducing phaseless or other biases.\cite{he2019finite} However, to simulate to even lower temperatures, ways of addressing the phase problem even beyond the phaseless approximation will be required.  
%\Edits{such that important low-temperature physics such as the spin-reorientation phase transition below 10 K in rare-earth ferroborates \cite{vasiliev2006rare} could readily be modeled.} 
%Thanks, yes, you're right. This should be more positive. 
Accessing the finite temperature physics of some of the most fascinating 3$d$, 4$d$, and 5$d$ transition metal oxides will additionally necessitate careful treatments of spin-orbit coupling, other relativistic effects, electron-phonon coupling, and pseudopotentials, not to mention ways of further accelerating the algorithm to accommodate the larger numbers of orbitals required, leaving open many further algorithmic opportunities for the community to pursue.  
%scaling of our method to plain cubic scaling or sub-cubic scaling with system size.

This all said, we believe that this work demonstrates that fully \emph{ab initio} modeling of finite temperature phase diagrams is within reach. Given increased computational power and proper consideration of the points discussed above, our method can be directly generalized to the study of metal-insulator and magnetic phase transitions in two- and three-dimensional materials, potentially someday including the cuprates. We therefore eagerly look forward to our technique's future maturation and application.

\begin{acknowledgement}
The authors thank Yang Yu, Hongxia Hao, James Shepherd, Emanuel Gull, and Richard Stratt for insightful discussions. B.R. acknowledges funding from the U.S. Department of Energy, Office of Science, Basic Energy Sciences, Materials Sciences and Engineering Division, as part of the Computational Materials Sciences Program and Center for Predictive Simulation of Functional Materials. Y.L. has graciously been supported by the Brown Presidential Fellows program, while T.S. has received partial support from the Brown Open Graduate Education program. H.Z. contributed to this work while a part of the International Exchange Program for Excellent Students of University of Science and Technology of China. All calculations were conducted using computational
resources and services at the Brown University Center for Computation and Visualization.
\end{acknowledgement}

\begin{suppinfo}
Additional theoretical details and tabulated data.
\end{suppinfo}

\bibstyle{achemso}
\bibliography{ref}

\providecommand{\latin}[1]{#1}
\makeatletter
\providecommand{\doi}
  {\begingroup\let\do\@makeother\dospecials
  \catcode`\{=1 \catcode`\}=2 \doi@aux}
\providecommand{\doi@aux}[1]{\endgroup\texttt{#1}}
\makeatother
\providecommand*\mcitethebibliography{\thebibliography}
\csname @ifundefined\endcsname{endmcitethebibliography}
  {\let\endmcitethebibliography\endthebibliography}{}
\begin{mcitethebibliography}{93}
\providecommand*\natexlab[1]{#1}
\providecommand*\mciteSetBstSublistMode[1]{}
\providecommand*\mciteSetBstMaxWidthForm[2]{}
\providecommand*\mciteBstWouldAddEndPuncttrue
  {\def\EndOfBibitem{\unskip.}}
\providecommand*\mciteBstWouldAddEndPunctfalse
  {\let\EndOfBibitem\relax}
\providecommand*\mciteSetBstMidEndSepPunct[3]{}
\providecommand*\mciteSetBstSublistLabelBeginEnd[3]{}
\providecommand*\EndOfBibitem{}
\mciteSetBstSublistMode{f}
\mciteSetBstMaxWidthForm{subitem}{(\alph{mcitesubitemcount})}
\mciteSetBstSublistLabelBeginEnd
  {\mcitemaxwidthsubitemform\space}
  {\relax}
  {\relax}

\bibitem[Ramirez and Falicov(1971)Ramirez, and Falicov]{Ramirez_PRB_1971}
Ramirez,~R.; Falicov,~L.~M. Theory of the
  $\ensuremath{\alpha}\ensuremath{-}\ensuremath{\gamma}$ Phase Transition in
  Metallic Cerium. \emph{Phys. Rev. B} \textbf{1971}, \emph{3},
  2425--2430\relax
\mciteBstWouldAddEndPuncttrue
\mciteSetBstMidEndSepPunct{\mcitedefaultmidpunct}
{\mcitedefaultendpunct}{\mcitedefaultseppunct}\relax
\EndOfBibitem
\bibitem[Zylbersztejn and Mott(1975)Zylbersztejn, and Mott]{Mott_VO_PRB_1975}
Zylbersztejn,~A.; Mott,~N.~F. Metal-insulator transition in vanadium dioxide.
  \emph{Phys. Rev. B} \textbf{1975}, \emph{11}, 4383--4395\relax
\mciteBstWouldAddEndPuncttrue
\mciteSetBstMidEndSepPunct{\mcitedefaultmidpunct}
{\mcitedefaultendpunct}{\mcitedefaultseppunct}\relax
\EndOfBibitem
\bibitem[Imada \latin{et~al.}(1998)Imada, Fujimori, and Tokura]{Imada_RMP_1998}
Imada,~M.; Fujimori,~A.; Tokura,~Y. Metal-insulator transitions. \emph{Rev.
  Mod. Phys.} \textbf{1998}, \emph{70}, 1039--1263\relax
\mciteBstWouldAddEndPuncttrue
\mciteSetBstMidEndSepPunct{\mcitedefaultmidpunct}
{\mcitedefaultendpunct}{\mcitedefaultseppunct}\relax
\EndOfBibitem
\bibitem[Lichtenstein \latin{et~al.}(2001)Lichtenstein, Katsnelson, and
  Kotliar]{Lichtenstein_PRL_2001}
Lichtenstein,~A.~I.; Katsnelson,~M.~I.; Kotliar,~G. Finite-Temperature
  Magnetism of Transition Metals: An ab initio Dynamical Mean-Field Theory.
  \emph{Phys. Rev. Lett.} \textbf{2001}, \emph{87}, 067205\relax
\mciteBstWouldAddEndPuncttrue
\mciteSetBstMidEndSepPunct{\mcitedefaultmidpunct}
{\mcitedefaultendpunct}{\mcitedefaultseppunct}\relax
\EndOfBibitem
\bibitem[Orenstein and Millis(2000)Orenstein, and
  Millis]{Orenstein_Science_2000}
Orenstein,~J.; Millis,~A.~J. Advances in the Physics of High-Temperature
  Superconductivity. \emph{Science} \textbf{2000}, \emph{288}, 468--474\relax
\mciteBstWouldAddEndPuncttrue
\mciteSetBstMidEndSepPunct{\mcitedefaultmidpunct}
{\mcitedefaultendpunct}{\mcitedefaultseppunct}\relax
\EndOfBibitem
\bibitem[Georges \latin{et~al.}(2013)Georges, Medici, and
  Mravlje]{Georges_AnnRev_2013}
Georges,~A.; Medici,~L.~d.; Mravlje,~J. Strong Correlations from Hund‚Äôs
  Coupling. \emph{Annu. Rev. Condens. Matter Phys.} \textbf{2013}, \emph{4},
  137--178\relax
\mciteBstWouldAddEndPuncttrue
\mciteSetBstMidEndSepPunct{\mcitedefaultmidpunct}
{\mcitedefaultendpunct}{\mcitedefaultseppunct}\relax
\EndOfBibitem
\bibitem[{Zhu} \latin{et~al.}(2013){Zhu}, {Albers}, {Haule}, {Kotliar}, and
  {Wills}]{Zhu_NatComm_2013}
{Zhu},~J.-X.; {Albers},~R.~C.; {Haule},~K.; {Kotliar},~G.; {Wills},~J.~M.
  {Site-selective electronic correlation in {\ensuremath{\alpha}}-plutonium
  metal}. \emph{Nat. Commun.} \textbf{2013}, \emph{4}, 2644\relax
\mciteBstWouldAddEndPuncttrue
\mciteSetBstMidEndSepPunct{\mcitedefaultmidpunct}
{\mcitedefaultendpunct}{\mcitedefaultseppunct}\relax
\EndOfBibitem
\bibitem[Wen and Li(2011)Wen, and Li]{Wen_AnnRev_2011}
Wen,~H.-H.; Li,~S. Materials and Novel Superconductivity in Iron Pnictide
  Superconductors. \emph{Annu. Rev. Condens. Matter Phys.} \textbf{2011},
  \emph{2}, 121--140\relax
\mciteBstWouldAddEndPuncttrue
\mciteSetBstMidEndSepPunct{\mcitedefaultmidpunct}
{\mcitedefaultendpunct}{\mcitedefaultseppunct}\relax
\EndOfBibitem
\bibitem[{Maeno} \latin{et~al.}(1994){Maeno}, {Hashimoto}, {Yoshida},
  {Nishizaki}, {Fujita}, {Bednorz}, and {Lichtenberg}]{Maeno_Nature_1994}
{Maeno},~Y.; {Hashimoto},~H.; {Yoshida},~K.; {Nishizaki},~S.; {Fujita},~T.;
  {Bednorz},~J.~G.; {Lichtenberg},~F. {Superconductivity in a layered
  perovskite without copper}. \emph{Nature} \textbf{1994}, \emph{372},
  532--534\relax
\mciteBstWouldAddEndPuncttrue
\mciteSetBstMidEndSepPunct{\mcitedefaultmidpunct}
{\mcitedefaultendpunct}{\mcitedefaultseppunct}\relax
\EndOfBibitem
\bibitem[{Drozdov} \latin{et~al.}(2015){Drozdov}, {Eremets}, {Troyan},
  {Ksenofontov}, and {Shylin}]{Drozdov_Nature_2015}
{Drozdov},~A.~P.; {Eremets},~M.~I.; {Troyan},~I.~A.; {Ksenofontov},~V.;
  {Shylin},~S.~I. {Conventional superconductivity at 203 kelvin at high
  pressures in the sulfur hydride system}. \emph{Nature} \textbf{2015},
  \emph{525}, 73--76\relax
\mciteBstWouldAddEndPuncttrue
\mciteSetBstMidEndSepPunct{\mcitedefaultmidpunct}
{\mcitedefaultendpunct}{\mcitedefaultseppunct}\relax
\EndOfBibitem
\bibitem[Giustino \latin{et~al.}(2007)Giustino, Yates, Souza, Cohen, and
  Louie]{Louie_PRL_2007}
Giustino,~F.; Yates,~J.~R.; Souza,~I.; Cohen,~M.~L.; Louie,~S.~G.
  Electron-Phonon Interaction via Electronic and Lattice Wannier Functions:
  Superconductivity in Boron-Doped Diamond Reexamined. \emph{Phys. Rev. Lett.}
  \textbf{2007}, \emph{98}, 047005\relax
\mciteBstWouldAddEndPuncttrue
\mciteSetBstMidEndSepPunct{\mcitedefaultmidpunct}
{\mcitedefaultendpunct}{\mcitedefaultseppunct}\relax
\EndOfBibitem
\bibitem[Kozik \latin{et~al.}(2010)Kozik, Houcke, Gull, Pollet, Prokof'ev,
  Svistunov, and Troyer]{Kozik_2010}
Kozik,~E.; Houcke,~K.~V.; Gull,~E.; Pollet,~L.; Prokof'ev,~N.; Svistunov,~B.;
  Troyer,~M. Diagrammatic Monte Carlo for correlated fermions. \emph{{EPL}}
  \textbf{2010}, \emph{90}, 10004\relax
\mciteBstWouldAddEndPuncttrue
\mciteSetBstMidEndSepPunct{\mcitedefaultmidpunct}
{\mcitedefaultendpunct}{\mcitedefaultseppunct}\relax
\EndOfBibitem
\bibitem[Zhang(1999)]{zhang1999finite}
Zhang,~S. Finite-temperature monte carlo calculations for systems with
  fermions. \emph{Phys. Rev. Lett.} \textbf{1999}, \emph{83}, 2777\relax
\mciteBstWouldAddEndPuncttrue
\mciteSetBstMidEndSepPunct{\mcitedefaultmidpunct}
{\mcitedefaultendpunct}{\mcitedefaultseppunct}\relax
\EndOfBibitem
\bibitem[Wang and Xiang(1997)Wang, and Xiang]{Wang_PRB_1997}
Wang,~X.; Xiang,~T. Transfer-matrix density-matrix renormalization-group theory
  for thermodynamics of one-dimensional quantum systems. \emph{Phys. Rev. B}
  \textbf{1997}, \emph{56}, 5061--5064\relax
\mciteBstWouldAddEndPuncttrue
\mciteSetBstMidEndSepPunct{\mcitedefaultmidpunct}
{\mcitedefaultendpunct}{\mcitedefaultseppunct}\relax
\EndOfBibitem
\bibitem[Georges \latin{et~al.}(1996)Georges, Kotliar, Krauth, and
  Rozenberg]{Georges_RMP_1996}
Georges,~A.; Kotliar,~G.; Krauth,~W.; Rozenberg,~M.~J. Dynamical mean-field
  theory of strongly correlated fermion systems and the limit of infinite
  dimensions. \emph{Rev. Mod. Phys.} \textbf{1996}, \emph{68}, 13--125\relax
\mciteBstWouldAddEndPuncttrue
\mciteSetBstMidEndSepPunct{\mcitedefaultmidpunct}
{\mcitedefaultendpunct}{\mcitedefaultseppunct}\relax
\EndOfBibitem
\bibitem[Maier \latin{et~al.}(2005)Maier, Jarrell, Pruschke, and
  Hettler]{Maier_RMP_2005}
Maier,~T.; Jarrell,~M.; Pruschke,~T.; Hettler,~M.~H. Quantum cluster theories.
  \emph{Rev. Mod. Phys.} \textbf{2005}, \emph{77}, 1027--1080\relax
\mciteBstWouldAddEndPuncttrue
\mciteSetBstMidEndSepPunct{\mcitedefaultmidpunct}
{\mcitedefaultendpunct}{\mcitedefaultseppunct}\relax
\EndOfBibitem
\bibitem[Savrasov and Kotliar(2004)Savrasov, and Kotliar]{Savrasov_PRB_2004}
Savrasov,~S.~Y.; Kotliar,~G. Spectral density functionals for electronic
  structure calculations. \emph{Phys. Rev. B} \textbf{2004}, \emph{69},
  245101\relax
\mciteBstWouldAddEndPuncttrue
\mciteSetBstMidEndSepPunct{\mcitedefaultmidpunct}
{\mcitedefaultendpunct}{\mcitedefaultseppunct}\relax
\EndOfBibitem
\bibitem[Kotliar \latin{et~al.}(2006)Kotliar, Savrasov, Haule, Oudovenko,
  Parcollet, and Marianetti]{Kotliar_RMP_2006}
Kotliar,~G.; Savrasov,~S.~Y.; Haule,~K.; Oudovenko,~V.~S.; Parcollet,~O.;
  Marianetti,~C.~A. Electronic structure calculations with dynamical mean-field
  theory. \emph{Rev. Mod. Phys.} \textbf{2006}, \emph{78}, 865--951\relax
\mciteBstWouldAddEndPuncttrue
\mciteSetBstMidEndSepPunct{\mcitedefaultmidpunct}
{\mcitedefaultendpunct}{\mcitedefaultseppunct}\relax
\EndOfBibitem
\bibitem[Sun and Kotliar(2002)Sun, and Kotliar]{Sun_PRB_2002}
Sun,~P.; Kotliar,~G. Extended dynamical mean-field theory and $\mathrm{GW}$
  method. \emph{Phys. Rev. B} \textbf{2002}, \emph{66}, 085120\relax
\mciteBstWouldAddEndPuncttrue
\mciteSetBstMidEndSepPunct{\mcitedefaultmidpunct}
{\mcitedefaultendpunct}{\mcitedefaultseppunct}\relax
\EndOfBibitem
\bibitem[Welden \latin{et~al.}(2016)Welden, Rusakov, and Zgid]{Walden_JCP_2016}
Welden,~A.~R.; Rusakov,~A.~A.; Zgid,~D. Exploring connections between
  statistical mechanics and Green's functions for realistic systems:
  Temperature dependent electronic entropy and internal energy from a
  self-consistent second-order Green's function. \emph{J. Chem. Phys.}
  \textbf{2016}, \emph{145}, 204106\relax
\mciteBstWouldAddEndPuncttrue
\mciteSetBstMidEndSepPunct{\mcitedefaultmidpunct}
{\mcitedefaultendpunct}{\mcitedefaultseppunct}\relax
\EndOfBibitem
\bibitem[Kananenka \latin{et~al.}(2016)Kananenka, Phillips, and
  Zgid]{Kananenka_JCTC_2016}
Kananenka,~A.~A.; Phillips,~J.~J.; Zgid,~D. Efficient Temperature-Dependent
  Green's Functions Methods for Realistic Systems: Compact Grids for Orthogonal
  Polynomial Transforms. \emph{J. Chem. Theory Comput.} \textbf{2016},
  \emph{12}, 564--571, PMID: 26735685\relax
\mciteBstWouldAddEndPuncttrue
\mciteSetBstMidEndSepPunct{\mcitedefaultmidpunct}
{\mcitedefaultendpunct}{\mcitedefaultseppunct}\relax
\EndOfBibitem
\bibitem[Kananenka \latin{et~al.}(2016)Kananenka, Welden, Lan, Gull, and
  Zgid]{Kananenka_JCTC_2016_2}
Kananenka,~A.~A.; Welden,~A.~R.; Lan,~T.~N.; Gull,~E.; Zgid,~D. Efficient
  Temperature-Dependent Green's Function Methods for Realistic Systems: Using
  Cubic Spline Interpolation to Approximate Matsubara Green's Functions.
  \emph{J. Chem. Theory Comput.} \textbf{2016}, \emph{12}, 2250--2259, PMID:
  27049642\relax
\mciteBstWouldAddEndPuncttrue
\mciteSetBstMidEndSepPunct{\mcitedefaultmidpunct}
{\mcitedefaultendpunct}{\mcitedefaultseppunct}\relax
\EndOfBibitem
\bibitem[{Sun} \latin{et~al.}(2019){Sun}, {Ray}, {Cui}, {Stoudenmire},
  {Ferrero}, and {Kin-Lic Chan}]{Sun_arXiv_2019}
{Sun},~C.; {Ray},~U.; {Cui},~Z.-H.; {Stoudenmire},~M.; {Ferrero},~M.; {Kin-Lic
  Chan},~G. {Finite temperature density matrix embedding theory}. \emph{arXiv
  e-prints} \textbf{2019}, arXiv:1911.07439\relax
\mciteBstWouldAddEndPuncttrue
\mciteSetBstMidEndSepPunct{\mcitedefaultmidpunct}
{\mcitedefaultendpunct}{\mcitedefaultseppunct}\relax
\EndOfBibitem
\bibitem[Marsman \latin{et~al.}(2009)Marsman, Gruneis, Paier, and
  Kresse]{Marsman_JCP_2009}
Marsman,~M.; Gruneis,~A.; Paier,~J.; Kresse,~G. Second-order Muller‚ Plesset
  perturbation theory applied to extended systems. I. Within the
  projector-augmented-wave formalism using a plane wave basis set. \emph{J.
  Chem. Phys.} \textbf{2009}, \emph{130}, 184103\relax
\mciteBstWouldAddEndPuncttrue
\mciteSetBstMidEndSepPunct{\mcitedefaultmidpunct}
{\mcitedefaultendpunct}{\mcitedefaultseppunct}\relax
\EndOfBibitem
\bibitem[McClain \latin{et~al.}(2017)McClain, Sun, Chan, and
  Berkelbach]{McClain_JCTC_2017}
McClain,~J.; Sun,~Q.; Chan,~G. K.-L.; Berkelbach,~T.~C. Gaussian-Based
  Coupled-Cluster Theory for the Ground-State and Band Structure of Solids.
  \emph{J. Chem. Theory Comput.} \textbf{2017}, \emph{13}, 1209--1218, PMID:
  28218843\relax
\mciteBstWouldAddEndPuncttrue
\mciteSetBstMidEndSepPunct{\mcitedefaultmidpunct}
{\mcitedefaultendpunct}{\mcitedefaultseppunct}\relax
\EndOfBibitem
\bibitem[Booth \latin{et~al.}(2013)Booth, Gr{\"u}neis, Kresse, and
  Alavi]{Booth_Nature_2013}
Booth,~G.~H.; Gr{\"u}neis,~A.; Kresse,~G.; Alavi,~A. Towards an exact
  description of electronic wavefunctions in real solids. \emph{Nature}
  \textbf{2013}, \emph{493}, 365--370\relax
\mciteBstWouldAddEndPuncttrue
\mciteSetBstMidEndSepPunct{\mcitedefaultmidpunct}
{\mcitedefaultendpunct}{\mcitedefaultseppunct}\relax
\EndOfBibitem
\bibitem[{Jha} and {Hirata}(2019){Jha}, and {Hirata}]{Jha_arXiv_2019}
{Jha},~P.~K.; {Hirata},~S. {Finite-Temperature Many-Body Perturbation Theory in
  the Canonical Ensemble}. \emph{arXiv e-prints} \textbf{2019},
  arXiv:1910.07628\relax
\mciteBstWouldAddEndPuncttrue
\mciteSetBstMidEndSepPunct{\mcitedefaultmidpunct}
{\mcitedefaultendpunct}{\mcitedefaultseppunct}\relax
\EndOfBibitem
\bibitem[White and Chan(2018)White, and Chan]{White_JCTC_2018}
White,~A.~F.; Chan,~G. K.-L. A Time-Dependent Formulation of Coupled-Cluster
  Theory for Many-Fermion Systems at Finite Temperature. \emph{J. Chem. Theory
  Comput.} \textbf{2018}, \emph{14}, 5690--5700, PMID: 30260642\relax
\mciteBstWouldAddEndPuncttrue
\mciteSetBstMidEndSepPunct{\mcitedefaultmidpunct}
{\mcitedefaultendpunct}{\mcitedefaultseppunct}\relax
\EndOfBibitem
\bibitem[Hummel(2018)]{hummel2018finite}
Hummel,~F. Finite temperature coupled cluster theories for extended systems.
  \emph{J. Chem. Theory Comput.} \textbf{2018}, \emph{14}, 6505--6514\relax
\mciteBstWouldAddEndPuncttrue
\mciteSetBstMidEndSepPunct{\mcitedefaultmidpunct}
{\mcitedefaultendpunct}{\mcitedefaultseppunct}\relax
\EndOfBibitem
\bibitem[Harsha \latin{et~al.}(2019)Harsha, Henderson, and
  Scuseria]{harsha2019thermofield}
Harsha,~G.; Henderson,~T.~M.; Scuseria,~G.~E. Thermofield theory for
  finite-temperature coupled cluster. \emph{J. Chem. Theory Comput.}
  \textbf{2019}, \emph{15}, 6127--6136\relax
\mciteBstWouldAddEndPuncttrue
\mciteSetBstMidEndSepPunct{\mcitedefaultmidpunct}
{\mcitedefaultendpunct}{\mcitedefaultseppunct}\relax
\EndOfBibitem
\bibitem[White and Chan(2020)White, and Chan]{white2020finite}
White,~A.~F.; Chan,~G.~K. Finite-temperature coupled cluster: Efficient
  implementation and application to prototypical systems. \emph{arXiv preprint
  arXiv:2004.01729} \textbf{2020}, \relax
\mciteBstWouldAddEndPunctfalse
\mciteSetBstMidEndSepPunct{\mcitedefaultmidpunct}
{}{\mcitedefaultseppunct}\relax
\EndOfBibitem
\bibitem[Petras \latin{et~al.}(2020)Petras, Ramadugu, Malone, and
  Shepherd]{petras2020using}
Petras,~H.~R.; Ramadugu,~S.~K.; Malone,~F.~D.; Shepherd,~J.~J. Using Density
  Matrix Quantum Monte Carlo for Calculating Exact-on-Average Energies for ab
  Initio Hamiltonians in a Finite Basis Set. \emph{J. Chem. Theory Comput.}
  \textbf{2020}, \emph{16}, 1029--1038\relax
\mciteBstWouldAddEndPuncttrue
\mciteSetBstMidEndSepPunct{\mcitedefaultmidpunct}
{\mcitedefaultendpunct}{\mcitedefaultseppunct}\relax
\EndOfBibitem
\bibitem[Militzer and Driver(2015)Militzer, and Driver]{Militzer_PRL_2015}
Militzer,~B.; Driver,~K.~P. Development of Path Integral Monte Carlo
  Simulations with Localized Nodal Surfaces for Second-Row Elements.
  \emph{Phys. Rev. Lett.} \textbf{2015}, \emph{115}, 176403\relax
\mciteBstWouldAddEndPuncttrue
\mciteSetBstMidEndSepPunct{\mcitedefaultmidpunct}
{\mcitedefaultendpunct}{\mcitedefaultseppunct}\relax
\EndOfBibitem
\bibitem[Liu \latin{et~al.}(2018)Liu, Cho, and Rubenstein]{Liu_JCTC_2018}
Liu,~Y.; Cho,~M.; Rubenstein,~B. Ab Initio Finite Temperature Auxiliary Field
  Quantum Monte Carlo. \emph{J. Chem. Theory Comput.} \textbf{2018}, \emph{14},
  4722--4732, PMID: 30102856\relax
\mciteBstWouldAddEndPuncttrue
\mciteSetBstMidEndSepPunct{\mcitedefaultmidpunct}
{\mcitedefaultendpunct}{\mcitedefaultseppunct}\relax
\EndOfBibitem
\bibitem[Rubenstein \latin{et~al.}(2012)Rubenstein, Zhang, and
  Reichman]{rubenstein2012finite}
Rubenstein,~B.~M.; Zhang,~S.; Reichman,~D.~R. Finite-temperature
  auxiliary-field quantum Monte Carlo technique for Bose-Fermi mixtures.
  \emph{Phys. Rev. A} \textbf{2012}, \emph{86}, 053606\relax
\mciteBstWouldAddEndPuncttrue
\mciteSetBstMidEndSepPunct{\mcitedefaultmidpunct}
{\mcitedefaultendpunct}{\mcitedefaultseppunct}\relax
\EndOfBibitem
\bibitem[Szabo and Ostlund(1996)Szabo, and Ostlund]{szabo1996modern}
Szabo,~A.; Ostlund,~N. \emph{Modern Quantum Chemistry: Introduction to Advanced
  Electronic Structure Theory}; Dover Books on Chemistry; Dover Publications,
  1996\relax
\mciteBstWouldAddEndPuncttrue
\mciteSetBstMidEndSepPunct{\mcitedefaultmidpunct}
{\mcitedefaultendpunct}{\mcitedefaultseppunct}\relax
\EndOfBibitem
\bibitem[Stella \latin{et~al.}(2011)Stella, Attaccalite, Sorella, and
  Rubio]{Stella_PRB_2011}
Stella,~L.; Attaccalite,~C.; Sorella,~S.; Rubio,~A. Strong electronic
  correlation in the hydrogen chain: A variational Monte Carlo study.
  \emph{Phys. Rev. B} \textbf{2011}, \emph{84}, 245117\relax
\mciteBstWouldAddEndPuncttrue
\mciteSetBstMidEndSepPunct{\mcitedefaultmidpunct}
{\mcitedefaultendpunct}{\mcitedefaultseppunct}\relax
\EndOfBibitem
\bibitem[Hachmann \latin{et~al.}(2006)Hachmann, Cardoen, and
  Chan]{Hachman_2006}
Hachmann,~J.; Cardoen,~W.; Chan,~G. K.-L. Multireference correlation in long
  molecules with the quadratic scaling density matrix renormalization group.
  \emph{J. Chem. Phys.} \textbf{2006}, \emph{125}, 144101\relax
\mciteBstWouldAddEndPuncttrue
\mciteSetBstMidEndSepPunct{\mcitedefaultmidpunct}
{\mcitedefaultendpunct}{\mcitedefaultseppunct}\relax
\EndOfBibitem
\bibitem[Motta \latin{et~al.}(2019)Motta, Genovese, Ma, Cui, Sawaya, Chan,
  Chepiga, Helms, Jimenez-Hoyos, Millis, \latin{et~al.}
  others]{motta2019ground}
Motta,~M.; Genovese,~C.; Ma,~F.; Cui,~Z.-H.; Sawaya,~R.; Chan,~G.~K.;
  Chepiga,~N.; Helms,~P.; Jimenez-Hoyos,~C.; Millis,~A.~J., \latin{et~al.}
  Ground-state properties of the hydrogen chain: insulator-to-metal transition,
  dimerization, and magnetic phases. \emph{arXiv preprint arXiv:1911.01618}
  \textbf{2019}, \relax
\mciteBstWouldAddEndPunctfalse
\mciteSetBstMidEndSepPunct{\mcitedefaultmidpunct}
{}{\mcitedefaultseppunct}\relax
\EndOfBibitem
\bibitem[Motta \latin{et~al.}(2017)Motta, Ceperley, Chan, Gomez, Gull, Guo,
  Jim{\'e}nez-Hoyos, Lan, Li, Ma, \latin{et~al.} others]{motta2017towards}
Motta,~M.; Ceperley,~D.~M.; Chan,~G. K.-L.; Gomez,~J.~A.; Gull,~E.; Guo,~S.;
  Jim{\'e}nez-Hoyos,~C.~A.; Lan,~T.~N.; Li,~J.; Ma,~F., \latin{et~al.}  Towards
  the solution of the many-electron problem in real materials: equation of
  state of the hydrogen chain with state-of-the-art many-body methods.
  \emph{Phys. Rev. X} \textbf{2017}, \emph{7}, 031059\relax
\mciteBstWouldAddEndPuncttrue
\mciteSetBstMidEndSepPunct{\mcitedefaultmidpunct}
{\mcitedefaultendpunct}{\mcitedefaultseppunct}\relax
\EndOfBibitem
\bibitem[Sinitskiy \latin{et~al.}(2010)Sinitskiy, Greenman, and
  Mazziotti]{Sinitskiy_JCP_2010}
Sinitskiy,~A.~V.; Greenman,~L.; Mazziotti,~D.~A. Strong correlation in hydrogen
  chains and lattices using the variational two-electron reduced density matrix
  method. \emph{J. Chem. Phys.} \textbf{2010}, \emph{133}, 014104\relax
\mciteBstWouldAddEndPuncttrue
\mciteSetBstMidEndSepPunct{\mcitedefaultmidpunct}
{\mcitedefaultendpunct}{\mcitedefaultseppunct}\relax
\EndOfBibitem
\bibitem[Rusakov and Zgid(2016)Rusakov, and Zgid]{rusakov2016self}
Rusakov,~A.~A.; Zgid,~D. Self-consistent second-order Green’s function
  perturbation theory for periodic systems. \emph{J. Chem. Phys.}
  \textbf{2016}, \emph{144}, 054106\relax
\mciteBstWouldAddEndPuncttrue
\mciteSetBstMidEndSepPunct{\mcitedefaultmidpunct}
{\mcitedefaultendpunct}{\mcitedefaultseppunct}\relax
\EndOfBibitem
\bibitem[Welden \latin{et~al.}(2016)Welden, Rusakov, and
  Zgid]{welden2016exploring}
Welden,~A.~R.; Rusakov,~A.~A.; Zgid,~D. Exploring connections between
  statistical mechanics and Green’s functions for realistic systems:
  Temperature dependent electronic entropy and internal energy from a
  self-consistent second-order Green’s function. \emph{J. Chem. Phys.}
  \textbf{2016}, \emph{145}, 204106\relax
\mciteBstWouldAddEndPuncttrue
\mciteSetBstMidEndSepPunct{\mcitedefaultmidpunct}
{\mcitedefaultendpunct}{\mcitedefaultseppunct}\relax
\EndOfBibitem
\bibitem[Lieb and Wu(1994)Lieb, and Wu]{lieb1994absence}
Lieb,~E.~H.; Wu,~F.~Y. \emph{Exactly Solvable Models Of Strongly Correlated
  Electrons}; World Scientific, 1994; pp 9--12\relax
\mciteBstWouldAddEndPuncttrue
\mciteSetBstMidEndSepPunct{\mcitedefaultmidpunct}
{\mcitedefaultendpunct}{\mcitedefaultseppunct}\relax
\EndOfBibitem
\bibitem[Schulz(1993)]{schulz1993wigner}
Schulz,~H. Wigner crystal in one dimension. \emph{Phys. Rev. Lett.}
  \textbf{1993}, \emph{71}, 1864\relax
\mciteBstWouldAddEndPuncttrue
\mciteSetBstMidEndSepPunct{\mcitedefaultmidpunct}
{\mcitedefaultendpunct}{\mcitedefaultseppunct}\relax
\EndOfBibitem
\bibitem[Gebhard \latin{et~al.}(1994)Gebhard, Girndt, and
  Ruckenstein]{gebhard1994charge}
Gebhard,~F.; Girndt,~A.; Ruckenstein,~A.~E. Charge-and spin-gap formation in
  exactly solvable Hubbard chains with long-range hopping. \emph{Phys. Rev. B}
  \textbf{1994}, \emph{49}, 10926\relax
\mciteBstWouldAddEndPuncttrue
\mciteSetBstMidEndSepPunct{\mcitedefaultmidpunct}
{\mcitedefaultendpunct}{\mcitedefaultseppunct}\relax
\EndOfBibitem
\bibitem[Hirsch(1984)]{hirsch1984charge}
Hirsch,~J. Charge-density-wave to spin-density-wave transition in the extended
  Hubbard model. \emph{Phys. Rev. Lett.} \textbf{1984}, \emph{53}, 2327\relax
\mciteBstWouldAddEndPuncttrue
\mciteSetBstMidEndSepPunct{\mcitedefaultmidpunct}
{\mcitedefaultendpunct}{\mcitedefaultseppunct}\relax
\EndOfBibitem
\bibitem[Kuroki \latin{et~al.}(1994)Kuroki, Kusakabe, and
  Aoki]{kuroki1994phase}
Kuroki,~K.; Kusakabe,~K.; Aoki,~H. Phase diagram of the extended attractive
  Hubbard model in one dimension. \emph{Phys. Rev. B} \textbf{1994}, \emph{50},
  575\relax
\mciteBstWouldAddEndPuncttrue
\mciteSetBstMidEndSepPunct{\mcitedefaultmidpunct}
{\mcitedefaultendpunct}{\mcitedefaultseppunct}\relax
\EndOfBibitem
\bibitem[Glocke \latin{et~al.}(2007)Glocke, Kl{\"u}mper, and
  Sirker]{glocke2007half}
Glocke,~S.; Kl{\"u}mper,~A.; Sirker,~J. Half-filled one-dimensional extended
  Hubbard model: Phase diagram and thermodynamics. \emph{Phys. Rev. B}
  \textbf{2007}, \emph{76}, 155121\relax
\mciteBstWouldAddEndPuncttrue
\mciteSetBstMidEndSepPunct{\mcitedefaultmidpunct}
{\mcitedefaultendpunct}{\mcitedefaultseppunct}\relax
\EndOfBibitem
\bibitem[Sun \latin{et~al.}(2017)Sun, Berkelbach, McClain, and
  Chan]{Sun_JCP_2017}
Sun,~Q.; Berkelbach,~T.~C.; McClain,~J.~D.; Chan,~G. K.-L. Gaussian and
  plane-wave mixed density fitting for periodic systems. \emph{J. Chem. Phys.}
  \textbf{2017}, \emph{147}, 164119\relax
\mciteBstWouldAddEndPuncttrue
\mciteSetBstMidEndSepPunct{\mcitedefaultmidpunct}
{\mcitedefaultendpunct}{\mcitedefaultseppunct}\relax
\EndOfBibitem
\bibitem[White \latin{et~al.}(1988)White, Sugar, and
  Scalettar]{white1988algorithm}
White,~S.; Sugar,~R.; Scalettar,~R. Algorithm for the simulation of
  many-electron systems at low temperatures. \emph{Phys. Rev. B} \textbf{1988},
  \emph{38}, 11665\relax
\mciteBstWouldAddEndPuncttrue
\mciteSetBstMidEndSepPunct{\mcitedefaultmidpunct}
{\mcitedefaultendpunct}{\mcitedefaultseppunct}\relax
\EndOfBibitem
\bibitem[Motta and Zhang(2017)Motta, and Zhang]{Motta_Review_2018}
Motta,~M.; Zhang,~S. Ab Initio Computations of Molecular Systems by the
  Auxiliary-Field Quantum Monte Carlo Method. \emph{arXiv:1711.02242}
  \textbf{2017}, \relax
\mciteBstWouldAddEndPunctfalse
\mciteSetBstMidEndSepPunct{\mcitedefaultmidpunct}
{}{\mcitedefaultseppunct}\relax
\EndOfBibitem
\bibitem[Hirsch(1985)]{Hirsch_PRB_1985}
Hirsch,~J.~E. Two-dimensional Hubbard model: Numerical simulation study.
  \emph{Phys. Rev. B} \textbf{1985}, \emph{31}, 4403--4419\relax
\mciteBstWouldAddEndPuncttrue
\mciteSetBstMidEndSepPunct{\mcitedefaultmidpunct}
{\mcitedefaultendpunct}{\mcitedefaultseppunct}\relax
\EndOfBibitem
\bibitem[Hirsch(1983)]{Hirsch_PRB_1983}
Hirsch,~J.~E. Discrete Hubbard-Stratonovich transformation for fermion lattice
  models. \emph{Phys. Rev. B} \textbf{1983}, \emph{28}, 4059--4061\relax
\mciteBstWouldAddEndPuncttrue
\mciteSetBstMidEndSepPunct{\mcitedefaultmidpunct}
{\mcitedefaultendpunct}{\mcitedefaultseppunct}\relax
\EndOfBibitem
\bibitem[Hirsch \latin{et~al.}(1982)Hirsch, Sugar, Scalapino, and
  Blankenbecler]{hirsch1982monte}
Hirsch,~J.~E.; Sugar,~R.~L.; Scalapino,~D.~J.; Blankenbecler,~R. Monte Carlo
  simulations of one-dimensional fermion systems. \emph{Phys. Rev. B}
  \textbf{1982}, \emph{26}, 5033\relax
\mciteBstWouldAddEndPuncttrue
\mciteSetBstMidEndSepPunct{\mcitedefaultmidpunct}
{\mcitedefaultendpunct}{\mcitedefaultseppunct}\relax
\EndOfBibitem
\bibitem[Zhang and Krakauer(2003)Zhang, and Krakauer]{Zhang_PRL_2003}
Zhang,~S.; Krakauer,~H. Quantum Monte Carlo Method using Phase-Free Random
  Walks with Slater Determinants. \emph{Phys. Rev. Lett.} \textbf{2003},
  \emph{90}, 136401\relax
\mciteBstWouldAddEndPuncttrue
\mciteSetBstMidEndSepPunct{\mcitedefaultmidpunct}
{\mcitedefaultendpunct}{\mcitedefaultseppunct}\relax
\EndOfBibitem
\bibitem[McClain \latin{et~al.}(2017)McClain, Sun, Chan, and
  Berkelbach]{mcclain2017gaussian}
McClain,~J.; Sun,~Q.; Chan,~G. K.-L.; Berkelbach,~T.~C. Gaussian-based
  coupled-cluster theory for the ground-state and band structure of solids.
  \emph{J. Chem. Theory Comput.} \textbf{2017}, \emph{13}, 1209--1218\relax
\mciteBstWouldAddEndPuncttrue
\mciteSetBstMidEndSepPunct{\mcitedefaultmidpunct}
{\mcitedefaultendpunct}{\mcitedefaultseppunct}\relax
\EndOfBibitem
\bibitem[Martin(2004)]{martin2004electronic}
Martin,~R. \emph{Electronic Structure: Basic Theory and Practical Methods};
  Cambridge University Press, 2004\relax
\mciteBstWouldAddEndPuncttrue
\mciteSetBstMidEndSepPunct{\mcitedefaultmidpunct}
{\mcitedefaultendpunct}{\mcitedefaultseppunct}\relax
\EndOfBibitem
\bibitem[Sun \latin{et~al.}(2017)Sun, Berkelbach, McClain, and
  Chan]{sun2017gaussian}
Sun,~Q.; Berkelbach,~T.~C.; McClain,~J.~D.; Chan,~G. K.-L. Gaussian and
  plane-wave mixed density fitting for periodic systems. \emph{J. Chem. Phys.}
  \textbf{2017}, \emph{147}, 164119\relax
\mciteBstWouldAddEndPuncttrue
\mciteSetBstMidEndSepPunct{\mcitedefaultmidpunct}
{\mcitedefaultendpunct}{\mcitedefaultseppunct}\relax
\EndOfBibitem
\bibitem[Predescu \latin{et~al.}(2003)Predescu, Sabo, Doll, and
  Freeman]{predescu2003heat}
Predescu,~C.; Sabo,~D.; Doll,~J.; Freeman,~D.~L. Heat capacity estimators for
  random series path-integral methods by finite-difference schemes. \emph{J.
  Chem. Phys.} \textbf{2003}, \emph{119}, 12119--12128\relax
\mciteBstWouldAddEndPuncttrue
\mciteSetBstMidEndSepPunct{\mcitedefaultmidpunct}
{\mcitedefaultendpunct}{\mcitedefaultseppunct}\relax
\EndOfBibitem
\bibitem[Fetter and Walecka(2003)Fetter, and Walecka]{fetter2012quantum}
Fetter,~A.~L.; Walecka,~J.~D. \emph{Quantum theory of many-particle systems};
  Dover Publications, Inc., 2003\relax
\mciteBstWouldAddEndPuncttrue
\mciteSetBstMidEndSepPunct{\mcitedefaultmidpunct}
{\mcitedefaultendpunct}{\mcitedefaultseppunct}\relax
\EndOfBibitem
\bibitem[Callen \latin{et~al.}(1985)Callen, Callen, of~Australia.
  Research~Division, and Sons]{callen1985thermodynamics}
Callen,~H.; Callen,~H.; of~Australia. Research~Division,~N. F. R.~C.;
  Sons,~W.~. \emph{Thermodynamics and an Introduction to Thermostatistics};
  Wiley, 1985\relax
\mciteBstWouldAddEndPuncttrue
\mciteSetBstMidEndSepPunct{\mcitedefaultmidpunct}
{\mcitedefaultendpunct}{\mcitedefaultseppunct}\relax
\EndOfBibitem
\bibitem[Malone \latin{et~al.}(2016)Malone, Blunt, Brown, Lee, Spencer,
  Foulkes, and Shepherd]{Malone_PRL_2016}
Malone,~F.~D.; Blunt,~N.~S.; Brown,~E.~W.; Lee,~D. K.~K.; Spencer,~J.~S.;
  Foulkes,~W. M.~C.; Shepherd,~J.~J. Accurate Exchange-Correlation Energies for
  the Warm Dense Electron Gas. \emph{Phys. Rev. Lett.} \textbf{2016},
  \emph{117}, 115701\relax
\mciteBstWouldAddEndPuncttrue
\mciteSetBstMidEndSepPunct{\mcitedefaultmidpunct}
{\mcitedefaultendpunct}{\mcitedefaultseppunct}\relax
\EndOfBibitem
\bibitem[Toldin and Assaad(2018)Toldin, and Assaad]{toldin2018entanglement}
Toldin,~F.~P.; Assaad,~F.~F. Entanglement Hamiltonian of interacting fermionic
  models. \emph{Phys. Rev. Lett.} \textbf{2018}, \emph{121}, 200602\relax
\mciteBstWouldAddEndPuncttrue
\mciteSetBstMidEndSepPunct{\mcitedefaultmidpunct}
{\mcitedefaultendpunct}{\mcitedefaultseppunct}\relax
\EndOfBibitem
\bibitem[Blunt \latin{et~al.}(2014)Blunt, Rogers, Spencer, and
  Foulkes]{Blunt_PRB_2014}
Blunt,~N.~S.; Rogers,~T.~W.; Spencer,~J.~S.; Foulkes,~W. M.~C. Density-matrix
  quantum Monte Carlo method. \emph{Phys. Rev. B} \textbf{2014}, \emph{89},
  245124\relax
\mciteBstWouldAddEndPuncttrue
\mciteSetBstMidEndSepPunct{\mcitedefaultmidpunct}
{\mcitedefaultendpunct}{\mcitedefaultseppunct}\relax
\EndOfBibitem
\bibitem[Sun \latin{et~al.}(2018)Sun, Berkelbach, Blunt, Booth, Guo, Li, Liu,
  McClain, Sayfutyarova, Sharma, \latin{et~al.} others]{sun2018pyscf}
Sun,~Q.; Berkelbach,~T.~C.; Blunt,~N.~S.; Booth,~G.~H.; Guo,~S.; Li,~Z.;
  Liu,~J.; McClain,~J.~D.; Sayfutyarova,~E.~R.; Sharma,~S., \latin{et~al.}
  PySCF: the Python-based simulations of chemistry framework. \emph{Wiley
  Interdiscip. Rev. Comput. Mol. Sci.} \textbf{2018}, \emph{8}, e1340\relax
\mciteBstWouldAddEndPuncttrue
\mciteSetBstMidEndSepPunct{\mcitedefaultmidpunct}
{\mcitedefaultendpunct}{\mcitedefaultseppunct}\relax
\EndOfBibitem
\bibitem[Liu(2020. https://doi.org/10.26300/hvdk-6836)]{BDR-integrals}
Liu,~Y. \emph{Integral Files. Brown University Open Data Collection}; Brown
  Digital Repository. Brown University Library, 2020.
  https://doi.org/10.26300/hvdk-6836\relax
\mciteBstWouldAddEndPuncttrue
\mciteSetBstMidEndSepPunct{\mcitedefaultmidpunct}
{\mcitedefaultendpunct}{\mcitedefaultseppunct}\relax
\EndOfBibitem
\bibitem[Liu(2020. https://doi.org/10.26300/0m74-xp13)]{BDR-figures-data}
Liu,~Y. \emph{Figures Data. Brown University Open Data Collection}; Brown
  Digital Repository. Brown University Library, 2020.
  https://doi.org/10.26300/0m74-xp13\relax
\mciteBstWouldAddEndPuncttrue
\mciteSetBstMidEndSepPunct{\mcitedefaultmidpunct}
{\mcitedefaultendpunct}{\mcitedefaultseppunct}\relax
\EndOfBibitem
\bibitem[Hubbard(1963)]{hubbard1963electron}
Hubbard,~J. Electron correlations in narrow energy bands. \emph{Proc. R. Soc.
  A} \textbf{1963}, \emph{276}, 238--257\relax
\mciteBstWouldAddEndPuncttrue
\mciteSetBstMidEndSepPunct{\mcitedefaultmidpunct}
{\mcitedefaultendpunct}{\mcitedefaultseppunct}\relax
\EndOfBibitem
\bibitem[Ghosh(1970)]{ghosh1970electron}
Ghosh,~D. Electron correlation in an extended Hubbard model. \emph{Phys. Lett.
  A} \textbf{1970}, \emph{32}, 429--431\relax
\mciteBstWouldAddEndPuncttrue
\mciteSetBstMidEndSepPunct{\mcitedefaultmidpunct}
{\mcitedefaultendpunct}{\mcitedefaultseppunct}\relax
\EndOfBibitem
\bibitem[Rusakov \latin{et~al.}(2018)Rusakov, Iskakov, Tran, and
  Zgid]{rusakov2018self}
Rusakov,~A.~A.; Iskakov,~S.; Tran,~L.~N.; Zgid,~D. Self-energy embedding theory
  (SEET) for periodic systems. \emph{J. Chem. Theory Comput.} \textbf{2018},
  \emph{15}, 229--240\relax
\mciteBstWouldAddEndPuncttrue
\mciteSetBstMidEndSepPunct{\mcitedefaultmidpunct}
{\mcitedefaultendpunct}{\mcitedefaultseppunct}\relax
\EndOfBibitem
\bibitem[Shiba(1972)]{shiba1972thermodynamic}
Shiba,~H. Thermodynamic Properties of the One-Dimensional Half-Filled-Band
  Hubbard Model. II: Application of the Grand Canonical Method. \emph{Prog.
  Theor. Phys.} \textbf{1972}, \emph{48}, 2171--2186\relax
\mciteBstWouldAddEndPuncttrue
\mciteSetBstMidEndSepPunct{\mcitedefaultmidpunct}
{\mcitedefaultendpunct}{\mcitedefaultseppunct}\relax
\EndOfBibitem
\bibitem[Mott(2004)]{mott2004metal}
Mott,~N. \emph{Metal-Insulator Transitions}; CRC Press, 2004\relax
\mciteBstWouldAddEndPuncttrue
\mciteSetBstMidEndSepPunct{\mcitedefaultmidpunct}
{\mcitedefaultendpunct}{\mcitedefaultseppunct}\relax
\EndOfBibitem
\bibitem[Chitra and Kotliar(2000)Chitra, and Kotliar]{chitra2000effect}
Chitra,~R.; Kotliar,~G. Effect of long range coulomb interactions on the mott
  transition. \emph{Phys. Rev. Lett.} \textbf{2000}, \emph{84}, 3678\relax
\mciteBstWouldAddEndPuncttrue
\mciteSetBstMidEndSepPunct{\mcitedefaultmidpunct}
{\mcitedefaultendpunct}{\mcitedefaultseppunct}\relax
\EndOfBibitem
\bibitem[Poilblanc \latin{et~al.}(1997)Poilblanc, Yunoki, Maekawa, and
  Dagotto]{poilblanc1997insulator}
Poilblanc,~D.; Yunoki,~S.; Maekawa,~S.; Dagotto,~E. Insulator-metal transition
  in one dimension induced by long-range electronic interactions. \emph{Phys.
  Rev. B} \textbf{1997}, \emph{56}, R1645\relax
\mciteBstWouldAddEndPuncttrue
\mciteSetBstMidEndSepPunct{\mcitedefaultmidpunct}
{\mcitedefaultendpunct}{\mcitedefaultseppunct}\relax
\EndOfBibitem
\bibitem[{J{\"o}rdens} \latin{et~al.}(2008){J{\"o}rdens}, {Strohmaier},
  {G{\"u}nter}, {Moritz}, and {Esslinger}]{Jordens_Nature}
{J{\"o}rdens},~R.; {Strohmaier},~N.; {G{\"u}nter},~K.; {Moritz},~H.;
  {Esslinger},~T. {A Mott insulator of fermionic atoms in an optical lattice}.
  \emph{Nature} \textbf{2008}, \emph{455}, 204--207\relax
\mciteBstWouldAddEndPuncttrue
\mciteSetBstMidEndSepPunct{\mcitedefaultmidpunct}
{\mcitedefaultendpunct}{\mcitedefaultseppunct}\relax
\EndOfBibitem
\bibitem[Scarola \latin{et~al.}(2009)Scarola, Pollet, Oitmaa, and
  Troyer]{Scarola_PRL_2009}
Scarola,~V.~W.; Pollet,~L.; Oitmaa,~J.; Troyer,~M. Discerning Incompressible
  and Compressible Phases of Cold Atoms in Optical Lattices. \emph{Phys. Rev.
  Lett.} \textbf{2009}, \emph{102}, 135302\relax
\mciteBstWouldAddEndPuncttrue
\mciteSetBstMidEndSepPunct{\mcitedefaultmidpunct}
{\mcitedefaultendpunct}{\mcitedefaultseppunct}\relax
\EndOfBibitem
\bibitem[Georges and Krauth(1992)Georges, and Krauth]{georges1992numerical}
Georges,~A.; Krauth,~W. Numerical solution of the d=$\infty$ Hubbard model:
  Evidence for a Mott transition. \emph{Phys. Rev. Lett.} \textbf{1992},
  \emph{69}, 1240\relax
\mciteBstWouldAddEndPuncttrue
\mciteSetBstMidEndSepPunct{\mcitedefaultmidpunct}
{\mcitedefaultendpunct}{\mcitedefaultseppunct}\relax
\EndOfBibitem
\bibitem[Georges and Krauth(1993)Georges, and Krauth]{georges1993physical}
Georges,~A.; Krauth,~W. Physical properties of the half-filled Hubbard model in
  infinite dimensions. \emph{Phys. Rev. B} \textbf{1993}, \emph{48}, 7167\relax
\mciteBstWouldAddEndPuncttrue
\mciteSetBstMidEndSepPunct{\mcitedefaultmidpunct}
{\mcitedefaultendpunct}{\mcitedefaultseppunct}\relax
\EndOfBibitem
\bibitem[Sciolla \latin{et~al.}(2013)Sciolla, Tokuno, Uchino, Barmettler,
  Giamarchi, and Kollath]{sciolla2013competition}
Sciolla,~B.; Tokuno,~A.; Uchino,~S.; Barmettler,~P.; Giamarchi,~T.; Kollath,~C.
  Competition of spin and charge excitations in the one-dimensional Hubbard
  model. \emph{Phys. Rev. A} \textbf{2013}, \emph{88}, 063629\relax
\mciteBstWouldAddEndPuncttrue
\mciteSetBstMidEndSepPunct{\mcitedefaultmidpunct}
{\mcitedefaultendpunct}{\mcitedefaultseppunct}\relax
\EndOfBibitem
\bibitem[Shen \latin{et~al.}(1991)Shen, List, Dessau, Wells, Jepsen, Arko,
  Barttlet, Shih, Parmigiani, Huang, \latin{et~al.} others]{shen1991electronic}
Shen,~Z.-X.; List,~R.; Dessau,~D.; Wells,~B.; Jepsen,~O.; Arko,~A.;
  Barttlet,~R.; Shih,~C.-K.; Parmigiani,~F.; Huang,~J., \latin{et~al.}
  Electronic structure of {NiO}: Correlation and band effects. \emph{Phys. Rev.
  B} \textbf{1991}, \emph{44}, 3604\relax
\mciteBstWouldAddEndPuncttrue
\mciteSetBstMidEndSepPunct{\mcitedefaultmidpunct}
{\mcitedefaultendpunct}{\mcitedefaultseppunct}\relax
\EndOfBibitem
\bibitem[Kune{\v{s}} \latin{et~al.}(2007)Kune{\v{s}}, Anisimov, Skornyakov,
  Lukoyanov, and Vollhardt]{kunevs2007nio}
Kune{\v{s}},~J.; Anisimov,~V.; Skornyakov,~S.; Lukoyanov,~A.; Vollhardt,~D.
  {NiO}: correlated band structure of a charge-transfer insulator. \emph{Phys.
  Rev. Lett.} \textbf{2007}, \emph{99}, 156404\relax
\mciteBstWouldAddEndPuncttrue
\mciteSetBstMidEndSepPunct{\mcitedefaultmidpunct}
{\mcitedefaultendpunct}{\mcitedefaultseppunct}\relax
\EndOfBibitem
\bibitem[Voit(1992)]{voit1992phase}
Voit,~J. Phase diagram and correlation functions of the half-filled extended
  Hubbard model in one dimension. \emph{Phys. Rev. B} \textbf{1992}, \emph{45},
  4027\relax
\mciteBstWouldAddEndPuncttrue
\mciteSetBstMidEndSepPunct{\mcitedefaultmidpunct}
{\mcitedefaultendpunct}{\mcitedefaultseppunct}\relax
\EndOfBibitem
\bibitem[L{\"o}wdin(1955)]{lowdin1955quantum}
L{\"o}wdin,~P.-O. Quantum theory of many-particle systems. I. Physical
  interpretations by means of density matrices, natural spin-orbitals, and
  convergence problems in the method of configurational interaction.
  \emph{Phys. Rev.} \textbf{1955}, \emph{97}, 1474\relax
\mciteBstWouldAddEndPuncttrue
\mciteSetBstMidEndSepPunct{\mcitedefaultmidpunct}
{\mcitedefaultendpunct}{\mcitedefaultseppunct}\relax
\EndOfBibitem
\bibitem[Mermin and Wagner(1966)Mermin, and Wagner]{mermin1966absence}
Mermin,~N.~D.; Wagner,~H. Absence of ferromagnetism or antiferromagnetism in
  one-or two-dimensional isotropic Heisenberg models. \emph{Phys. Rev. Lett.}
  \textbf{1966}, \emph{17}, 1133\relax
\mciteBstWouldAddEndPuncttrue
\mciteSetBstMidEndSepPunct{\mcitedefaultmidpunct}
{\mcitedefaultendpunct}{\mcitedefaultseppunct}\relax
\EndOfBibitem
\bibitem[Ghosh(1971)]{ghosh1971nonexistence}
Ghosh,~D.~K. Nonexistence of magnetic ordering in the one-and two-dimensional
  Hubbard model. \emph{Phys. Rev. Lett.} \textbf{1971}, \emph{27}, 1584\relax
\mciteBstWouldAddEndPuncttrue
\mciteSetBstMidEndSepPunct{\mcitedefaultmidpunct}
{\mcitedefaultendpunct}{\mcitedefaultseppunct}\relax
\EndOfBibitem
\bibitem[Malone \latin{et~al.}(2015)Malone, Blunt, Shepherd, Lee, Spencer, and
  Foulkes]{Malone_Foulkes_JCP_2015}
Malone,~F.~D.; Blunt,~N.~S.; Shepherd,~J.~J.; Lee,~D. K.~K.; Spencer,~J.~S.;
  Foulkes,~W. M.~C. Interaction picture density matrix quantum Monte Carlo.
  \emph{J. Chem. Phys.} \textbf{2015}, \emph{143}, 044116\relax
\mciteBstWouldAddEndPuncttrue
\mciteSetBstMidEndSepPunct{\mcitedefaultmidpunct}
{\mcitedefaultendpunct}{\mcitedefaultseppunct}\relax
\EndOfBibitem
\bibitem[Kwee \latin{et~al.}(2008)Kwee, Zhang, and Krakauer]{Kwee_PRL_2008}
Kwee,~H.; Zhang,~S.; Krakauer,~H. Finite-Size Correction in Many-Body
  Electronic Structure Calculations. \emph{Phys. Rev. Lett.} \textbf{2008},
  \emph{100}, 126404\relax
\mciteBstWouldAddEndPuncttrue
\mciteSetBstMidEndSepPunct{\mcitedefaultmidpunct}
{\mcitedefaultendpunct}{\mcitedefaultseppunct}\relax
\EndOfBibitem
\bibitem[Ma \latin{et~al.}(2011)Ma, Zhang, and Krakauer]{Ma_PRB_2011}
Ma,~F.; Zhang,~S.; Krakauer,~H. Finite-size correction in many-body electronic
  structure calculations of magnetic systems. \emph{Phys. Rev. B}
  \textbf{2011}, \emph{84}, 155130\relax
\mciteBstWouldAddEndPuncttrue
\mciteSetBstMidEndSepPunct{\mcitedefaultmidpunct}
{\mcitedefaultendpunct}{\mcitedefaultseppunct}\relax
\EndOfBibitem
\bibitem[Dornheim \latin{et~al.}(2016)Dornheim, Groth, Sjostrom, Malone,
  Foulkes, and Bonitz]{dornheim2016ab}
Dornheim,~T.; Groth,~S.; Sjostrom,~T.; Malone,~F.~D.; Foulkes,~W.; Bonitz,~M.
  Ab initio Quantum Monte Carlo simulation of the warm dense electron gas in
  the thermodynamic limit. \emph{Phys. Rev. Lett.} \textbf{2016}, \emph{117},
  156403\relax
\mciteBstWouldAddEndPuncttrue
\mciteSetBstMidEndSepPunct{\mcitedefaultmidpunct}
{\mcitedefaultendpunct}{\mcitedefaultseppunct}\relax
\EndOfBibitem
\bibitem[Brown \latin{et~al.}(2013)Brown, Clark, DuBois, and
  Ceperley]{brown2013path}
Brown,~E.~W.; Clark,~B.~K.; DuBois,~J.~L.; Ceperley,~D.~M. Path-integral Monte
  Carlo simulation of the warm dense homogeneous electron gas. \emph{Phys. Rev.
  Lett.} \textbf{2013}, \emph{110}, 146405\relax
\mciteBstWouldAddEndPuncttrue
\mciteSetBstMidEndSepPunct{\mcitedefaultmidpunct}
{\mcitedefaultendpunct}{\mcitedefaultseppunct}\relax
\EndOfBibitem
\bibitem[He \latin{et~al.}(2019)He, Qin, Shi, Lu, and Zhang]{he2019finite}
He,~Y.-Y.; Qin,~M.; Shi,~H.; Lu,~Z.-Y.; Zhang,~S. Finite-temperature
  auxiliary-field quantum Monte Carlo: Self-consistent constraint and
  systematic approach to low temperatures. \emph{Phys. Rev. B} \textbf{2019},
  \emph{99}, 045108\relax
\mciteBstWouldAddEndPuncttrue
\mciteSetBstMidEndSepPunct{\mcitedefaultmidpunct}
{\mcitedefaultendpunct}{\mcitedefaultseppunct}\relax
\EndOfBibitem
\end{mcitethebibliography}


\begin{thebibliography}{10}%
\makeatletter
\providecommand \@ifxundefined [1]{%
 \@ifx{#1\undefined}
}%
\providecommand \@ifnum [1]{%
 \ifnum #1\expandafter \@firstoftwo
 \else \expandafter \@secondoftwo
 \fi
}%
\providecommand \@ifx [1]{%
 \ifx #1\expandafter \@firstoftwo
 \else \expandafter \@secondoftwo
 \fi
}%
\providecommand \natexlab [1]{#1}%
\providecommand \enquote  [1]{``#1''}%
\providecommand \bibnamefont  [1]{#1}%
\providecommand \bibfnamefont [1]{#1}%
\providecommand \citenamefont [1]{#1}%
\providecommand \href@noop [0]{\@secondoftwo}%
\providecommand \href [0]{\begingroup \@sanitize@url \@href}%
\providecommand \@href[1]{\@@startlink{#1}\@@href}%
\providecommand \@@href[1]{\endgroup#1\@@endlink}%
\providecommand \@sanitize@url [0]{\catcode `\\12\catcode `\$12\catcode
  `\&12\catcode `\#12\catcode `\^12\catcode `\_12\catcode `\%12\relax}%
\providecommand \@@startlink[1]{}%
\providecommand \@@endlink[0]{}%
\providecommand \url  [0]{\begingroup\@sanitize@url \@url }%
\providecommand \@url [1]{\endgroup\@href {#1}{\urlprefix }}%
\providecommand \urlprefix  [0]{URL }%
\providecommand \Eprint [0]{\href }%
\providecommand \doibase [0]{http://dx.doi.org/}%
\providecommand \selectlanguage [0]{\@gobble}%
\providecommand \bibinfo  [0]{\@secondoftwo}%
\providecommand \bibfield  [0]{\@secondoftwo}%
\providecommand \translation [1]{[#1]}%
\providecommand \BibitemOpen [0]{}%
\providecommand \bibitemStop [0]{}%
\providecommand \bibitemNoStop [0]{.\EOS\space}%
\providecommand \EOS [0]{\spacefactor3000\relax}%
\providecommand \BibitemShut  [1]{\csname bibitem#1\endcsname}%
\let\auto@bib@innerbib\@empty
%</preamble>
\bibitem [{\citenamefont {Liu}\ \emph {et~al.}(2018)\citenamefont {Liu},
  \citenamefont {Cho},\ and\ \citenamefont {Rubenstein}}]{Liu_JCTC_2018}%
  \BibitemOpen
  \bibfield  {author} {\bibinfo {author} {\bibfnamefont {Y.}~\bibnamefont
  {Liu}}, \bibinfo {author} {\bibfnamefont {M.}~\bibnamefont {Cho}}, \ and\
  \bibinfo {author} {\bibfnamefont {B.}~\bibnamefont {Rubenstein}},\ }\href
  {\doibase 10.1021/acs.jctc.8b00569} {\bibfield  {journal} {\bibinfo
  {journal} {J. Chem. Theory Comput.}\ }\textbf {\bibinfo {volume} {14}},\
  \bibinfo {pages} {4722} (\bibinfo {year} {2018})},\ \bibinfo {note} {pMID:
  30102856},\ \Eprint
  {http://arxiv.org/abs/https://doi.org/10.1021/acs.jctc.8b00569}
  {https://doi.org/10.1021/acs.jctc.8b00569} \BibitemShut {NoStop}%
\bibitem [{\citenamefont {L{\"o}wdin}(1950)}]{lowdin1950non}%
  \BibitemOpen
  \bibfield  {author} {\bibinfo {author} {\bibfnamefont {P.-O.}\ \bibnamefont
  {L{\"o}wdin}},\ }\href {https://aip.scitation.org/doi/abs/10.1063/1.1747632}
  {\bibfield  {journal} {\bibinfo  {journal} {J. Chem. Phys.}\ }\textbf
  {\bibinfo {volume} {18}},\ \bibinfo {pages} {365} (\bibinfo {year}
  {1950})}\BibitemShut {NoStop}%
\bibitem [{\citenamefont {McClain}\ \emph {et~al.}(2017)\citenamefont
  {McClain}, \citenamefont {Sun}, \citenamefont {Chan},\ and\ \citenamefont
  {Berkelbach}}]{McClain_JCTC_2017}%
  \BibitemOpen
  \bibfield  {author} {\bibinfo {author} {\bibfnamefont {J.}~\bibnamefont
  {McClain}}, \bibinfo {author} {\bibfnamefont {Q.}~\bibnamefont {Sun}},
  \bibinfo {author} {\bibfnamefont {G.~K.-L.}\ \bibnamefont {Chan}}, \ and\
  \bibinfo {author} {\bibfnamefont {T.~C.}\ \bibnamefont {Berkelbach}},\ }\href
  {\doibase 10.1021/acs.jctc.7b00049} {\bibfield  {journal} {\bibinfo
  {journal} {J. Chem. Theory Comput.}\ }\textbf {\bibinfo {volume} {13}},\
  \bibinfo {pages} {1209} (\bibinfo {year} {2017})},\ \bibinfo {note} {pMID:
  28218843},\ \Eprint
  {http://arxiv.org/abs/https://doi.org/10.1021/acs.jctc.7b00049}
  {https://doi.org/10.1021/acs.jctc.7b00049} \BibitemShut {NoStop}%
\bibitem [{\citenamefont {Malone}\ \emph {et~al.}(2015)\citenamefont {Malone},
  \citenamefont {Blunt}, \citenamefont {Shepherd}, \citenamefont {Lee},
  \citenamefont {Spencer},\ and\ \citenamefont
  {Foulkes}}]{Malone_Foulkes_JCP_2015}%
  \BibitemOpen
  \bibfield  {author} {\bibinfo {author} {\bibfnamefont {F.~D.}\ \bibnamefont
  {Malone}}, \bibinfo {author} {\bibfnamefont {N.~S.}\ \bibnamefont {Blunt}},
  \bibinfo {author} {\bibfnamefont {J.~J.}\ \bibnamefont {Shepherd}}, \bibinfo
  {author} {\bibfnamefont {D.~K.~K.}\ \bibnamefont {Lee}}, \bibinfo {author}
  {\bibfnamefont {J.~S.}\ \bibnamefont {Spencer}}, \ and\ \bibinfo {author}
  {\bibfnamefont {W.~M.~C.}\ \bibnamefont {Foulkes}},\ }\href {\doibase
  10.1063/1.4927434} {\bibfield  {journal} {\bibinfo  {journal} {J. Chem.
  Phys.}\ }\textbf {\bibinfo {volume} {143}},\ \bibinfo {pages} {044116}
  (\bibinfo {year} {2015})},\ \Eprint
  {http://arxiv.org/abs/https://doi.org/10.1063/1.4927434}
  {https://doi.org/10.1063/1.4927434} \BibitemShut {NoStop}%
\bibitem [{\citenamefont {Malone}\ \emph {et~al.}(2016)\citenamefont {Malone},
  \citenamefont {Blunt}, \citenamefont {Brown}, \citenamefont {Lee},
  \citenamefont {Spencer}, \citenamefont {Foulkes},\ and\ \citenamefont
  {Shepherd}}]{Malone_PRL_2016}%
  \BibitemOpen
  \bibfield  {author} {\bibinfo {author} {\bibfnamefont {F.~D.}\ \bibnamefont
  {Malone}}, \bibinfo {author} {\bibfnamefont {N.~S.}\ \bibnamefont {Blunt}},
  \bibinfo {author} {\bibfnamefont {E.~W.}\ \bibnamefont {Brown}}, \bibinfo
  {author} {\bibfnamefont {D.~K.~K.}\ \bibnamefont {Lee}}, \bibinfo {author}
  {\bibfnamefont {J.~S.}\ \bibnamefont {Spencer}}, \bibinfo {author}
  {\bibfnamefont {W.~M.~C.}\ \bibnamefont {Foulkes}}, \ and\ \bibinfo {author}
  {\bibfnamefont {J.~J.}\ \bibnamefont {Shepherd}},\ }\href {\doibase
  10.1103/PhysRevLett.117.115701} {\bibfield  {journal} {\bibinfo  {journal}
  {Phys. Rev. Lett.}\ }\textbf {\bibinfo {volume} {117}},\ \bibinfo {pages}
  {115701} (\bibinfo {year} {2016})}\BibitemShut {NoStop}%
\bibitem [{\citenamefont {White}\ and\ \citenamefont
  {Chan}(2018)}]{White_JCTC_2018}%
  \BibitemOpen
  \bibfield  {author} {\bibinfo {author} {\bibfnamefont {A.~F.}\ \bibnamefont
  {White}}\ and\ \bibinfo {author} {\bibfnamefont {G.~K.-L.}\ \bibnamefont
  {Chan}},\ }\href {\doibase 10.1021/acs.jctc.8b00773} {\bibfield  {journal}
  {\bibinfo  {journal} {J. Chem. Theory Comput.}\ }\textbf {\bibinfo {volume}
  {14}},\ \bibinfo {pages} {5690} (\bibinfo {year} {2018})},\ \bibinfo {note}
  {pMID: 30260642},\ \Eprint
  {http://arxiv.org/abs/https://doi.org/10.1021/acs.jctc.8b00773}
  {https://doi.org/10.1021/acs.jctc.8b00773} \BibitemShut {NoStop}%
\bibitem [{\citenamefont {Spencer}\ and\ \citenamefont
  {Alavi}(2008)}]{spencer2008efficient}%
  \BibitemOpen
  \bibfield  {author} {\bibinfo {author} {\bibfnamefont {J.}~\bibnamefont
  {Spencer}}\ and\ \bibinfo {author} {\bibfnamefont {A.}~\bibnamefont
  {Alavi}},\ }\href
  {https://journals.aps.org/prb/abstract/10.1103/PhysRevB.77.193110} {\bibfield
   {journal} {\bibinfo  {journal} {Physical Review B}\ }\textbf {\bibinfo
  {volume} {77}},\ \bibinfo {pages} {193110} (\bibinfo {year}
  {2008})}\BibitemShut {NoStop}%
\bibitem [{\citenamefont {Sun}\ \emph {et~al.}(2018)\citenamefont {Sun},
  \citenamefont {Berkelbach}, \citenamefont {Blunt}, \citenamefont {Booth},
  \citenamefont {Guo}, \citenamefont {Li}, \citenamefont {Liu}, \citenamefont
  {McClain}, \citenamefont {Sayfutyarova}, \citenamefont {Sharma} \emph
  {et~al.}}]{sun2018pyscf}%
  \BibitemOpen
  \bibfield  {author} {\bibinfo {author} {\bibfnamefont {Q.}~\bibnamefont
  {Sun}}, \bibinfo {author} {\bibfnamefont {T.~C.}\ \bibnamefont {Berkelbach}},
  \bibinfo {author} {\bibfnamefont {N.~S.}\ \bibnamefont {Blunt}}, \bibinfo
  {author} {\bibfnamefont {G.~H.}\ \bibnamefont {Booth}}, \bibinfo {author}
  {\bibfnamefont {S.}~\bibnamefont {Guo}}, \bibinfo {author} {\bibfnamefont
  {Z.}~\bibnamefont {Li}}, \bibinfo {author} {\bibfnamefont {J.}~\bibnamefont
  {Liu}}, \bibinfo {author} {\bibfnamefont {J.~D.}\ \bibnamefont {McClain}},
  \bibinfo {author} {\bibfnamefont {E.~R.}\ \bibnamefont {Sayfutyarova}},
  \bibinfo {author} {\bibfnamefont {S.}~\bibnamefont {Sharma}},  \emph
  {et~al.},\ }\href {https://onlinelibrary.wiley.com/doi/abs/10.1002/wcms.1340}
  {\bibfield  {journal} {\bibinfo  {journal} {Wiley Interdiscip. Rev. Comput.
  Mol. Sci.}\ }\textbf {\bibinfo {volume} {8}},\ \bibinfo {pages} {e1340}
  (\bibinfo {year} {2018})}\BibitemShut {NoStop}%
\bibitem [{\citenamefont {Stanescu}\ and\ \citenamefont
  {Phillips}(2001)}]{stanescu2001local}%
  \BibitemOpen
  \bibfield  {author} {\bibinfo {author} {\bibfnamefont {T.~D.}\ \bibnamefont
  {Stanescu}}\ and\ \bibinfo {author} {\bibfnamefont {P.}~\bibnamefont
  {Phillips}},\ }\href
  {https://journals.aps.org/prb/abstract/10.1103/PhysRevB.64.235117} {\bibfield
   {journal} {\bibinfo  {journal} {Physical Review B}\ }\textbf {\bibinfo
  {volume} {64}},\ \bibinfo {pages} {235117} (\bibinfo {year}
  {2001})}\BibitemShut {NoStop}%
\bibitem [{\citenamefont {Shiba}(1972)}]{shiba1972thermodynamic}%
  \BibitemOpen
  \bibfield  {author} {\bibinfo {author} {\bibfnamefont {H.}~\bibnamefont
  {Shiba}},\ }\href
  {https://academic.oup.com/ptp/article-abstract/48/6/2171/1857344} {\bibfield
  {journal} {\bibinfo  {journal} {Prog. Theor. Phys.}\ }\textbf {\bibinfo
  {volume} {48}},\ \bibinfo {pages} {2171} (\bibinfo {year}
  {1972})}\BibitemShut {NoStop}%
\end{thebibliography}%

\begin{figure}[ht]
\includegraphics[width=\textwidth]{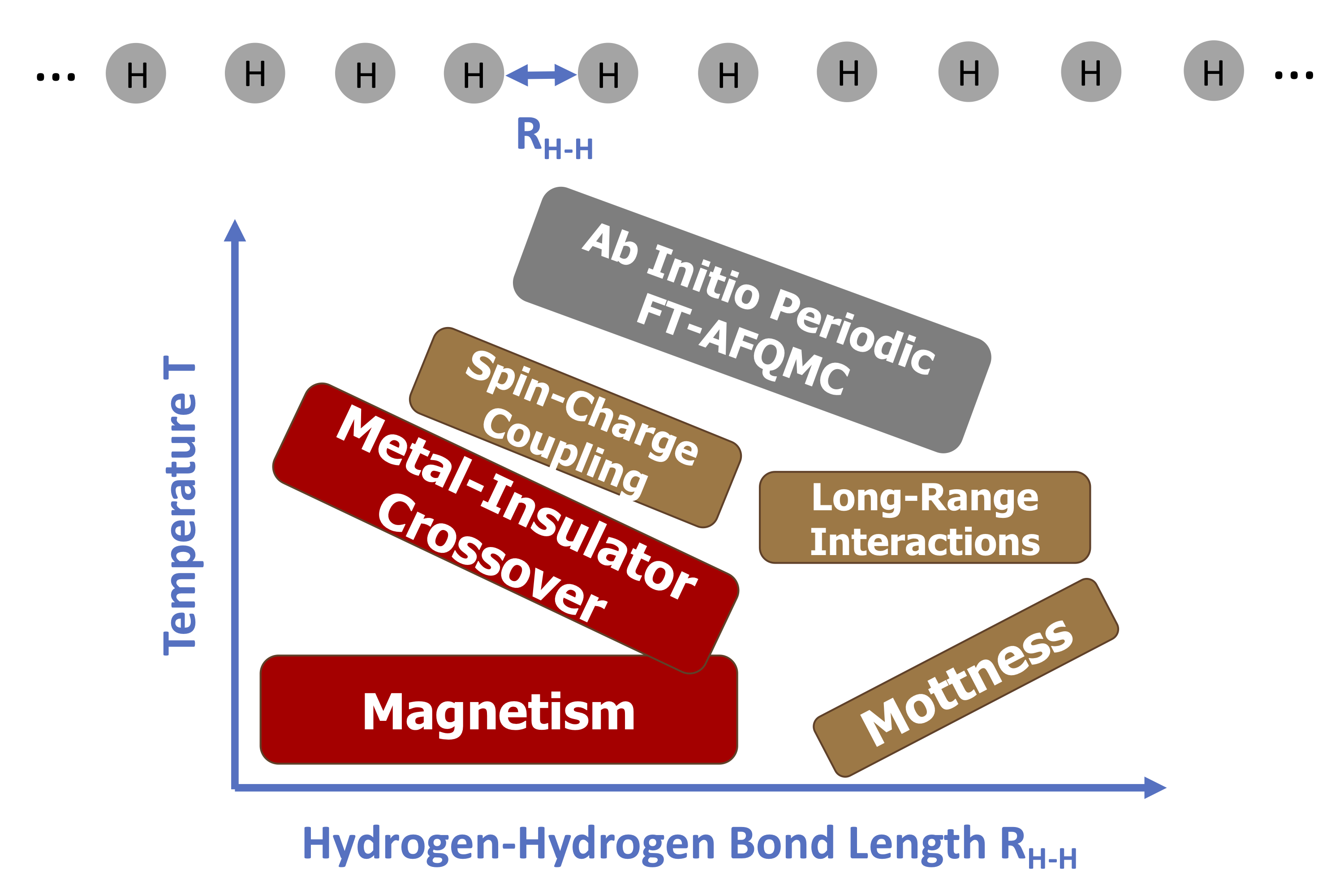}
\caption{Table of Contents (TOC)}
\end{figure}

\end{document}

%%Not Final Titles, Just Some Starters We Can Play With Depending Upon Final Results
\title{Supplemental Information for ``Unveiling the Finite Temperature Physics of Hydrogen Chains via Auxiliary Field Quantum Monte Carlo"}

\author{Yuan Liu}
\affiliation{Department of Chemistry, Brown University, Providence, RI 02912}
\author{Tong Shen}
\affiliation{Department of Chemistry, Brown University, Providence, RI 02912}
\author{Hang Zhang}
\affiliation{Department of Chemistry, Princeton University, Princeton, NJ 08544}
\author{Brenda Rubenstein}
\email{brenda\_rubenstein@brown.edu}
\affiliation{Department of Chemistry, Brown University, Providence, RI 02912}
%\date{\today}

\maketitle

% --------------------------------------------------------------------------------------

\section{Mean Field Hamiltonian and Mean Field Expressions for Thermodynamic Observables}

In order to obtain our mean field Hamiltonian and related mean field density matrices and observables, we follow the same approach described in our previous work.\cite{Liu_JCTC_2018} However, we provide a detailed discussion here in order to ease reproducibility. 
In conventional mean field theories, the mean field Hamiltonian is obtained by approximating all two-body contributions to the Hamiltonian in terms of products of mean-field and one-body contributions. That said, we approximate the two body portion of the \textit{ab initio} Hamiltonian given in Equation \eqref{Hamiltonian} in the main text by first rearranging the potential contribution into
\begin{equation}
\frac{1}{2} \sum'_{\substack{p\bm{k_p}q\bm{k_q} \\ m\bm{k_m}n\bm{k_n}\\ \alpha \beta}} V_{p\bm{k_p},q\bm{k_q},m\bm{k_m},n\bm{k_n}}^{\alpha\beta\alpha\beta} \hat{c}^{\dagger}_{p\bm{k_p}\alpha} \hat{c}^{\dagger}_{q\bm{k_q}\beta} \hat{c}_{n\bm{k_n}\beta} \hat{c}_{m\bm{k_m}\alpha} 
= \frac{1}{2} \sum'_{\substack{p\bm{k_p}q\bm{k_q} \\ m\bm{k_m}n\bm{k_n}\\ \alpha \beta}} V_{p\bm{k_p},q\bm{k_q},m\bm{k_m},n\bm{k_n}}^{\alpha\beta\alpha\beta} \hat{c}^{\dagger}_{p\bm{k_p}\alpha} \hat{c}_{m\bm{k_m}\alpha} \hat{c}^{\dagger}_{q\bm{k_q}\beta} \hat{c}_{n\bm{k_n}\beta}-\frac{1}{2}\sum_{\alpha}\hat{\rho}_{0}^{\alpha}, 
\end{equation}
where  
\begin{align}
    \hat{\rho}_{0}^{\alpha}=\sum_{m\bm{k_m}n\bm{k_n}} (\sum_{p\bm{k_p}} V_{m\bm{k_m},p\bm{k_p},p\bm{k_p},n\bm{k_n}}^{\alpha\alpha\alpha\alpha}) \hat{c}_{m\bm{k_m}\alpha}^{\dagger} \hat{c}_{n\bm{k_n}\alpha}.
\end{align}
A mean field density matrix, $\bar{D}_{p\bm{k_p}m\bm{k_m}}^{\alpha} = \langle \hat{c}_{p {\bm{k_p}} \alpha}^{\dagger} \hat{c}_{m {\bm{k_{m}}} \alpha} \rangle$, may then be subtracted from each corresponding pair of creation and annihilation operators in the potential to yield
\begin{align}
&\frac{1}{2} \sum'_{\substack{p\bm{k_p}q\bm{k_q} \\ m\bm{k_m}n\bm{k_n}\\ \alpha \beta}} V_{p\bm{k_p},q\bm{k_q},m\bm{k_m},n\bm{k_n}}^{\alpha\beta\alpha\beta} \hat{c}^{\dagger}_{p\bm{k_p}\alpha} \hat{c}_{m\bm{k_m}\alpha} \hat{c}^{\dagger}_{q\bm{k_q}\beta} \hat{c}_{n\bm{k_n}\beta} \nonumber \\
=&\frac{1}{2} \sum'_{\substack{p\bm{k_p}q\bm{k_q} \\ m\bm{k_m}n\bm{k_n}\\ \alpha \beta}} V_{p\bm{k_p},q\bm{k_q},m\bm{k_m},n\bm{k_n}}^{\alpha\beta\alpha\beta} (\hat{c}^{\dagger}_{p\bm{k_p}\alpha} \hat{c}_{m\bm{k_m}\alpha} - \bar{D}_{p\bm{k_p}m\bm{k_m}}^{\alpha}) (\hat{c}^{\dagger}_{q\bm{k_q}\beta} \hat{c}_{n\bm{k_n}\beta} - \bar{D}_{q\bm{k_q}n\bm{k_n}}^{\beta}) +\frac{1}{2}\sum_{\alpha}(\hat{\rho}_{MF1}^{\alpha}+\hat{\rho}_{MF2}^{\alpha})-\frac{1}{2}\hat{\rho}_{MF0}. 
\label{bgs}
\end{align}
In the above, $\hat{\rho}_{MF1}^{\alpha}$ and $\hat{\rho}_{MF2}^{\alpha}$ are one-body terms and $\hat{\rho}_{MF0}$ is a constant term
\begin{align}
\hat{\rho}_{MF1}^{\alpha}&=\sum_{p\bm{k_p}m\bm{k_m}}\bigg(\sum_{q\bm{k_q}n\bm{k_n}}\sum_{\beta}V_{p\bm{k_q},q\bm{k_q},m\bm{k_m},n\bm{k_n}}^{\alpha\beta\alpha\beta}\bar{D}_{q\bm{k_q}n\bm{k_n}}^{\beta}\bigg)\hat{c}_{p\bm{k_p}\alpha}^{\dagger} \hat{c}_{m\bm{k_m}\alpha},  \\
\hat{\rho}_{MF2}^{\alpha}&=\sum_{p\bm{k_p}m\bm{k_m}}\bigg(\sum_{q\bm{k_q}n\bm{k_n}}\sum_{\beta}V_{q\bm{k_q},p\bm{k_q},n\bm{k_n},m\bm{k_m}}^{\beta\alpha\beta\alpha}\bar{D}_{q\bm{k_q}n\bm{k_n}}^{\beta}\bigg)\hat{c}_{p\bm{k_p}\alpha}^{\dagger}\hat{c}_{m\bm{k_m}\alpha}; \\
\hat{\rho}_{MF0}&=\sum_{\alpha\beta}\sum_{\substack{p\bm{k_p}q\bm{k_q} \\ m\bm{k_m}n\bm{k_n}}} V_{p\bm{k_p},q\bm{k_q},m\bm{k_m},n\bm{k_n}}^{\alpha\beta\alpha\beta} \bar{D}_{p\bm{k_p}m\bm{k_m}}^{\alpha} \bar{D}_{q\bm{k_q}n\bm{k_n}}^{\beta}.
\end{align}

Invoking the mean field approximation means that we ignore the two-body term (the first term) in Equation \eqref{bgs}, which is a small contribution representing the deviation of the system from the mean field solution. Substituting the remaining terms for the second, potential energy term of Equation \eqref{Hamiltonian} in the main text yields the final mean field Hamiltonian
\begin{equation}
    \hat{H}_{MF}
    =\sum_{m\bm{k_m},n\bm{k_n},\alpha} T_{m\bm{k_m}\alpha,n\bm{k_n}\alpha} \hat{c}^{\dagger}_{m\bm{k_m}\alpha} \hat{c}_{n\bm{k_n}\alpha}
    -\frac{1}{2}\sum_{\alpha}\hat{\rho}_{0}^{\alpha}
    +\frac{1}{2}\sum_{\alpha}\big(\hat{\rho}_{MF1}^{\alpha}
    +\hat{\rho}_{MF2}^{\alpha}\big)-\frac{1}{2}\hat{\rho}_{MF0}.
\label{hmf}
\end{equation}
The one-body terms in the above may be combined with the one-body kinetic operator, while the constant term can be added at the end of any total energy calculation.  

This mean field Hamiltonian can be solved at finite temperature by starting with initial guesses for the mean field density matrices in the Hamiltonian, using them to construct a new Hamiltonian, $\hat{H}_{MF}$ given by Equation \eqref{hmf}, and then self-consistently solving for new mean field density matrices until they converge. Once self-consistency is achieved, we can use the resulting trial density matrices for performing background subtraction, or for directly calculating the mean field observables plotted in the text. Note that at low temperatures, the self-consistency loop may be difficult to converge. We use a linear density mixing scheme to improve convergence.

%----------------------------------------------------------------------------------
\section{Density Matrix Transformations from the Molecular Orbital to Orthogonal Atomic Orbital Basis}

The molecular orbital (MO) basis typically serves as a natural starting point for post-Hartree-Fock (HF) calculations, not only because molecular orbitals are eigenstates of the Hartree Fock Hamiltonian whose eigenvalues correspond to orbital energies, but also because they are orthonormal to each other. The energetic information MOs carry may be used as a criterion of the selection of active spaces in many post-Hartree Fock methods. However, to compute obersvables that are localized in real space, such as double occupancies, it is useful to represent them in a localized atomic orbital (AO) basis. The MOs are obtained by forming a linear combination of atomic orbitals (LCAO), where the MO coefficients can be used to transform the density matrix in the MO basis to the initial nonorthogonal AO basis (nAOs). Under these nAOs, however, it is difficult to represent local observables, such as site densities and correlation functions between local degrees of freedom, so in this work, we adopt a further transformation by L\"owdin \cite{lowdin1950non} to orthogonalize the nAOs, and calculate the site double occupancies and correlation functions in the orthogonal AO basis (oAOs). In this section, we give a brief description of the transformation from MOs to oAOs we employed. 

Let the $i$-th normalized nAOs, oAOs, and orthonormalized MOs be denoted by $\{\phi_i\}$, $\{\varphi_i\}$ and $\{\Phi_i\}$, respectively, where $i$ runs from 1 to $N$, and $N$ is the total number of basis functions. We further denote the overlap matrix between the nAOs $S_{ij}$ as
\begin{align}
    \langle \phi_i | \phi_j \rangle = S_{ij}. \label{overlap}
\end{align}
In the case of the orthogonal bases, this overlap matrix reduces to the identity matrix.

Due to their nonorthogonality, the identity operator for the nAOs becomes
\begin{align}
    \hat{I} = \sum_{ij} S^{-1}_{ij} |\phi_i \rangle \langle \phi_j|. \label{idnAO} 
\end{align}
%There are generally two ways to deal with nonorthogonal basis in second quantization, the direct basis approach and the dual basis approach. Here we adopt the %direct basis approach, where we modify the anticommutation relation between the creation and annihilation operators as
%\begin{align}
%    \{\hat{a}_{i}, \hat{a}_{j}^{\dagger}\} = S_{ij},
%    \label{commutation}
%\end{align}
%while preserving their Hermitian properties $(\hat{a}_i^{\dagger})^{\dagger} = \hat{a}_i$ which guarantees that the number operator $\hat{n}_i = %\hat{a}_{i}^{\dagger} \hat{a}_i$ is still Hermitian \cite{hu2015direct}. Here, $\hat{a}_{i}$ ($\hat{a}_{i}^{\dagger}$) annihilates (creates) one electron in AO %basis $\phi_{i}$.

By solving the self-consistent HF equations, we obtain the MOs as a linear combinations of the nAOs. The coefficient $C_{ij}$ for the $i$-th element of the $j$-th MO is defined by 
\begin{align}
    |\Phi_j \rangle = \sum_{i=1}^{N} C_{ij} |\phi_i \rangle. \label{lcao}
\end{align}

We now consider how to transform the matrix elements of the density matrix operator $\hat{\rho}$ in the MO basis into that in the nAO basis. By definition, we have
\begin{align}
    \hat{\rho} \equiv \sum_{ij} \rho_{ij}^{MO} | \Phi_i \rangle \langle \Phi_j|, \label{rho-mo}
\end{align}
and
\begin{align}
    \hat{\rho} \equiv \sum_{ij} \rho_{ij}^{nAO} | \phi_i \rangle \langle \phi_j|. \label{rho-nAO}
\end{align}
Inserting the identity operator defined by Equation \eqref{idnAO} into Equation \eqref{rho-mo} twice, yields
\begin{align}
    \hat{\rho} &= \sum_{ij} \rho_{ij}^{MO} \sum_{kl} S^{-1}_{kl} |\phi_k \rangle \langle \phi_l | \Phi_i \rangle \langle \Phi_j| \sum_{pq} S^{-1}_{pq} |\phi_p \rangle \langle \phi_q |.
\end{align}
Further substituting Equation \eqref{lcao} into the above, gives
\begin{align}
    \hat{\rho} = \sum_{ijklpq} \rho_{ij}^{MO} S^{-1}_{kl} S^{-1}_{pq} \langle \phi_l| \sum_{\mu} C_{\mu i} |\phi_{\mu} \rangle \sum_{\nu} C_{\nu j}^{*} \langle \phi_{\nu} | \phi_p \rangle |\phi_k \rangle \langle \phi_q |. \label{temp1}
\end{align}
By using the definition of the overlap matrix in Equation \eqref{overlap} as well as the identity $SS^{-1} = I$, after renaming the dummy indices, Equation \eqref{temp1} simplifies to
\begin{align}
    \hat{\rho} = \sum_{ij} (C\rho^{MO}C^{\dagger})_{ij} |\phi_i \rangle \langle \phi_j |. \label{rho-mo2nAO1}
\end{align}

Comparing Equation \eqref{rho-mo2nAO1} to Equation \eqref{rho-nAO}, we immediately arrive at
\begin{align}
    \rho^{nAO} = C\rho^{MO}C^{\dagger}. \label{rho-mo2nAO}
\end{align}
The above equation transforms the density matrix elements from the MO basis to the nAO basis.  

Next, we illustrate how to transform the density matrix element from the nAO basis to the oAO basis. We begin by defining a new set of basis functions  $\{\varphi_{i}\}$ by the transformation due to L\"owdin \cite{lowdin1950non}, 
\begin{align}
    |\varphi_{j} \rangle = \sum_{i} (S^{-\frac{1}{2}})_{ij} |\phi_{i} \rangle. \label{lowdin}
\end{align}

Based upon Equation \eqref{overlap}, it is easy to verify that the $\{\varphi_{i}\}$ defined in this way are orthonormal to each other, so we name them oAOs. The inverse transformation is given by
\begin{align}
    |\phi_{j} \rangle = \sum_{i} (S^{\frac{1}{2}})_{ij} |\varphi_{i} \rangle. \label{inverse-lowdin}
\end{align}

Substituting Equation \eqref{inverse-lowdin} into Equation \eqref{rho-nAO}, we obtain 
\begin{align}
    \hat{\rho} &\equiv \sum_{ij} \rho_{ij}^{nAO} \sum_{l} (S^{\frac{1}{2}})_{li} |\varphi_{l} \rangle \sum_{m} (S^{\frac{1}{2}})_{mj}^{*} \langle \varphi_{m} | \nonumber \\
               &=  \sum_{lm} \big[S^{\frac{1}{2}} \rho^{nAO} (S^{\frac{1}{2}})^{\dagger}\big]_{lm} |\varphi_{l} \rangle \langle \varphi_{m}|,
               \label{rho-nAO2oAO1}
\end{align}
where we used the definition of matrix multiplication.

Following Equation \eqref{rho-mo}, we can likewise define the density matrix operator in the oAO basis as
\begin{align}
    \hat{\rho} \equiv \sum_{ij} \rho_{ij}^{oAO} | \varphi_i \rangle \langle \varphi_j|. \label{rho-oAO}
\end{align}

Comparing Equations \eqref{rho-nAO2oAO1} and \eqref{rho-oAO}, we immediately obtain
\begin{align}
    \rho^{oAO} = S^{\frac{1}{2}} \rho^{nAO} (S^{\frac{1}{2}})^{\dagger}. \label{rho-nAO2oAO}
\end{align}

Lastly, by combining Equations \eqref{rho-mo2nAO} and \eqref{rho-nAO2oAO}, we arrive at the density matrix transformation relation between the MO and the oAO density matrices
\begin{align}
    \rho^{oAO} = [S^{\frac{1}{2}} C] \rho^{MO} [C^{\dagger} (S^{\frac{1}{2}})^{\dagger}]. \label{rho-MO2oAO}
\end{align}
It is this expression that was used throughout this work to compute local observables. 

%----------------------------------------------------------------------------------
\section{Brillouin Zone Sampling and Finite Size Effects}

Because we employ a p-GTO basis, our results are sensitive to how well we sample the Brillouin zone. There are two key sources of errors that arise from undersampling the Brillouin zone. 

Firstly, capturing the full variation of a system's charge density distribution, necessitates integrating over a system's full Brillouin zone; sampling a finite set of k-points will thus inherently introduce errors. At zero temperature, this error is completely determined by the ground state properties of the system. At finite temperature, excited states often contribute significantly to the overall properties of the system. If the excited states possess more nodes on their wave functions, this will lead to larger charge density variations. As a result, at finite, but low temperatures, the contribution to the charge density from the excited states may increase the Brillouin zone sampling error. On the other hand, thermal fluctuations tend to increase the kinetic energy of the electrons and smear out variations in the charge distribution. 
%\BR{the nodal structures of the charge distribution}. 
It is thus expected that with increasing temperature, thermal fluctuations will eventually lead to a more or less uniform charge distribution and reduced Brillouin zone sampling errors, even though many excited states may be involved. 

Secondly, the Hamiltonian matrix elements in our calculation are evaluated in the p-GTO basis by a Fourier transform over the crystal momentum $\bm{G}$ of the periodic hydrogen chains, as in Equations \eqref{pgto-ft1} and \eqref{pgto-ft2} in the main text. The matrix elements of the electron Coulomb repulsion and the electron-nuclear interactions both separately diverge at $\bm{G}=0$, yet sum to zero in the limit of infinite k-point sampling. It is therefore safe to ignore the contributions from the $\bm{G}=0$ term in this limit. However, when only a finite set of k-points is sampled, the cancellation at $\bm{G}=0$ will not be exact, resulting in an $O(N_k^{-1/3})$ finite size error in the exchange energy and an $O(N_k^{-1})$ error in the MP2 correlation energy at zero temperature as discussed in Reference \onlinecite{McClain_JCTC_2017}. \YL{Thorough studies of finite size corrections at finite temperature have thus far been conscribed to the uniform electron gas.\cite{Malone_Foulkes_JCP_2015,Malone_PRL_2016,White_JCTC_2018} These studies have demonstrated that finite size corrections are temperature-dependent, thus also suggesting that the errors that stem from our omission of the $\bm{G}=0$ term are temperature-dependent.} 

\YL{Because of this temperature-dependence}, accelerating Brillouin zone sampling at finite temperatures is not as simple as repurposing techniques used for the ground state.\cite{spencer2008efficient, sun2018pyscf} For this reason, we do not apply any finite size corrections in this work beyond increasing the number of k-points we sample. \YL{In this work, we uniformaly sample k-points in the first Brillouin zone to obtain the Monkhorst-Pack mesh. For our supercell containing 10 hydrogen atoms with an H-H bond length of a, the first Brillouin zone spans $[-\frac{\pi}{10a}, \frac{\pi}{10a}]$. Thus, the 5 k-points we sampled are $-\frac{\pi}{10a},-\frac{\pi}{20a},0,\frac{\pi}{20a},\frac{\pi}{10a}$. Sampling an odd number of k-points in this manner thus readily ensures that the Gamma point is sampled. In order to accommodate an even number of points, as were used to generate Figure \ref{fig:convergence}, we shifted the original Monkhorst-Pack grid so as to center it at the Gamma point. Due to this shift when even numbers of k-points are sampled, the negative and positive parts of the first Brillouin zone will not be sampled equally, leading to different convergence behaviors in the odd and even cases.} 

We leave an exposition of the correct finite size corrections for finite temperature simulations to a future publication.

\section{Expressions for the Spin and Charge Correlation Functions}
Here, we provide a detailed derivation of how our spin-spin and charge-charge correlation functions, $C_{ss}$ and $C_{cc}$, can be expressed in terms of the one-body density matrix in the oAO basis. From the definition of $C_{ss}$ given in the main text, we have
\begin{align}
    C_{ss}(i) &= \frac{1}{N_k^2} \sum'_{\substack{\bm{kk'} \\ \bm{k''k'''}}} \langle (\hat{n}_{0\bm{kk'}}^{\uparrow} - \hat{n}_{0\bm{kk'}}^{\downarrow})(\hat{n}_{i\bm{k''k'''}}^{\uparrow} - \hat{n}_{i\bm{k''k'''}}^{\downarrow}) \rangle \nonumber \\
    &= \frac{1}{N_k^2} \sum'_{\substack{\bm{kk'} \\ \bm{k''k'''}}} \bigg( \langle \hat{n}_{0\bm{kk'}}^{\uparrow} \hat{n}_{i\bm{k''k'''}}^{\uparrow} \rangle 
    + \langle \hat{n}_{0\bm{kk'}}^{\downarrow} \hat{n}_{i\bm{k''k'''}}^{\downarrow} \rangle
    - \langle \hat{n}_{0\bm{kk'}}^{\uparrow} \hat{n}_{i\bm{k''k'''}}^{\downarrow} \rangle 
    - \langle \hat{n}_{0\bm{kk'}}^{\downarrow} \hat{n}_{i\bm{k''k'''}}^{\uparrow} \rangle \bigg) \nonumber \\
    &= \frac{1}{N_k^2} \sum'_{\substack{\bm{kk'} \\ \bm{k''k'''}}}
    \bigg( 
          D_{0\bm{k},0\bm{k'}}^{\uparrow} D_{i\bm{k''},i\bm{k'''}}^{\uparrow} 
        + D_{0\bm{k},i\bm{k'''}}^{\uparrow}  G_{0\bm{k'},i\bm{k''}}^{\uparrow} 
        + D_{0\bm{k},0\bm{k'}}^{\downarrow} D_{i\bm{k''},i\bm{k'''}}^{\downarrow}
        + D_{0\bm{k},i\bm{k'''}}^{\downarrow}  G_{0\bm{k'},i\bm{k''}}^{\downarrow} \nonumber \\
        &- D_{0\bm{k},0\bm{k'}}^{\uparrow} D_{i\bm{k''},i\bm{k'''}}^{\downarrow}
        - D_{0\bm{k},0\bm{k'}}^{\downarrow} D_{i\bm{k''},i\bm{k'''}}^{\uparrow}
        \bigg) \nonumber \\
        &= \frac{1}{N_k^2} \sum'_{\substack{\bm{kk'} \\ \bm{k''k'''}}} \Big [
           (D_{0\bm{k},0\bm{k'}}^{\uparrow} - D_{0\bm{k},0\bm{k'}}^{\downarrow})
           (D_{i\bm{k''},i\bm{k'''}}^{\uparrow} - D_{i\bm{k''},i\bm{k'''}}^{\downarrow})
          + D_{0\bm{k},i\bm{k'''}}^{\uparrow}   G_{0\bm{k'},i\bm{k''}}^{\uparrow}
          + D_{0\bm{k},i\bm{k'''}}^{\downarrow} G_{0\bm{k'},i\bm{k''}}^{\downarrow}
          \Big ], 
\end{align}
where $D_{i\bm{k}j\bm{k'}}^{\sigma}$ and $G_{i\bm{k}j\bm{k'}}^{\sigma}$ are the matrix elements of the one-body density matrix and equal time Green's function,  respectively, as defined in the main text, and 
\begin{align}
    G_{i\bm{k}j\bm{k'}}^{\sigma} = \delta_{i\bm{k},j\bm{k'}} - D_{j\bm{k'}i\bm{k}}^{\sigma}.
\end{align}
Note that, because the atomic orbitals here have been orthogonalized, there is no ambiguity regarding the anti-commutation relation and the application of Wick's theorem here.

Similarly, for the first term in the charge-charge correlation function, $C_{cc}$, we have
\begin{align}
    C_{cc}(i)&= \frac{1}{N_k^2} \sum'_{\substack{\bm{kk'} \\ \bm{k''k'''}}} \langle (\hat{n}_{0\bm{kk'}}^{\uparrow} + \hat{n}_{0\bm{kk'}}^{\downarrow})(\hat{n}_{i\bm{k''k'''}}^{\uparrow} + \hat{n}_{i\bm{k''k'''}}^{\downarrow}) \rangle \nonumber \\
   &= \frac{1}{N_k^2} \sum'_{\substack{\bm{kk'} \\ \bm{k''k'''}}} \Big [
           (D_{0\bm{k},0\bm{k'}}^{\uparrow} + D_{0\bm{k},0\bm{k'}}^{\downarrow})
           (D_{i\bm{k''},i\bm{k'''}}^{\uparrow} + D_{i\bm{k''},i\bm{k'''}}^{\downarrow})
          + D_{0\bm{k},i\bm{k'''}}^{\uparrow}   G_{0\bm{k'},i\bm{k''}}^{\uparrow}
          + D_{0\bm{k},i\bm{k'''}}^{\downarrow} G_{0\bm{k'},i\bm{k''}}^{\downarrow}
          \Big ]. 
\end{align}

\section{\label{sec:cv-error} Error Propagation of the Heat Capacity}
\YL{
The heat capacity in Fig.~\ref{fig:cv} of the main text is obtained by interpolating the internal energy (as depicted in Fig.~\ref{fig:eng} of the main text) using a cubic spline and then taking derivatives of the interpolated analytical energy function with respect to temperature. The small fluctuations of the resulting heat capacity may originate from either the interpolation process or the statistical errors on our Monte Carlo-generated internal. 

To pinpoint the source of these small heat capacity fluctuations, we interpolated our internal energy data using different order polynomials and then calculated the corresponding heat capacity. Illustrative results for hydrogen chains with three different bond lengths are shown in Figs.~\ref{fig:cv0d5}-\ref{fig:cv1d5}. For each chain, we used first- through fourth-order polynomials (four total polynomials) to perform the interpolations. We can see that a linear (first-order) interpolation of the internal energy produces step-like behavior in the heat capacity plots. In contrast, the higher-order (second- through fourth-order) polynomial interpolations, result in small fluctuations in the heat capacities. These fluctuations are consistently reproduced regardless of the order of the polynomials used to interpolate. This suggests that the fluctuations we observe in our heat capacity curves are not artifacts of the interpolation process, but rather stem from the intrinsic Monte Carlo noise on the internal energy data.  

To better estimate the statistical error bars on the heat capacity curve, we thus need to propagate the errors on the internal energy data. This is accomplished by a convolution process. The question can be formulated in the following way. Consider two adjacent internal energy points at $T_1$ and $T_2$ in Fig.~\ref{fig:eng} as two random variables $X_1$ and $X_2$ both satisfying normal distributions $N(U_1, \sigma_1)$ and $N(U_2, \sigma_2)$. Then, the heat capacity $C_v$ in $[T_1, T_2]$, defined as their derivatives with respect to temperature, is given by 
\begin{align}
    C_v (T) = \frac{X_2 - X_1}{T_2 - T_1}, ~~T_1 < T < T_2.
\end{align}
$C_v$ is thus a new random variable and the question becomes: given the probability distributions of $X_1$ and $X_2$, how can the probability distribution of $C_v$, $P_{C_{v}}$, be determined? From probability theory, we know that the probability distribution of $X_2 - X_1$ is given by the convolution of the probability distributions of $X_1$ and $X_2$. In our case, this means the probability distribution $X_{2} - X_{1}$ is given by the convolution of $N(U_2, \sigma_2)$ and $N(-U_1, \sigma_1)$ as
\begin{align}
    N(U_2, \sigma_2) \circledast N(-U_1, \sigma_1) = N \bigg( U_2 - U_1, \sqrt{\sigma_1^2 + \sigma_2^2} \bigg).
\end{align}
To obtain $P_{C_{v}}$, this distribution must be divided by $T_{2}-T_{1}$
\begin{align}
    P_{C_v} = N \bigg( \frac{U_2 - U_1}{T_2 - T_1}, \frac{\sqrt{\sigma_1^2 + \sigma_2^2}}{T_2 - T_1} \bigg).
\end{align}
Note that in the above equation, the standard deviation is inversely proportional to the magnitude of $T_{2}-T_{1}$, meaning that it will diverge for infinitesimally small temperature intervals. It is based upon the above equation that we estimate the errors on the heat capacity curves given in Fig~\ref{fig:cv}. 
}

\section{Supplementary Tables}

\subsection{Errors in Mean Field Internal Energies} 
\FloatBarrier
\begin{table*}[h]
\begin{tabular}{C{0.8cm}|C{1.5cm}|C{1.5cm}|C{1.5cm}|C{1.5cm}|C{1.5cm}|C{1.5cm}|C{1.5cm}|C{1.5cm}|C{1.5cm}}
\hline \hline
{\bf T}  & {\bf 0.50 \AA} & {\bf 0.75 \AA} & {\bf 1.00 \AA} & {\bf 1.25 \AA} & {\bf 1.50 \AA} & {\bf 1.75 \AA} & {\bf 2.00 \AA} & {\bf 2.25 \AA} & {\bf 2.50 \AA} \\ \hline 
 {\bf 10}   & 0.0193 & 0.042  &  0.7254 & 0.0466 & 0.0295 & 0.0211 & 0.0232 & 0.0224 & 0.0238 \\            
 {\bf 9}    & 0.0212 & 0.045  &  0.6763 & 0.0513 & 0.0325 & 0.0231 & 0.0261 & 0.0250 & 0.0261 \\           
 {\bf 8}    & 0.0250 & 0.053  &  0.6368 & 0.0561 & 0.0355 & 0.0253 & 0.0288 & 0.0273 & 0.0293 \\           
 {\bf 7}    & 0.0281 & 0.062  &  0.5990 & 0.0625 & 0.0404 & 0.0290 & 0.0328 & 0.0309 & 0.0322 \\           
 {\bf 6}    & 0.0349 & 0.075  &  0.5623 & 0.0712 & 0.0460 & 0.0327 & 0.0376 & 0.0352 & 0.0372 \\           
 {\bf 5}    & 0.0428 & 0.092  &  0.5286 & 0.0821 & 0.0528 & 0.0388 & 0.0431 & 0.0406 & 0.0433 \\           
 {\bf 4.5}  & /      & 0.106  &  0.5146 & 0.0898 & 0.0578 & 0.0421 & 0.0467 & 0.0447 & 0.0476 \\           
 {\bf 4}    & 0.0576 & 0.123  &  0.4927 & 0.0974 & 0.0637 & 0.0462 & 0.0518 & 0.0492 & 0.0520 \\           
 {\bf 3.5}  & /      & 0.150  &  0.4894 & 0.1061 & 0.0712 & 0.0518 & 0.0586 & 0.0551 & 0.0575 \\           
 {\bf 3}    & 0.0844 & 0.187  &  0.4626 & 0.1173 & 0.0785 & 0.0581 & 0.0652 & 0.0612 & 0.0651 \\           
 {\bf 2.5}  & 0.1072 & 0.251  &  0.4381 & 0.1309 & 0.0908 & 0.0664 & 0.0747 & 0.0702 & 0.0742 \\           
 {\bf 2}    & 0.1470 & 0.404  &  0.4139 & 0.1499 & 0.1055 & 0.0783 & 0.0889 & 0.0828 & 0.0877 \\           
{\bf 1.6}   & 0.2049 & 0.792  &  0.3990 & 0.1679 & 0.1206 & 0.0901 & 0.1024 & 0.0952 & 0.1025 \\           
{\bf 1.3}   & 0.2987 & 3.431  &  0.3785 & 0.1841 & 0.1377 & 0.1033 & 0.1162 & 0.1103 & 0.1163 \\
{\bf 1.2}   & 0.3448 & 18.900 &  /      & /      &  /     &  /     &    /   &    /   &   /    \\
{\bf 1.1}   & 0.4379 & 2.482  &  /      & /      &  /     &  /     &    /   &    /   &   /    \\
{\bf 1.0}   & 0.5707 & 1.234  &  0.3354 & 0.1992 & 0.1575 & 0.1189 & 0.1361 & 0.1284 & 0.1353 \\           
{\bf 0.9}   & 0.8909 & 0.796  &  0.3225 & 0.2054 & 0.1621 & 0.1263 & 0.1450 & 0.1376 & 0.1433 \\           
{\bf 0.8}   & 2.1813 & 0.569  &  0.3049 & 0.2063 & 0.1723 & 0.1327 & 0.1539 & 0.1458 & 0.1534 \\           
{\bf 0.7}   & 2.7833 & 0.438  &  0.2905 & 0.2170 & 0.1799 & 0.1418 & 0.1645 & 0.1564 & 0.1627 \\           
{\bf 0.6}   & 0.7432 & 0.327  &  0.2692 & 0.2126 & 0.1894 & 0.1499 & 0.1760 & 0.1667 & 0.1753 \\           
{\bf 0.5}   & 0.3880 & 0.242  &  0.2361 & 0.2094 & 0.1935 & 0.1557 & 0.1848 & 0.1814 & 0.1857 \\           
{\bf 0.4}   & 0.2328 & 0.181  &  0.2030 & 0.1975 & 0.1941 & 0.1655 & 0.1952 & 0.1941 & 0.1951 \\           
{\bf 0.3}   & 0.1370 & 0.125  &  0.1517 & 0.1730 & 0.1946 & 0.1643 & 0.2035 & 0.1976 & 0.2066 \\           
{\bf 0.2}   & 0.0719 & 0.077  &  0.1004 & 0.1295 & 0.1536 & 0.1476 & 0.2019 & 0.1992 & 0.2090 \\           
{\bf 0.1}   & 0.0343 & 0.036  &  0.0491 & 0.0643 & 0.0931 & 0.1019 & 0.1687 & 0.1845 & 0.2038 \\           
{\bf 0.05}  & 0.0317 & 0.023  &  0.0276 & 0.0377 & 0.0553 & 0.0926 & 0.1316 & 0.1184 & 0.1130 \\
\hline \hline
\end{tabular}
\caption{Relative errors of the internal energy calculated using mean field theory relative to the exact FT-AFQMC energies, $|\frac{U_{\text{MFT}} - U_{\text{AFQMC}}}{U_{\text{AFQMC}}}|$, where $U_{\text{MFT}}$ and $U_{\text{AFQMC}}$ are the internal energies caluclalated from MFT and FT-AFQMC, respectively.} 
\label{tab:mft-error}
\end{table*}
\FloatBarrier

\subsection{Characteristic Spin and Charge Excitation Temperatures} 
\FloatBarrier
\begin{table*}[h]
\begin{tabular}{c|c|c}
\hline \hline
{\bf U/t}  & {\bf $T_{spin}^{*}/t$} & {\bf $T_{charge}^{*}/t$ }\\ \hline
1.0  &  /   & 0.61 \\
2.0  &  /   & 0.60 \cite{stanescu2001local} \\
4.0  & 0.37 & 1.0 \footnote{The spin and charge excitation peak are too close to be distinguished, the value 1.0 is a reasonable estimation.}   \\
6.0  & 0.23 & 1.18 \\
8.0  & 0.21 & 1.61 \\
12.0 & 0.14 & 2.50 \\
\hline \hline
\end{tabular}
\caption{Characteristic temperatures of the spin and charge excitations in the one-dimensional Hubbard model. Unless otherwise specified, all data is taken from Ref.  \cite{shiba1972thermodynamic}}
\end{table*}
\FloatBarrier

\section{Supplementary Figures}

\YL{
\subsection{Eigenvalue Spectrum of the One-Body Mean Field Hamiltonian for the Spin Up Sector}
\begin{figure}[H]
\centering{ 
\includegraphics[width=0.55\textwidth]{sfigure_eig-mft-T0.5.jpg}}
\caption{\color{black} Eigenvalue spectrum of the one-body Hamiltonian for the spin up sector from mean field theory at all bond lengths studied in the main text. The spin down eigenvalue spectrum is similar.}
\label{fig:eig-mft-T0.5}
\end{figure}

\subsection{Ground State Hartree-Fock Energy Convergence with Respect to the Vacuum Space in the Unit Cell}
\begin{figure}[H]
\centering{ 
\includegraphics[width=0.55\textwidth]{sfigure_vac-conv-hf.jpg}}
\caption{\color{black} Convergence of the ground state Hartree-Fock energy with respect to the vacuum space along the lateral dimensions of the $R=1.0$ \AA~ hydrogen chain unit cell. The energy appears to converge with around 10 \AA~of vacuum. While our calculations were performed at finite temperatures, this plot is a reasonable indicator of the vacuum space needed within our cells.}
\label{fig:convergence-vacuum-HF}
\end{figure}
}

\subsection{Convergence of the Internal Energy and Double Occupancy as a Function of the Number of k-Points}

\begin{figure}[H]
\centering{ 
\includegraphics[width=0.55\textwidth]{sfigure_kconv_T1.jpg}}
\caption{Internal energy and double occupancy of the hydrogen chain at $T = 1$ Hartree/$k_{B}$ with H-H bond lengths of 0.75 (red empty squares) and 1.75 (blue dot crosses) \AA~ as a function of the number of k-points sampled in the first Brillouin zone. The overall convergence rate is slow and the convergence is not fully established with 7 k-points. }
\label{fig:convergence-T1}
\end{figure}

\YL{
\begin{figure}[H]
\centering{ 
\includegraphics[width=0.55\textwidth]{sfigure_kconv-T0d1.jpg}}
\caption{\color{black} Internal energy and double occupancy of the hydrogen chain at $T = 0.1$ Hartree/$k_{B}$ with H-H bond lengths of 0.75 (red) and 1.75 (blue) \AA~ as a function of the number of k-points sampled in the first Brillouin zone. We have drawn separate lines through the data produced using odd and even numbers of k-points to illustrate the  convergence properties of odd and even k-point data.}
\label{fig:convergence-T0d1-oddeven}
\end{figure}
}

\subsection{Internal Energy {vs.} Temperature from Finite Temperature AFQMC}
\begin{figure}[H]
\centering{
\includegraphics[width=0.55\textwidth]{sfigure_eng-afqmc.jpg}}
\caption{Internal energy of the hydrogen chains for different H-H bond lengths over a range of temperatures from FT-AFQMC. Different colors correspond to different bond lengths. Note that the horizontal temperature axis is plotted on a log scale in order to emphasize the low-temperature behavior. The inset highlights the crossover regime at intermediate temperatures and is plotted on a linear scale.}
\label{fig:eng-afqmc}
\end{figure}

\subsection{Internal Energy {vs.} Temperature from Mean Field Theory}
\begin{figure}[H]
\centering{ 
\includegraphics[width=0.55\textwidth]{sfigure_eng-mft.jpg}}
\caption{Internal energy of the hydrogen chains for different H-H bond lengths over a range of temperatures from MFT. Different colors correspond to different bond lengths. Note that the horizontal temperature axis is plotted on a log scale in order to emphasize the low-temperature behavior. The inset highlights the crossover regime at intermediate temperatures and is plotted on a linear scale.}
\label{fig:eng-mft}
\end{figure}

\YL{
\subsection{Correlation Energy {vs.} Temperature}
\begin{figure}[H]
\centering{ 
\includegraphics[width=0.55\textwidth]{sfigure_eng-corr.jpg}}
\caption{\color{black} Correlation energy of the hydrogen chains for different H-H bond lengths over a range of temperatures. We define the correlation energy in this study as the difference between the mean field and AFQMC internal energies, $U_{\text{MFT}} - U_{\text{AFQMC}}$. Note that the horizontal temperature axis is plotted on a log scale in order to emphasize the low-temperature behavior. It is evident that the largest discrepancies occur at intermediate temperatures.}
\label{fig:eng-corr}
\end{figure}
}

\YL{
\subsection{Heat Capacity Obtained from Interpolating Internal Energies Using Polynomial Splines of Different Orders}
\label{ssec:cv-interp-compare}

\begin{figure}[H]
\centering{
\includegraphics[width=0.7\textwidth]{sfigure_cv0d5.jpg}}
\caption{\color{black}Heat capacity of the 0.5 \AA~chain obtained by interpolating the internal energy using 1st- through 4th- order polynomial splines.}
\label{fig:cv0d5}
\end{figure}

\begin{figure}[H]
\centering{
\includegraphics[width=0.7\textwidth]{sfigure_cv1d25.jpg}}
\caption{\color{black} Heat capacity of the 1.25 \AA~chain obtained by interpolating the internal energy using 1st- through 4th- order polynomial splines.}
\label{fig:cv1d25}
\end{figure}

\begin{figure}[H]
\centering{
\includegraphics[width=0.7\textwidth]{sfigure_cv1d5.jpg}}
\caption{\color{black} Heat capacity of the 1.5 \AA~chain obtained by interpolating the internal energy using 1st- through 4th- order polynomial splines.}
\label{fig:cv1d5}
\end{figure}
}

\subsection{Spin-Spin Correlation Function from Mean Field Theory}
\begin{figure}[H]
\centering{
\includegraphics[width=0.8\textwidth]{sfigure_ss_mft.jpg}}
\caption{Spin-spin correlation function $C_{ss}$ obtained from mean field theory at $T = 0.05$, 0.1, and 1 Hartree for various bond lengths. Note the unphysical behavior of the spin-spin correlation function in the strongly correlated regime for H-H distances greater than 1.75 \AA.}
\label{fig:ss-mft}
\end{figure}

\subsection{Charge-Charge Correlation Function from Mean Field Theory}
\begin{figure}[H]
\centering{ 
\includegraphics[width=0.8\textwidth]{sfigure_cc_mft.jpg}}
\caption{Charge-charge correlation function $C_{cc}$ obtained from mean field theory at $T$ = 0.05, 0.1, and 1 Hartree for various bond lengths. Note the unphysical behavior of the charge-charge correlation function in the strongly correlated regime for H-H distances greater than 1.75 \AA.}
\label{fig:cc-mft}
\end{figure}

\bibstyle{apsrev4-1}
\bibliography{ref}